\pdfoutput=1 
\newif\ifdraft\draftfalse
\newif\ifanon\anonfalse
\newif\ifaclanonsubmission\aclanonsubmissionfalse
\documentclass[11pt]{article}
\usepackage[preprint]{acl}

\usepackage{times}
\usepackage{latexsym}
\usepackage[T1]{fontenc}
\usepackage[utf8]{inputenc}
\usepackage{microtype}
\usepackage{inconsolata}
\usepackage{graphicx}
\usepackage[inline]{enumitem}

\makeatletter
\@ifpackagewith{acl}{preprint}{
    
    \anonfalse
}{}

\@ifpackagewith{acl}{final}{
    \errmessage{Do not use until paper is accepted.}
    \anonfalse
}{}

\@ifpackagewith{acl}{review}{
    
    \anontrue
    \aclanonsubmissiontrue
}{}
\makeatother

\usepackage{xspace}
\usepackage{cleveref}
\usepackage{booktabs} 
\usepackage{listings}
\usepackage{siunitx} 
\usepackage[normalem]{ulem} 
\usepackage{tikz}
\usepackage{circledsteps}
\usepackage{subcaption} 
\usepackage{colortbl} 
\usepackage{amsfonts} 

\lstdefinestyle{codeblock}{
    basicstyle=\ttfamily\footnotesize,
    commentstyle=\color{red!60!black},
    keywordstyle=\color{green!50!black},
    stringstyle=\color{red!60!black},
    basicstyle=\ttfamily\footnotesize,
    breakatwhitespace=false,         
    breaklines=true,                 
    captionpos=b,                    
    keepspaces=true,                 
    showspaces=false,                
    showstringspaces=false,
    showtabs=false,   
    aboveskip=0pt,
    belowskip=0pt,
    tabsize=2,
    numbers=none,
    escapechar={@},
}

\definecolor{wellesleyblue}{RGB}{0, 39, 118}
\definecolor{oberlinred}{RGB}{207,16,45}
\definecolor{brownbrown}{RGB}{78,54,41}

\newcommand{\studenteval}{\textsc{StudentEval}\xspace}

\Crefname{section}{\S}{\S\S}
\crefformat{section}{\S#2#1#3}
\Crefname{figure}{Figure}{Figures}
\crefformat{figure}{Figure #2#1#3}
\Crefname{Figure}{Figure}{Figures}
\Crefname{Table}{Table}{Tables}
\crefformat{table}{Table #2#1#3}

\newcommand{\Exec}{\textsc{Exec}}
\newcommand{\ok}{o_{\textsc{ok}}}
\newcommand{\Nat}{\mathbb{N}}

\title{Substance Beats Style: Why Beginning Students Fail to Code with LLMs}
\author{Francesca Lucchetti \\ Northeastern University
\And Zixuan Wu \\ Wellesley College \\
\AND Arjun Guha \\ Northeastern University \\
\And Molly Q Feldman \\ Oberlin College \\
\And Carolyn Jane Anderson \\ Wellesley College}

\begin{document}
\maketitle
\begin{abstract}

Although LLMs are increasing the productivity of professional programmers, existing work shows that beginners struggle to prompt LLMs to solve text-to-code tasks~\cite{nguyen_how_2024,prather_widening_2024,mordechai_novicode_2024}. Why is this the case? This paper explores two competing hypotheses about the cause of student-LLM miscommunication: (1)~students simply lack the technical vocabulary needed to write good prompts, and (2)~students do not understand the extent of information that LLMs need to solve code generation tasks. We study (1) with a causal intervention experiment on technical vocabulary and (2) by analyzing graphs that abstract how students edit prompts and the different failures that they encounter. We find that substance beats style: a poor grasp of technical vocabulary is merely correlated with prompt failure; that the information content of prompts predicts success; that students get stuck making trivial edits; and more. Our findings have implications for the use of LLMs in programming education, and for efforts to make computing more accessible with LLMs.

\end{abstract}

\section{Introduction}

There is a growing body of evidence that large language models (LLMs) are increasing the productivity of professional programmers~\cite{etsenake_understanding_2024}. At the same time, previous work shows that students struggle to leverage LLMs in programming across a variety of tasks and models~\cite{nguyen_how_2024, prather_widening_2024, mordechai_novicode_2024}. But why is this the case?

Prior work has reported on students' and instructors' perception of why student-LLM interactions go wrong, positing many explanations including unfamiliarity with technical vocabulary~\citep{nguyen_how_2024,feldman_non-expert_2024,mordechai_novicode_2024, prather_widening_2024}, model non-determinism~\citep{lau_ban_2023,vadaparty_cs1_2024}, and trouble understanding LLM output~\citep{nguyen_how_2024,vadaparty_cs1_2024}. However, there is little quantitative evidence about these  potential sources of miscommunication.

In this paper, we test two competing hypotheses about the cause of student-LLM miscommunication. One possibility is that students provide all of the information that the model needs, but use language that models cannot understand. Non-expert programmers talk about code differently than experts, leading to problems for models trained largely on expert code. A second possibility is that students do not understand what information a model needs to solve a given problem. Writing prompts involves decisions about what information the model may be able to infer from pretraining versus what information must be stated directly in the prompt. These decisions may be more challenging for students to make, since they do not yet have a strong sense of what information code typically contains. 

This paper tests the impact of these potential error sources in two sets of experiments on a dataset of 1,749 prompts authored by 80 students~\citep{babe_studenteval_2024}. To isolate the effect of linguistic variation, we conduct a causal analysis of lexical choices for technical terminology by replacing them with near-synonyms used by students. To study information selection, 
we annotate series of prompts in student problem-solving attempts with problem-specific ``clues,'' or information that describes the intended behavior of generated code. 

Overall, our findings reveal that student-LLM coding difficulties spring from challenges in selecting relevant information rather than challenges with technical vocabulary. Our study of the information content of prompts shows that prompts with missing clues almost always fail. Moreover, students typically get ``stuck'' in cycles because they make trivial edits to prompts instead of changing their information content. Our causal analysis of prompt wording finds relatively weak effects of modifying technical terminology. Although certain substitutions can hurt prompt success rates, correcting non-standard terminology rarely improves them. This suggests that the relationship between technical vocabulary and prompt success is more correlational than causal.

Taken together, our results provide empirical evidence that the information content of student prompts is what matters, rather than their (mis)use of technical vocabulary. 
These findings have strong implications for the use of LLMs in programming education and, more broadly, for efforts to broaden the accessibility of computing with LLMs.

\section{Related Work}

As the use of LLMs for programming has become widespread, the question of prompt wording has become increasingly important. Early work revealed high sensitivity to prompt wording on programs~\citep{white_prompt_2023,doderlein_piloting_2023}, which has efficiency implications~\cite{mozannar_reading_2024}.
Several techniques address prompt wording~\cite{strobelt_interactive_2022, oppenlaender_taxonomy_2023, zamfirescu-pereira_why_2023,ma_what_2024}. \citet{liu_what_2023} take a user-centered approach to teaching strategies for prompting. \citet{doderlein_piloting_2023} study keyword removal and replacement. \citet{zhu-tian_sketch_2024} generate program sketches from Python keywords in prompts. \citet{xia2024top} automatically reword existing task descriptions for more robust code generation benchmarks.

\paragraph{Novice Programmers and LLMs.} LLMs have have sparked much discussion in computing education~\citep{finnie-ansley_robots_2022}. 
There is a growing body of work studying how students use LLMs in computing classes~\cite{zamfirescu-pereira_why_2023, prather_its_2023, kazemitabaar_studying_2023,  denny_promptly_2023, mordechai_novicode_2024,vadaparty_cs1_2024}.
A convergent finding is that students struggle to leverage LLMs. Many potential explanations have been advanced: \citet{lau_ban_2023}'s study of CS educators discusses model non-determinism as a barrier; \citet{prather_interactions_2024} explores the cognitive load imposed by code suggestions; and \citet{kazemitabaar_how_2023} describe over-reliance on the model. Finally, multiple studies posit that technical language is a barrier between students and LLMs~\citep{nguyen_how_2024,feldman_non-expert_2024,mordechai_novicode_2024, prather_widening_2024}. 

This paper uses the dataset by \citet{babe_studenteval_2024}, which contains 1,749 prompts from students who have completed one college programming course. \citet{babe_studenteval_2024} turn their dataset into a benchmark 
to measure LLM performance on novice-written prompts. They report some correlations between technical terms and prompt success. \citet{nguyen_how_2024} study student experiences during the experiment, including students' self-perceptions of why the task is challenging: they highlight prompt wording as a key student-perceived barrier. This is reaffirmed in 
\citet{feldman_non-expert_2024}'s replication with students with no coding experience.



\begin{figure*}
\small
\centering
\begin{tabular}{l}
\textbf{Function signature} \\ 
\begin{lstlisting}[language=Python,style=codeblock]
def total_bill(grocery_list, sales_tax):
\end{lstlisting} \\ 
\textbf{Tests} \\
\begin{lstlisting}[language=Python,style=codeblock]
total_bill([['eggs', 6, 0.99], ['milk', 1, 1.49], ['bread', 2, 3.5]], 0.07) # 15.44
total_bill([['eggs', 6, 0.99], ['milk', 1, 1.49], ['bread', 2, 3.50]], 0.0) # 14.43
total_bill([['bread', 2, 3.50]], 0.5) # 10.5
\end{lstlisting} \\
\textbf{Docstring Attempt 1} (generated code fails some tests) \\
\begin{minipage}[t]{0.95\textwidth}
\textit{you will have two inputs a list of lists and the tax rate. for every list in the list of lists multiply the second and third item and add all of them and then multiply that by the sales tax plus 1}
\end{minipage} \\
\textbf{Docstring Attempt 2} (generated code passes all tests) \\ 
\begin{minipage}[t]{0.95\textwidth}
\textit{you will have two inputs a list of lists and the tax rate. for every list in the list of lists multiply the second and third item and add all of them and then multiply that by the sales tax plus 1\uline{. if the resulting number has more than two decimal places shorten it to two decimal places.}}
\end{minipage} \\
\end{tabular}
    
\caption{An example problem that a student solves in two attempts. Given the function signature and tests, they write the first docstring. The platform prompts the model to generate the function body from the function signature and docstring (not the tests), and then tests the generated code. From the failed tests, the student realizes that the model needs to be told to round to two decimal places. They add this \uline{clue} in the second prompt, which succeeds.}
\label{dataset-example}

\end{figure*}

\paragraph{Prompting Effects in Generative Models.}

There is a large set of existing work exploring the effect of different prompting techniques for LLMs more broadly. Prior work has shown that models are surprisingly robust to misleading, corrupted, or irrelevant prompts~\citep{webson-pavlick-2022-prompt,min-etal-2022-rethinking,madaan-etal-2023-makes,ye2022the,khashabi-etal-2022-prompt,wang-etal-2023-towards}. In this light, the documented issues that novice programmers experience when working with LLMs for programming are surprising. Our work may help to reconcile these two bodies of work by exploring the cause of student-LLM miscommunications.

\paragraph{Terminology in Other Generative Domains.}The impact of prompt terminology has been studied in non-code domains. For text generation, previous work has studied prompting techniques to control style \cite{yeh_ghostwriter_2024,raheja_coedit_2023}. 
Text-to-image models are very sensitive to choices in keywords~\citep{liu_design_2022}, limiting their usability for some applications~\citep{tseng_keyframer_2024} and users~\cite{chang_editscribe_2024}.  

\section{Dataset}
\label{dataset}

Our goal is to understand what it is about student-written prompts that makes them less effective for LLM code generation. We use the \studenteval{} dataset released by \citet{babe_studenteval_2024}, who use a subset of their data to benchmark LLMs for code generation. Unlike many datasets of programming prompts, this dataset contains many different prompts per task, including multiple submissions by the same author, allowing us to explore both wording choices and how the information content of prompts is edited during a prompting session.\footnote{The dataset contains sequences of prompt-edits, but their benchmark uses only the first/last prompt by each student.}

The dataset contains 1,749 prompts written by 80 students who had completed exactly one programming course. They were asked to complete problems drawn from a set of 48 CS1 programming tasks exercising a range of programming concepts. The dataset was collected in a prompting experiment that worked as follows (\cref{dataset-example}): (1)~the student was shown 3-5 test cases and asked to write a Python docstring for the function; (2)~the experimental platform prompted an LLM (\emph{code-davinci-002}) to generate a Python function, conditioned on the function signature and the student-written docstring; (3)~the experimental platform tested the generated function on the provided tests; and (4)~the student could try again or give up and move on to the next problem. Each student did 8 problems.


We use different subsets of the \studenteval{} dataset to explore our research questions. To study the effect of information content on prompt success,
we consider problems where at least five students submitted multiple times (33 tasks). To study the  effect of prompt wording, we select a lexically diverse subset by taking each student's first and last prompts per problem (953 prompts). 

\section{Methods}

Our work explores the impact of two potential causes of student-LLM miscommunication: how students word their prompts, and how students select information to include in their prompts.

\subsection{Measuring the Impact of Prompt Wording}
\label{methods_style}
To understand how students' wording of prompts affects model performance, we use a counterfactual causal inference approach.
We systematically measure the impact of wording related to what \citet{mordechai_novicode_2024} refer to as the ``structured language'' that experts use ``to describe the logical control flows within the desired program.'' We define a set of key programming concepts and systematically substitute alternative terms used by students to measure how the success of their prompt would have been impacted by alternative wording. 

\subsubsection{Tagging Concept References}
We select 12 key technical concepts that occur frequently in the \studenteval{} dataset, including references to data types (e.g., list, string, dictionary), operations on data (e.g., concatenate, append, typecast), and terms related to data flow and control flow (e.g., input, loop, return). 

For each concept, two expert annotators identified every lexical variation used to refer to these concepts in the prompts. The tag set includes tags for all morphological variants of a given lemma, to ensure that the substitutions match the capitalization and tense of the original terms. In addition, three sets of tags were used for terms referring to function input, to capture different syntactic structures. 
The full tag set contains 78 tags for 14 category lemmas. See \Cref{lexical-appendix} for the annotation procedure and all lemmas.

Overall, references to these concepts appear 4,262 times across the dataset. Collapsing variations of the same lemma within a prompt (e.g., ``string'',``strings''), we find that the median number of technical terms per prompt is three and the maximum is ten. 
\Cref{fig:substitution-example} shows an example of how three concept references in a prompt get tagged.

\begin{figure}
    \centering
    \begin{flushleft} 
    \small 
    \textbf{Original:}\textit{ 
    Convert the input into integers and check if it is a prime number. 
    } \\
    \textbf{Tagged:} \textit{
    \colorbox{blue!30}{\$Typecast:{\textcolor{blue}{Convert\$}}} the \colorbox{blue!30}{\$parameter:{\textcolor{blue}{input\$}}} into \colorbox{blue!30}{\$integers:{\textcolor{blue}{integers\$}}} and check if it is a prime number. 
    } \\
    \textbf{Substitution:}\textit{ 
    Convert the input into \colorbox{blue!30}{\textcolor{blue}{whole numbers}} and check if it is a prime number. 
    } \\
    \end{flushleft}
    \caption{An example of tagging and then substituting ``integer'' with ``whole number''.} 
    \label{fig:substitution-example}
\end{figure}

\subsubsection{Replacement Sets}
We identify the most common terms that students use to refer to to each concept category. An initial list was developed by reading through all prompts in the \citet{nguyen_how_2024} and \citet{feldman_non-expert_2024} datasets, to get the widest possible set of variations. We computed frequencies for terms in this initial list and selected terms used at least twice in \studenteval. This led to a final set of 65 substitution terms, with at least two substitutions for each of the 14 concept lemmas. 

\subsubsection{Causal Analysis}

We conducted term-by-term substitution experiments across 65 category-replacement pairs. For each category-replacement pair, we replaced all expressions tagged with the category using the replacement lemma. Terms tagged with other categories were left unchanged, with the category tags removed and the original terms restored.

\Cref{fig:substitution-example} shows an example of the term-by-term substitution on a tagged prompt, where we replace all terms tagged with category \emph{integer} with the replacement term \emph{whole number}. Our tagging retains information about the tenses, plurals, and capitalization of the original words. In this example, \emph{integers} tagged with \emph{integers} is replaced with \emph{whole numbers}. Terms \emph{Convert} and \emph{input} tagged with other categories are unchanged by the substitution.

Using Llama 3.1 8B and 70B \cite{llama_team_llama_2024}, we generate completions for all prompts before and after substitution. A completion is considered correct if it passes all tests for the problem. We compute a pass rate per problem by sampling 200 completions using common  hyperparameters for code generation.\footnote{Following \citet{chen:codex}, we use top-p sampling (0.95) and temperature (0.2) to calculate pass@1.}

\subsubsection{Significance Testing}\label{sec:stats}

We measure the statistical reliability of observed differences in pass rates using mixed-effects binary logistic regression models that include random effects for prompt ID and problem. The outcome variable is the pass@1 rate. 

\begin{figure*}
\centering
\begin{tikzpicture}


\node[draw=gray,text width=1.4in, anchor=north west] at (12.2,2.8) {\begin{minipage}{2in}
\tiny
\textbf{Clues:}\hspace{0pt}
\begin{enumerate}[itemsep=-1mm,nosep,left=0pt]
\item First input is a list
\item List structure explained
\item Second input is sales tax 
\item Multiply item price by quantity
\item Sum results
\item Apply sales tax
\item Round to two decimal places
\item Return total
\end{enumerate}
\end{minipage}
};

 \node[draw=gray, line width=0.5pt, text width=6.3cm, align=left, inner sep=1pt, font=\tiny\selectfont\itshape, anchor=north west] (prompt1)
    at (0,7.9) {\Circled{1} This function takes in a list of the item purchased, the price, the tax, and the overall sales tax. All of the prices and tax within the lists are added together. The sales tax is then multiplied by the outcome of the added prices, and then the result of the multiplication is added onto the total price. The total price is then returned as the output.
};
\node[draw=gray, line width=0.5pt, text width=3.5cm, align=left, inner sep=1pt, font=\tiny\itshape, anchor=north west] (prompt2) at (0,6) {\Circled{2} prices and \sout{tax} \uline{taxes} within the lists};

\node[draw=gray, line width=0.5pt, text width=2.6cm, align=left, inner sep=1pt, font=\tiny\itshape, anchor=north west] (prompt3) at (3.5,1.2) {\Circled{3} the prices and \sout{taxes} \uline{tax} within the lists\uline{, which is the last two components of the list}
};

\node[draw=gray, line width=0.5pt, text width=3cm, align=left, inner sep=1pt, font=\tiny\itshape, anchor=north west] (prompt4) at (0,1.6) {\Circled{4} \sout{All of the prices and tax within the lists are added together.} \uline{The amount purchased is multiplied with price for each item.}};

\node[anchor=south west, inner sep=0] (img) at (0,0) {\includegraphics[width=0.99\textwidth]{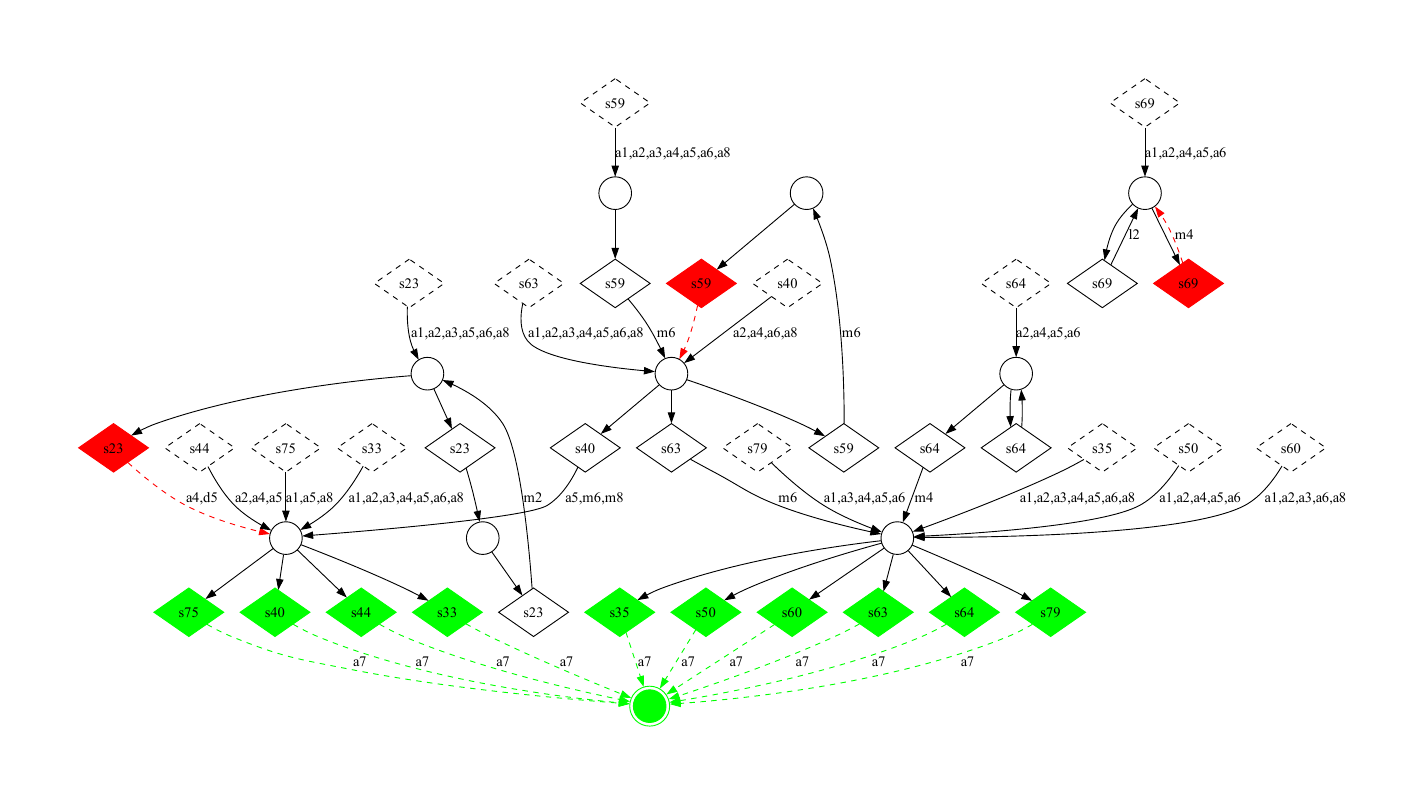}};

\draw[gray,thin,<-] (prompt1) -- (4.5,5.3);
\draw[gray,thin,<-,] (prompt2) -- (4.9,4.3);
\draw[gray,thin,<-,] (prompt3) -- (5.9,2.7);
\draw[gray,thin,<-,] (prompt4) -- (1.8,4.2);
\end{tikzpicture}
\caption{The graph of prompt trajectories for \textsc{total\_bill} (\cref{dataset-example}). We highlight the trajectory of S23 who ultimately fails: their first prompts \Circled{1} has most clues, but omits Clue~\#7 (bottom right of figure). Their next prompt \Circled{2} is a trivial change. \Circled{3} adds detail about the list structure (Clue~\#2), but it was already described well so they cycle back to a previous state. Finally, \Circled{4} adds the missing Clue~\#4 (and deletes Clue~\#5, but it isn't necessary to solve the problem). Here they give up and fail, but many others succeed from this state after adding Clue~\#7.}
\label{big-viz}
\end{figure*}
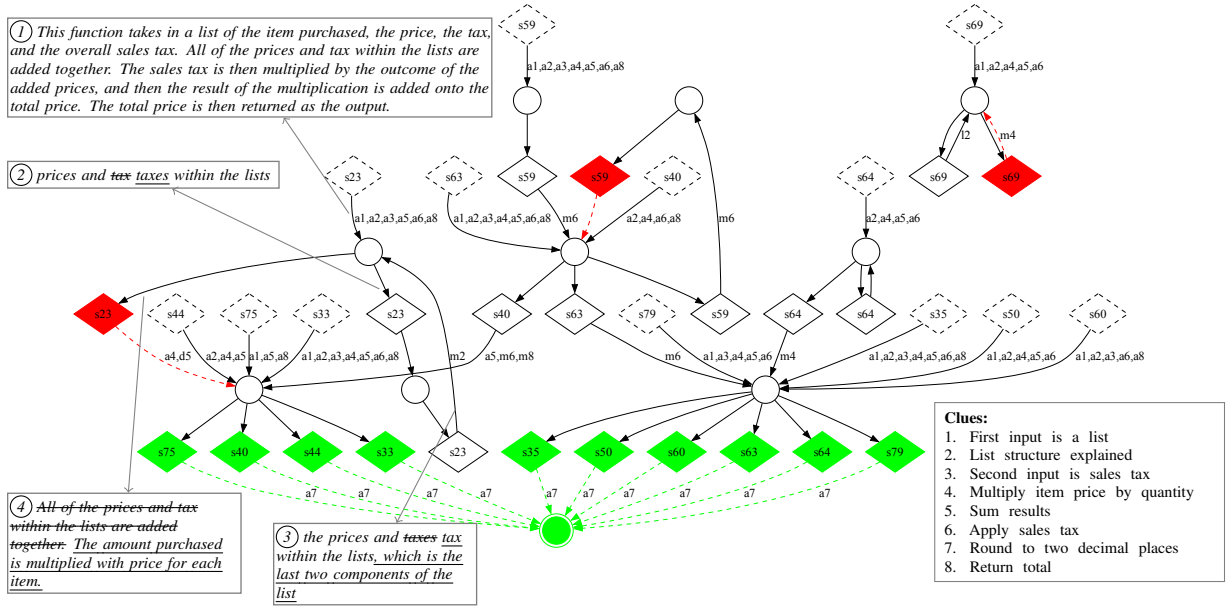

\subsection{Prompt Trajectories}
\label{prompting-trajectories}

Another possible source of error is the information content of  student prompts. Like other forms of communication, prompting involves a trade-off between communicative efficiency and likelihood of success. An effective prompter seeks to obtain correct results from the LLM while minimizing their own descriptive effort. 

A key part of effective prompting, therefore, is understanding the level of detail that is necessary to guide the model. An expert prompter may be able to quickly describe a task in a concise prompt. Novices, on the other hand, may struggle to distinguish cases that need to be specified (e.g., both branches of a conditional) from cases that pattern together, or atypical coding patterns from typical ones. This may be the case even when students fully understand the programming task, since efficient prompt-writing involves guessing what information models can infer without explicit direction.

We seek to understand how the information content of prompts changes over the course of a prompt trajectory. When a prompt fails, are students able to identify what information is missing? 
Prior work shows that students tend to write successively longer prompts~\citep{babe_studenteval_2024}; in this analysis, we seek to understand whether this additional verbiage contains useful information.

\subsubsection{Grouping LLM Outputs by Test Results}

When a prompt fails to generate correct code, a prompter must decide how to edit their prompt to improve their chances of success. An edit may add information about the intended behavior, remove information that is  distracting or wrong, or simply change how the information is described. By studying how and when students edit the information in their prompts, we gain insight into the relationship between information content and prompt success.

To do this, we study a set of 303 \emph{prompt trajectories}: sequences of prompts entered by a student for a particular task, starting from their first prompt and ending with a final prompt that may or may not succeed on the task.

Although prompts vary significantly in wording, we can group them based on their effect: when used to prompt a model, what is the behavior of the generated code?  Every problem has a single group of prompts where the tests produce the expected output (successes). In addition, there are multiple states where tests produce incorrect answers or throw exceptions. 
The $\circ$-nodes in \cref{big-viz} represent the ten states that students encounter on the \textsc{total\_bill} problem: the green node is the success state and the others are different failures.





\subsubsection{Information in Prompt Edits}


We use the notion of a \textbf{prompt clue} to study the information content of prompts. A clue is a piece of information about the function's intended behavior. For each problem, we identify a set of clues by examining the information that successful prompts tend to contain, as well as the expert-written prompts from the \studenteval{} dataset. We strive for sets of 3-6 clues per problem.

Expert annotators (experienced CS1 educators) developed the set of clues for each problem and used it to annotate each prompt trajectory. We tag the first prompt in each trajectory with the set of clues present. Subsequently, we tag each prompt edit in terms of its information change: adding a clue ($a$), deleting a clue ($d$), removing detail from a clue ($l$), or rewording a clue without removing detail ($m$). A null tag ($0$) is used to mark edits that do not change the information content of a prompt.

\Cref{big-viz} (bottom right) lists the eight clues for the \textsc{Total\_Bill} problem  (\cref{dataset-example}). Some of these clues describe the input and output types (Clues~\#1, \#3, and \#8). The remaining clues describe the computation. The edge labels in the graph show how students modify the clues present in their prompts.


\subsubsection{Prompt Trajectory Graphs}

We define a graph with alternating states of all prompt trajectories for a problem from the sequence of prompts, execution outputs, and expert annotations discussed above. For a given problem, let $s \in S$ be the set of students and $p_{s,i} \in P_{S,\Nat}$ be the set of prompts indexed by student and attempt number. Let $p_{s,i_\mathrm{max}}$ be the final prompt by $s$. Let $\Exec : P_{S,\Nat} \to O$ be the mapping from a prompt to its test  output, where there is a distinguished output $\ok \in O$ where all tests pass.

We construct a directed graph $G = (V, E)$ where $V = O \cup P_{S,\Nat}$.
The graph edges are:
\begin{itemize}[itemsep=0mm,nosep]
    \item $\langle p_{s,i}, o\rangle \in E$ where $\Exec(p_{s,i}) = o$
    \item $\langle o, p_{s, i+1} \rangle \in E$ if there exists $p'_{s,i} \in P$ and $\langle p'_{s,i}, o \rangle \in E$
\end{itemize}
A node $p_{s,i_\mathrm{max}}$ is a \emph{success node} if $\langle p_{s, i_\mathrm{max}}, \ok \rangle \in E$, and is otherwise a \emph{failure node}. We label edges $\langle p_{s,0}, o\rangle \in E$ with the initial clues for student $s$. For  $\langle p_{s,i-1}, o\rangle, \langle o, p_{s,i}\rangle \in E$, we label the edge $\langle p_{s,i}, o\rangle$ with the edits to the clues made from prompt $p_{s,i-1}$ to $p_{s,i}$.

In \Cref{big-viz}, the $\circ$-nodes are test result nodes and the $\diamond$-nodes are prompt edit nodes $p_{s,i}$. We label each $\diamond$-node with the student's identifier $s$.\footnote{The index $i$ can  be inferred, unless the student sees the same output 3+ times.} The $\diamond$-nodes with dashed edges represent a student's first prompt and the $\diamond$-nodes colored green or red represent their final prompt (success or failure, respectively). We label edges with clue edits. For convenience, we color each student's last edit edge green (success) or red (failure). 

The caption of \cref{big-viz} describes the prompt and clue edits by a student who ultimately fails the task. Other patterns can also be read from the graph. For instance, most students succeed in two attempts after adding a clue about rounding (Clue~\#7). The three students who never solve the problem get stuck in cycles. The graph also shows a disconnected failure state visited only by student s69, who struggled to describe the input list: the generated code assumes a triply-nested list.

We see the kinds of patterns described above in almost all problems, including longer loops and far more failures in the harder problems.  We analyze the structure of these graphs in \Cref{results-prompting-trajectories} to understand prompt trajectories in more depth.



\section{Results: Style Rarely Matters}

We measure the effect of prompt wording through a causal intervention experiment in which we explore a range of lexical substitutions for terms referring to 12 key programming concepts. If what hinders students is their lack of fluency with technical vocabulary, we should be able to improve the pass rate of their prompts by substituting more precise technical vocabulary for their non-canonical ways of referring to these concepts. We also measure the effect of word choice on high-quality prompts: by including substitution terms that are commonly used by students but less technically precise, we can test whether they decrease pass rates.

\begin{figure}[t]
    \centering
    \includegraphics[width=\linewidth]{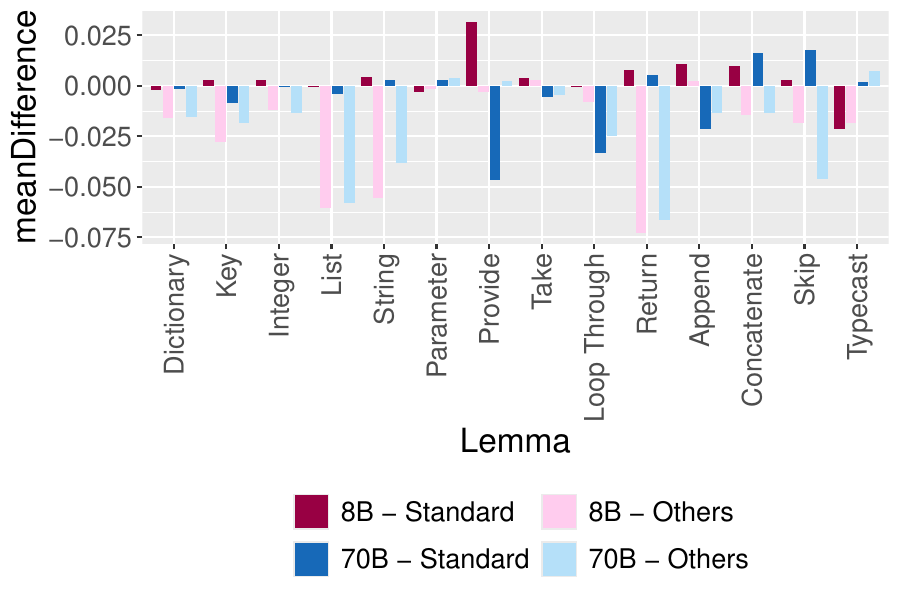}
    \caption{Differences between pass@1 rates before and after lexical substitutions. A negative mean difference represents a decrease in performance after substitution. }
    \label{fig:word_results}
    \vspace{-3mm}
\end{figure}

\begin{table}[t]
\footnotesize
    \centering
    \begin{tabular}{|l|l|c|c|}\hline
         Lemma&Substitution&8B&70B\\\hline
         String&character&$\downarrow$&$\downarrow$\\
         &phrase&$\downarrow$&-\\
         &set of characters&$\downarrow$&$\downarrow$\\
         &word&$\downarrow$&-\\
         List&brackets&$\downarrow$&$\downarrow$\\
         &set of brackets&$\downarrow$&$\downarrow$\\
         &set&$\downarrow$&$\downarrow$\\
         Key&attribute&$\downarrow$&$\downarrow$\\
         &entry&$\downarrow$&-\\
         &item&$\downarrow$&-\\
         &part&$\downarrow$&-\\
         &variable&$\downarrow$&-\\
         Parameter&argument&-&$\uparrow$\\
         Provide&provide&-&$\downarrow$\\
         Return&display&$\downarrow$&$\downarrow$\\
         &print&$\downarrow$&$\downarrow$\\
         Loop&go through&$\downarrow$&-\\
         &execute a for loop with&$\downarrow$&$\downarrow$\\
         &run a for loop through&$\downarrow$&$\downarrow$\\
         &iterate&-&$\downarrow$\\
         &loop through&-&$\downarrow$\\
         Concatenate&splice&$\downarrow$&-\\
         Skip&remove&$\downarrow$&-\\
         &avoid&$\downarrow$&-\\
         &ignore&-&$\downarrow$\\
         &neglect&-&$\downarrow$\\
         Typecast&cast&$\downarrow$&-\\
         &change&-&$\uparrow$\\\hline
    \end{tabular}
    \caption{Statistically reliable differences in pass@1 after lexical substitutions, Llama 3.1 8B and Llama 70B. $\downarrow$ denotes a reliably lower post-substitution pass@1; $\uparrow$ denotes a reliable increase; and - indicates no significant difference.}
    \label{tab:stats}
\end{table}


\subsection{How Much Does Style Matter?}

We perform lexical substitutions for the 12 concept categories, comparing the original and post-substitution prompt pass@1 rates using Llama 3.1 8B and 70B. We test each concept category separately, holding the rest of the prompt constant.

\Cref{fig:word_results} summarizes the results of the 65 lexical substitution experiments. Full model tables can be found in \Cref{app:stats}. In general, we observe only weak effects of lexical substitution across all categories. For 4 out of 14 concept lemmas, there are no statistically reliable differences between the pass rates for the reworded prompts and the originals; moreover, when there are statistically reliable differences, they tend to be small (\Cref{tab:stats}). Contrary to the perceptions of students reported in \citet{nguyen_how_2024}, technical vocabulary does not seem to have a strong impact on how well models are able to generate code from student prompts. 

\subsection{Can Rewording Help Failing Prompts?}

The overall results show little effect of lexical substitution. Since our substitution sets consist of terms commonly used by students, they include both standard and non-standard ways of referring to the target concepts. This means that some substitutions make a prompt less technically precise, while others make it more technically precise.

It is particularly important to understand how prompt wording impacts unsuccessful student prompts. If student word choice is a driving factor in the failure of their prompts, it would be relatively simple to intervene. There are two possible outcomes for low-quality prompts. If the student's vocabulary is causing the low pass rate, then substituting a more precise term should improve its pass rate. On the other hand, the use of non-standard terminology may simply be correlated with poor quality prompts; if this is the case, improving terminology may not lead to higher success rates. 

Unlike the analysis in \citet{babe_studenteval_2024}, the lexical substitution experiment enables us to distinguish these two scenarios. We find no evidence of significant gains from fixing terminology: across all categories, there are no statistically reliable gains from substituting standard terminology (see \Cref{app:stats}). 

\subsection{When Does Wording Matter?}

Our lexical substitution experiments reveal that correcting word choice does not significantly improve pass rates for prompts that use non-standard ways of referring to the target concepts. However, we do observe some statistically significant changes in pass rates: there are reliable negative effects from substituting certain non-standard terms.

We find particularly robust negative effects of diverse non-standard ways of referring to strings: substituting ``character'' and ``set of characters'' lower pass rates for string-referring prompts for both models. We also find negative effects of non-standard list terms (``brackets'', ``set'', ``set of brackets''). The largest magnitude effects are from ``set,'' likely because set is a distinct data type.

For concepts related to control flow, there are interesting differences between input and output concepts. Both models are robust to a range of ways of referring to a function's input. However, for return, substituting either ``print'' and ``display'' brings pass rates down. This is not surprising: since all of the tasks involve functions that return values, prompting the model to print or display instead is actively misleading. This finding also aligns with the correlational findings of \citet{babe_studenteval_2024}.


Overall, the lexical substitution experiments reveal only weak causal effects of prompt wording. Although substituting non-standard terminology can decrease success rates, correcting non-standard terminology does not seem to help weak prompts. This suggests that the interactions between word choice and prompt success reported in \citet{babe_studenteval_2024} were correlative, rather than causal: prompts that use non-standard terminology are weak for independent reasons. 

We view this finding as both surprising, given the body of prior work in which both students and educators identify technical vocabulary as a barrier to working with LLMs, and disappointing, since it would be easier to intervene into student terminology than other aspects of their prompting process.

\section{Results: Substance Matters}
\label{results-prompting-trajectories}

An alternative hypothesis about student-LLM miscommunication is that students struggle to select the right information for models. We explore this using prompt trajectory graphs (\cref{prompting-trajectories}) to understand prompt editing. What kinds of edits to information content do students make, and how do they effect the success of their prompts? We focus our discussion on high-level trends; \Cref{appendix-prompting-trajectories} contains graphs for each studied task.


\subsection{Successful Prompts Have All Clues}

We first examine the last prompt in every trajectory. We find that \emph{when all clues for the problem are present in the final prompt, the likelihood of success is 86\%.} Conversely, \emph{when even one clue is missing from the final prompt, the likelihood of success falls to 40\%}. This shows that information content is a main factor in the success of student prompts.

There are a few exceptions where students succeed even though their prompts omit clues. We manually inspect these exceptions, which we identify using the prompt trajectory graphs, and find that most fall into one of three cases: (1)~the prompt contains hardcoded answers that do not generalize beyond test cases; (2)~the function signature has informative names that subsume some clues; or (3) a clue may be technically missing, but duck typing allows the LLM to generate correct code (e.g., the student describes adding strings instead of lists, which uses the same operator in Python). 

Considering this, the number of success prompts that are missing one or more clues represents an upper bound on prompt success with partial information. This supports the conclusion that providing all the necessary clues about function behavior is typically what determines prompt success.

\subsection{Rewording Existing Clues Hardly Helps}

Prompt trajectory graphs illuminate the impact of edits that merely add/remove detail from existing clues, or make trivial edits (edges labelled $m$, $l$, or $0$ in the graphs). Out of all edges incident to nodes where all tests pass ($\ok$), we find (1)~28\% add  detail to an existing clue ($m$), (2)~11\% are trivial rewrites ($0$) and (3)~just 4\% remove detail from an existing clue ($l$). \emph{Rephrasing a prompt without adding a new clue leads to success less than half the time}. Moreover, of these edits, 65\% add detail to an existing clue.

Finally, when a prompt contains less than half the clues for a problem, we find that adding/removing detail leads to success only 11\% of the time. In other words, the fewer clues a prompt has, the harder it is to succeed by tweaking wording alone. Together, these findings show the impact of information content on prompt success.

\subsection{Cycles Involve Uninformative Edits}

Prior work shows that students often give up in frustration when their prompt edits do not produce different output~\citep{nguyen_how_2024}. We identify these cycles and measure how hard it is for students to escape them: \emph{when a prompt trajectory has a cycle, its likelihood of eventual success is 30\%, compared to 72\% without a cycle}. When the cycle exceeds three edges,  the likelihood of success drops to 14\%. We find a moderate negative correlation between success and cycle length ($\rho$ = -0.42).

Examining the edits in cycles, we find the majority (90\%) involve missing clues. Furthermore, most cycles edits (75\%) are exclusively rewrites ($l$, $m$, or $0$); of these, 54\% do not change the level of detail in any clues ($0$). This shows that students get stuck in a cycle of failing prompts when they are missing important information. 


How do students escape? Of the 44 prompt trajectories that manage to break out of a cycle, only 7 have trivial edits. Most escape by adding a new clue (13) or adding detail to existing clues (20). 
Taken together, our results show that the most successful strategy is adding information, but that most students in cycles simply try trivial wording changes. 

\subsection{When Does Style Matter, Revisited}
\label{exceptions}


Overall, our findings support the view that the information content of prompts is more important that wording. However, there are a handful of cases where prompts fail even with all clues.

\begin{figure}
\begin{minipage}{0.99\columnwidth}
\textbf{Prompt:} \itshape
  This function takes the input of a 
  dictionary. If the key is a planet, it
  takes the entry and  adds it to the
  total mass. The function outputs the
  total mass of all planets in the 
  dictionary.
\end{minipage}
\vspace{0.5em}
\begin{lstlisting}[language=Python,style=codeblock]
def planets_mass(planets):
  total_mass = 0
  for key in planets:
    if key in planets:
      total_mass += planets[key]["mass"]
  return total_mass
\end{lstlisting}
\caption{Variable/concept confusion.}
\label{fig:planets_mass_65}
\end{figure}

\Cref{fig:planets_mass_65} shows a prompt that succinctly states all clues for the problem. However, the model cannot disambiguate between ``planets'' as a parameter name and as a general concept, and ends up translating the instruction \emph{if the key is a planet} into \lstinline[language=Python,basicstyle=\ttfamily\footnotesize]|if key in planets|. In other cases, the model interprets language in a surprising way. Three students experienced the same model error in a task to capitalize every other letter in a string: the model produced code that followed their instructions, but also rearranged the string so that all the uppercase letters came first (\Cref{fig:altText_23} in the Appendix).

The remaining exceptions can be found in \Cref{appendix:form-beats-content}. Overall, we observe that these failures stem from ambiguity in natural language or model limitations rather than technical vocabulary issues. 

\section{Conclusion}

By investigating two commonly espoused concrete hypotheses about why students struggle to effectively prompt LLMs for code, our work sheds light on what it means for students to write ``good prompts.''  Our results suggest that it is the (lack of) information in prompts, rather than how the information is communicated, that causes student-LLM miscommunication. Although these findings imply that attempts to help student prompters by suggesting alternative wording are unlikely to be very useful, by providing the first empirical evidence of the source of student struggles, we hope our findings will guide future work on teaching prompting towards more impactful interventions. 

\section*{Limitations}

This work builds on the existing \studenteval dataset, which was collected from 80 students in early 2023. These students were selected from three institutions and all had taken only one programming course. \citet{babe_studenteval_2024} argue that they are representative of beginning students, but they are not representative of students with more programming experience. Our findings may not generalize to more advanced programmers.

The prompts we study were written by students using \textit{code-davinci-002}, which was state-of-the-art at the time, but is now an older model. A newer model, such as a chat model, would lead to different interactions. However, \citet{babe_studenteval_2024} show that their benchmark remains challenging for several newer models. We re-evaluate \studenteval{} using Llama 3.1 8B and 70B and also find that the prompts remain challenging.

The set of categories and terms we explore in our causal inference experiments are specific to the \citet{babe_studenteval_2024} and \citet{feldman_non-expert_2024} user populations. These students attend select US institutions, therefore their wording choices represent a certain level of English proficiency. The set of substitutions would differ with speakers of other natural languages, as might their effect.

The clues used to tag prompt trajectories represent an expert annotator's perception of the information that successful prompts typically contain. There may be other ways to formulate the same problem. However, we studied all exceptions to our finding and did not find cases where students appeared to use a different set of clues than what the expert annotator found (\cref{exceptions}).

\section*{Ethics Statement}

The main ethical concerns surrounding this work lie in its study of student interactions with LLMs. This work uses the public, fully anonymized version of the \studenteval dataset. Therefore, this work has no additional ethical considerations beyond those described in the ethics statement of \citet{babe_studenteval_2024}. The secondary analysis of existing data that we do is consistent with the intended use of the dataset, which is to study how students write prompts.


\bibliography{molly,custom,anthology}

\newpage

\appendix

\onecolumn

\section{Dataset and Code Availability}
\label{artifact-url-and-license}
The code and dataset for this submission is publicly available and licensed under the terms of the BSD 3 Clause license. The \studenteval{} dataset is licensed under under the terms of the OpenRAIL license.

\ifanon
The URL is omitted for review. 
\else
\fi

\section{Computing Resources}
\label{computing-resources}

The computational experiments for this paper were conducted with less than 1,000 hours of A100 GPU time. The models evaluated were Meta Llama 3.1 8B and 70B~\cite{llama_team_llama_2024}.

\section{Software Configuration}
\label{packages}

We use vLLM 0.6.2 for LLM inference~\citep{kwon:paged-attention}. We use spaCy 3.8.0 for lemmatization with the \verb|en_core_web_trf| pipeline.

\ifaclanonsubmission
\section{Use of AI Assistants}
\label{use-of-ai-assistants}

Some code for this paper was written with AI assistants enabled.
\fi

\section{Causal Analysis of Lexical Choices}
\label{lexical-appendix}

This section describes the procedure we used to perform the causal analysis of lexical choices and presents detailed results.

\subsection{Data Annotation Procedure}

The overall approach to data annotation is described in \Cref{methods_style}. We provide some additional detail below.

The process for tagging concept references proceeded as follows. First, we developed an automated script to perform tagging automatically. This approximated the set of necessary tags, but a manual pass was necessary for numerous reasons. For instance, some student terms (e.g., convert) occurred in numerous problems, but were either function names or parameters in some. In other cases grammatical features, such as prepositions, led the automated approach to be insufficient (e.g., \$takes:brings\$ should be tagged as \$takes:brings in\$).

The two expert annotators, who are both CS1 instructors, then proceeded to perform a manual review. During this review, care was taken to tag idiosyncratic references; for instance, when a participant mistakenly referred to an input dictionary as a list, this was tagged under the dictionary category, so that we could explore substitutions of a more accurate term. The goal of this process was a consistent tag set, thus the annotators ultimately came to consensus on all tags for all prompts. Inter-annotator reliability was not calculated due to the emphasis on consensus and the number/precision of tags per prompt. 

To gain insight into the range of terms used over problems, the annotators independently assessed two distinct prompts for each of the 48 problems, for a total of 96 problems. They then met to discuss their tagging edits. Out of this discussion, we made three main changes: (1) ``given'' was removed as a possible term, as it has too many possible use cases; (2) the Input concept was divided into the three lemmas of ``parameters'', ``take'', and ``provide;'' and (3) specific disambiguation for ``concatenate'' and ``insert'' was developed. 
The annotators then came to consensus on all tags for the 96 problems.

After this process, the above changes were made to the automatic tagging script and then the two annotators independent tagged the remainder of the problems in the dataset. They then met to discuss the tagging edits and determine the consensus decision. Most disagreements were easily resolved (e.g., missed tags, typos). The main substantive disagreement was regarding tags relevant to the String concept. Specifically, determining student meaning of character versus string was too challenging to tag consistently. Therefore, most mentions of character/s were removed from the String tag set. This was done retroactively to the original 96 problems as well. 

\subsection{Concepts, Expressions, and Interventions}
\label{appendix-details-on-concepts}

Table \ref{tab:subs} shows the lemmas for each concept category used in the lexical substitution experiments, along with the set of replacement terms and example expressions that students use to refer to them.

\begin{table*}
    \centering
    \begin{tabular}{|l|l|l|l|}\hline
    Concept&Lemma&Substitution Lemmas&Example Student Terms\\\hline\hline
    String&string&word, phrase, string, character,& string, word, string of text,\\
    &&set of characters& word or sentence, string of characters\\
    List&list&brackets, set of brackets, set, list,& list, array, set, arrangement,series,\\
    &&array list, array& collection, sequence\\
    Dictionary&dictionary&map, dictionary& dictionary, dict, object, array\\
    Integer&integer&integer, whole number, int&int, integer numbers, whole number\\
Key&key&key, item, entry, attribute, part,& key, key value, category\\
&&element, variable&element, variable, parameter\\\hline
    Input&parameter&parameter, argument, value provided,&input, parameter, value, component,\\
    &&input& input value, value inputted\\
    &take&take, bring in, accept, get, input& take, take in, take input of, get\\
    &provide&provide, enter, input& input\\
    Loop&loop through&go through, run through, iterate through,& loop, loop through, go through, parse,\\
    &&loop through, run a for loop through,& iterate through, run through\\
    &&look through, execute a for loop with&\\
    Output&return&return, output, print, produce, display&return, output, print, provide, out put\\\hline
    Concatenate&concatenate&concatenate, combine, splice, add&concatenate, append, add, combine\\
    Insert&insert&insert, add, append, attach& put, insert, input, add, give\\
    Skip&skip&skip, avoid, neglect, ignore, remove&skip, ignore, avoid, neglect\\
    Typecast&typecast&typecast, type cast, cast, convert,&typecast, convert, turn, change\\
    &&change&\\\hline
    \end{tabular}
    \caption{Concepts, Lemmas, and Substitution Terms for Causal Analysis Experiments}
    \label{tab:subs}
\end{table*}

\subsection{Experimental Method}
\label{hyperparameters}

For generations, we generated 200 completions for each model with temperature (0.2), top-p sampling (0.95), and a 512 token limit.

\subsection{Statistical Analysis}\label{app:stats}

Statistical significance results are from mixed-effects binary logistic regression models that include random effects for prompt ID and problem. The random effects structure for problem contains both random slopes and intercepts; due to issues with convergence, the random effects for prompt ID contain only random intercepts. 

The outcome variable is the pass@1 rate calculated with 200 samples. All models were fit in R using the lme4 library~\citep{lme4} with sample weights of 200 (the number of observations from which the proportion was computed).

\subsubsection{Type Concepts}

Tables \ref{tab:mem_ss}-\ref{tab:mem:kl} provide the full mixed-effects results for datatype concepts.

\begin{table}
\centering
\begin{tabular}[t]{lllll}
\toprule
Fixed effects & $\widehat{\beta}$ & SE & $z$ & $p$\\\hline
(Intercept) & -4.4 & 0.69 & -6.3 & \textbf{\textless 0.0001}\\
character & -1.1 & 0.32 & -3.6 & \textbf{0.0004}\\
phrase & -0.69 & 0.2 & -3.4 & \textbf{0.0006}\\
set of characters & -1.4 & 0.36 & -4.1 & \textbf{\textless 0.0001}\\
string & 0.048 & 0.054 & 0.87 & 0.38\\
word & -0.62 & 0.19 & -3.3 & \textbf{0.0009}\\
\bottomrule
\end{tabular}
\caption{Llama 8B mixed-effects model for String concept.}\label{tab:mem_ss}
\end{table}

\begin{table}
\centering
\begin{tabular}{lllll}
\toprule
Fixed effects & $\widehat{\beta}$ & SE & $z$ & $p$\\
\midrule
(Intercept) & -3.4 & 0.62 & -5.5 & \textbf{\textless 0.0001}\\
character & -0.59 & 0.19 & -3 & \textbf{0.002}\\
phrase & -0.29 & 0.18 & -1.6 & 0.1\\
set of characters & -1.1 & 0.22 & -4.8 & \textbf{\textless 0.0001}\\
string & 0.11 & 0.075 & 1.5 & 0.14\\
\addlinespace
word & -0.098 & 0.12 & -0.79 & 0.43\\
\bottomrule
\end{tabular}
\caption{Llama 70B mixed-effects model for String concept.}
\end{table}

\begin{table}
\centering
\begin{tabular}{lllll}
\toprule
Fixed effects & $\widehat{\beta}$ & SE & $z$ & $p$\\
\midrule
(Intercept) & -5.5 & 0.75 & -7.4 & \textbf{\textless 0.0001}\\
array & 0.1 & 0.087 & 1.2 & 0.24\\
array list & 0.11 & 0.11 & 1 & 0.31\\
brackets & -1 & 0.23 & -4.4 & \textbf{\textless 0.0001}\\
list & -0.0032 & 0.041 & -0.078 & 0.94\\
set & -2.4 & 0.51 & -4.7 & \textbf{\textless 0.0001}\\
set of brackets & -1.9 & 0.37 & -5.2 & \textbf{\textless 0.0001}\\
\bottomrule
\end{tabular}
\caption{Llama 8B mixed-effects model for List concept.}
\end{table}

\begin{table}
\centering
\begin{tabular}{lllll}
\toprule
Fixed effects & $\widehat{\beta}$ & SE & $z$ & $p$\\
\midrule
(Intercept) & -5.1 & 0.74 & -6.9 & \textbf{\textless 0.0001}\\
array & -0.057 & 0.083 & -0.69 & 0.49\\
array list & 0.11 & 0.098 & 1.1 & 0.28\\
brackets & -0.55 & 0.16 & -3.3 & \textbf{0.0009}\\
list & -0.074 & 0.057 & -1.3 & 0.19\\
\addlinespace
set & -2.3 & 0.55 & -4.1 & \textbf{\textless 0.0001}\\
set of brackets & -1.6 & 0.41 & -3.9 & \textbf{0.0001}\\
\bottomrule
\end{tabular}
\caption{Llama 70B mixed-effects model for List concept.}
\end{table}

\begin{table}
\centering
\begin{tabular}{lllll}
\toprule
Fixed effects & $\widehat{\beta}$ & SE & $z$ & $p$\\
\midrule
(Intercept) & -5.9 & 0.9 & -6.6 & \textbf{\textless 0.0001}\\
int & -0.1 & 0.091 & -1.2 & 0.25\\
integer & 0.057 & 0.038 & 1.5 & 0.13\\
whole number & -0.45 & 0.23 & -1.9 & 0.052\\
\bottomrule
\end{tabular}
\caption{Llama 8B mixed-effects model for Integer concept.}
\end{table}

\begin{table}
\centering
\begin{tabular}{lllll}
\toprule
Fixed effects & $\widehat{\beta}$ & SE & $z$ & $p$\\
\midrule
(Intercept) & -5.6 & 1.1 & -5.3 & \textbf{\textless 0.0001}\\
int & -0.096 & 0.14 & -0.71 & 0.48\\
integer & 0.14 & 0.18 & 0.77 & 0.44\\
whole number & -0.1 & 0.21 & -0.5 & 0.62\\
\bottomrule
\end{tabular}
\caption{Llama 70B mixed-effects model for Integer concept.}
\end{table}

\begin{table}
\centering
\begin{tabular}{lllll}
\toprule
Fixed effects & $\widehat{\beta}$ & SE & $z$ & $p$\\
\midrule
(Intercept) & -13 & 1.3 & -9.9 & \textbf{\textless 0.0001}\\
dictionary & -0.099 & 0.056 & -1.8 & 0.075\\
map & -0.066 & 0.41 & -0.16 & 0.87\\
\bottomrule
\end{tabular}
\caption{Llama 8B mixed-effects model for Dictionary concept.}
\end{table}

\begin{table}
\centering
\begin{tabular}{lllll}
\toprule
Fixed effects & $\widehat{\beta}$ & SE & $z$ & $p$\\
\midrule
(Intercept) & -12 & 1.4 & -9 & \textbf{\textless 0.0001}\\
dictionary & -0.1 & 0.14 & -0.73 & 0.46\\
map & -0.33 & 0.27 & -1.2 & 0.23\\
\bottomrule
\end{tabular}
\caption{Llama 70B mixed-effects model for Dictionary concept.}
\end{table}

\begin{table}
\centering
\begin{tabular}{lllll}
\toprule
Fixed effects & $\widehat{\beta}$ & SE & $z$ & $p$\\
\midrule
(Intercept) & -9.8 & 1.3 & -7.8 & \textbf{\textless 0.0001}\\
attribute & -0.7 & 0.32 & -2.2 & \textbf{0.028}\\
element & -0.25 & 0.15 & -1.7 & 0.087\\
entry & -0.39 & 0.14 & -2.8 & \textbf{0.0048}\\
item & -0.28 & 0.14 & -2 & \textbf{0.047}\\
\addlinespace
key & 0.046 & 0.15 & 0.3 & 0.77\\
part & -0.59 & 0.21 & -2.8 & \textbf{0.005}\\
variable & -0.56 & 0.14 & -3.9 & \textbf{\textless 0.0001}\\
\bottomrule
\end{tabular}
\caption{Llama 8B mixed-effects model for Key concept.}
\end{table}

\begin{table}
\centering
\begin{tabular}{lllll}
\toprule
Fixed effects & $\widehat{\beta}$ & SE & $z$ & $p$\\
\midrule
(Intercept) & -6.5 & 1.4 & -4.7 & \textbf{\textless 0.0001}\\
attribute & -0.87 & 0.43 & -2 & \textbf{0.04}\\
element & -0.29 & 0.16 & -1.8 & 0.065\\
entry & -0.28 & 0.19 & -1.5 & 0.14\\
item & -0.28 & 0.22 & -1.3 & 0.21\\
\addlinespace
key & -0.16 & 0.11 & -1.5 & 0.15\\
part & -0.48 & 0.36 & -1.3 & 0.18\\
variable & -0.85 & 0.5 & -1.7 & 0.092\\
\bottomrule
\end{tabular}
\caption{Llama 70B mixed-effects model for Key concept.}\label{tab:mem:kl}
\end{table}

\subsubsection{Control Flow Concepts}

Tables \ref{tab:mem_sr}-\ref{tab:mem_tl} provide the full mixed-effects results for control flow concepts.

\begin{table}
\centering
\begin{tabular}{lllll}
\toprule
Fixed effects & $\widehat{\beta}$ & SE & $z$ & $p$\\
\midrule
(Intercept) & -5 & 0.62 & -8.1 & \textbf{\textless 0.0001}\\
display & -1.3 & 0.25 & -5.1 & \textbf{\textless 0.0001}\\
output & -0.14 & 0.13 & -1.1 & 0.27\\
print & -2.9 & 0.37 & -7.8 & \textbf{\textless 0.0001}\\
produce & -0.16 & 0.12 & -1.3 & 0.2\\
\addlinespace
return & 0.11 & 0.061 & 1.8 & 0.068\\
\bottomrule
\end{tabular}
\caption{Llama 8B mixed-effects model for Return concept.}\label{tab:mem_sr}
\end{table}

\begin{table}
\centering
\begin{tabular}{lllll}
\toprule
Fixed effects & $\widehat{\beta}$ & SE & $z$ & $p$\\
\midrule
(Intercept) & -4.9 & 0.56 & -8.7 & \textbf{\textless 0.0001}\\
display & -0.87 & 0.28 & -3 & \textbf{0.002}\\
output & -0.071 & 0.17 & -0.42 & 0.67\\
print & -2.8 & 0.41 & -6.8 & \textbf{\textless 0.0001}\\
produce & 0.21 & 0.16 & 1.3 & 0.18\\
\addlinespace
return & 0.0061 & 0.096 & 0.063 & 0.95\\
\bottomrule
\end{tabular}
\caption{Llama 70B mixed-effects model for Return concept.}
\end{table}

\begin{table}
\centering
\begin{tabular}{lllll}
\toprule
Fixed effects & $\widehat{\beta}$ & SE & $z$ & $p$\\
\midrule
(Intercept) & -11 & 2.1 & -5 & \textbf{	\textless 0.0001}\\
execute a for loop with & -3.5 & 0.59 & -6 & \textbf{	\textless 0.0001}\\
go through & -0.56 & 0.15 & -3.8 & \textbf{0.0001}\\
iterate through & 0.0045 & 0.14 & 0.032 & 0.97\\
look through & -0.41 & 0.27 & -1.6 & 0.12\\
loop through & -0.38 & 0.27 & -1.4 & 0.16\\
run a for loop through & -2.9 & 0.48 & -6 & \textbf{	\textless 0.0001}\\
run through & -0.37 & 0.2 & -1.8 & 0.067\\
\bottomrule
\end{tabular}
\caption{Llama 8B mixed-effects model for Loop concept.}
\end{table}

\begin{table}
\centering
\begin{tabular}{lllll}
\toprule
Fixed effects & $\widehat{\beta}$ & SE & $z$ & $p$\\
\midrule
(Intercept) & -10 & 1.8 & -5.8 & \textbf{\textless 0.0001}\\
execute a for loop with & -4.7 & 1.3 & -3.5 & \textbf{0.0005}\\
go through & -0.92 & 0.56 & -1.6 & 0.1\\
iterate through & -1.3 & 0.33 & -4.1 & \textbf{\textless 0.0001}\\
look through & -0.62 & 0.45 & -1.4 & 0.16\\
loop through & -1.4 & 0.34 & -4.2 & \textbf{\textless 0.0001}\\
run a for loop through & -1.7 & 0.52 & -3.3 & \textbf{0.001}\\
run through & -0.34 & 0.48 & -0.71 & 0.48\\
\bottomrule
\end{tabular}
\caption{Llama 70B mixed-effects model for Loop concept.}
\end{table}

\begin{table}
\centering
\begin{tabular}{lllll}
\toprule
Fixed effects & $\widehat{\beta}$ & SE & $z$ & $p$\\
\midrule
(Intercept) & -15 & 4.1 & -3.6 & \textbf{0.0004}\\
enter & 0.31 & 0.36 & 0.87 & 0.38\\
input & 0.079 & 0.24 & 0.33 & 0.74\\
provide & 1.8 & 1.2 & 1.5 & 0.13\\
\bottomrule
\end{tabular}
\caption{Llama 8B mixed-effects model for Input - Provide lemma.}
\end{table}

\begin{table}
\centering
\begin{tabular}{lllll}
\toprule
Fixed effects & $\widehat{\beta}$ & SE & $z$ & $p$\\
\midrule
(Intercept) & -15 & 6.5 & -2.2 & \textbf{0.03}\\
enter & -0.27 & 0.43 & -0.63 & 0.53\\
input & 0.29 & 0.49 & 0.6 & 0.55\\
provide & -1.1 & 0.42 & -2.6 & 0.008\\
\bottomrule
\end{tabular}
\caption{Llama 70B mixed-effects model for Input - Provide lemma.}
\end{table}

\begin{table}
\centering
\begin{tabular}{lllll}
\toprule
Fixed effects & $\widehat{\beta}$ & SE & $z$ & $p$\\
\midrule
(Intercept) & -5.6 & 0.83 & -6.7 & \textbf{\textless 0.0001}\\
argument & -0.045 & 0.1 & -0.43 & 0.67\\
input & 0.088 & 0.06 & 1.5 & 0.14\\
parameter & -0.095 & 0.11 & -0.86 & 0.39\\
value provided & -0.17 & 0.13 & -1.3 & 0.19\\
\bottomrule
\end{tabular}
\caption{Llama 8B mixed-effects model for Input - Parameter lemma.}
\end{table}

\begin{table}
\centering
\begin{tabular}{lllll}
\toprule
Fixed effects & $\widehat{\beta}$ & SE & $z$ & $p$\\
\midrule
(Intercept) & -5.4 & 0.98 & -5.5 & \textbf{\textless 0.0001}\\
argument & 0.28 & 0.13 & 2.2 & \textbf{0.03}\\
input & -0.031 & 0.094 & -0.33 & 0.74\\
parameter & 0.23 & 0.14 & 1.6 & 0.1\\
value provided & 0.27 & 0.16 & 1.7 & 0.086\\
\bottomrule
\end{tabular}
\caption{Llama 70B mixed-effects model for Input - Parameter lemma.}
\end{table}

\begin{table}
\centering
\begin{tabular}{lllll}
\toprule
Fixed effects & $\widehat{\beta}$ & SE & $z$ & $p$\\
\midrule
(Intercept) & -5.7 & 0.97 & -5.9 & \textbf{\textless 0.0001}\\
accept & -0.061 & 0.078 & -0.79 & 0.43\\
bring in & -0.051 & 0.14 & -0.38 & 0.71\\
get & 0.0086 & 0.11 & 0.081 & 0.94\\
input & -0.15 & 0.1 & -1.4 & 0.15\\
take & 0.029 & 0.056 & 0.51 & 0.61\\
\bottomrule
\end{tabular}
\caption{Llama 8B mixed-effects model for Input - Take lemma.}
\end{table}

\begin{table}
\centering
\begin{tabular}{lllll}
\toprule
Fixed effects & $\widehat{\beta}$ & SE & $z$ & $p$\\
\midrule
(Intercept) & -5.2 & 1.1 & -4.6 & \textbf{\textless 0.0001}\\
accept & -0.024 & 0.14 & -0.17 & 0.87\\
bring in & -0.22 & 0.22 & -1 & 0.31\\
get & 0.14 & 0.14 & 0.96 & 0.34\\
input & 0.054 & 0.14 & 0.39 & 0.7\\
\addlinespace
take & -0.066 & 0.12 & -0.53 & 0.6\\
\bottomrule
\end{tabular}
\caption{Llama 70B mixed-effects model for Input - Take lemma.}\label{tab:mem_tl}
\end{table}

\clearpage

\subsubsection{Operation Concepts}

Tables \ref{tab:mem_sc}-\ref{tab:mem_lt} provide the full mixed-effects results for control flow concepts.

\begin{table}
\centering
\begin{tabular}{lllll}
\toprule
Fixed effects & $\widehat{\beta}$ & SE & $z$ & $p$\\
\midrule
(Intercept) & -4.4 & 1.5 & -2.8 & \textbf{0.005}\\
add & 0.11 & 0.27 & 0.39 & 0.7\\
combine & -0.14 & 0.13 & -1.1 & 0.29\\
concatenate & 0.24 & 0.14 & 1.7 & 0.081\\
splice & -0.56 & 0.19 & -3 & \textbf{0.003}\\
\bottomrule
\end{tabular}
\caption{Llama 8B mixed-effects model for Concatenate concept.}\label{tab:mem_sc}
\end{table}

\begin{table}
\centering
\begin{tabular}{lllll}
\toprule
Fixed effects & $\widehat{\beta}$ & SE & $z$ & $p$\\
\midrule
(Intercept) & -1.5 & 1.5 & -0.97 & 0.33\\
add & -0.024 & 0.21 & -0.12 & 0.91\\
combine & 0.37 & 0.43 & 0.87 & 0.38\\
concatenate & 0.28 & 0.35 & 0.82 & 0.41\\
splice & -0.31 & 0.51 & -0.61 & 0.55\\
\bottomrule
\end{tabular}
\caption{Llama 70B mixed-effects model for Concatenate concept.}
\end{table}

\begin{table}
\centering
\begin{tabular}{lllll}
\toprule
Fixed effects & $\widehat{\beta}$ & SE & $z$ & $p$\\
\midrule
(Intercept) & -5.5 & 1.8 & -3 & \textbf{0.003}\\
add & -0.091 & 0.12 & -0.73 & 0.47\\
append & -0.38 & 0.46 & -0.82 & 0.41\\
attach & -0.53 & 0.45 & -1.2 & 0.24\\
insert & -1.4 & 1.1 & -1.3 & 0.18\\
\bottomrule
\end{tabular}
\caption{Llama 8B mixed-effects model for Append concept.}
\end{table}

\begin{table}
\centering
\begin{tabular}{lllll}
\toprule
Fixed effects & $\widehat{\beta}$ & SE & $z$ & $p$\\
\midrule
(Intercept) & -9.9 & 1.9 & -5.3 & \textbf{\textless 0.0001}\\
add & -0.4 & 0.22 & -1.8 & 0.067\\
append & -0.51 & 0.33 & -1.5 & 0.12\\
attach & -0.11 & 0.34 & -0.32 & 0.75\\
insert & -0.76 & 0.45 & -1.7 & 0.089\\
\bottomrule
\end{tabular}
\caption{Llama 70B mixed-effects model for Append concept.}
\end{table}

\begin{table}
\centering
\begin{tabular}{lllll}
\toprule
Fixed effects & $\widehat{\beta}$ & SE & $z$ & $p$\\
\midrule
(Intercept) & -13 & 3.8 & -3.3 & \textbf{0.001}\\
avoid & -0.76 & 0.28 & -2.7 & \textbf{0.006}\\
ignore & 0.014 & 0.17 & 0.083 & 0.93\\
neglect & -0.18 & 0.28 & -0.62 & 0.54\\
remove & -4.2 & 2.1 & -2 & \textbf{0.046}\\
skip & -0.21 & 0.46 & -0.46 & 0.65\\
\bottomrule
\end{tabular}
\caption{Llama 8B mixed-effects model for Skip concept.}
\end{table}

\begin{table}
\centering
\begin{tabular}{lllll}
\toprule
Fixed effects & $\widehat{\beta}$ & SE & $z$ & $p$\\
\midrule
(Intercept) & -14 & 6.4 & -2.2 & \textbf{0.03}\\
avoid & 0.041 & 0.81 & 0.051 & 0.96\\
ignore & -0.98 & 0.22 & -4.4 & \textbf{\textless 0.0001}\\
neglect & -1.2 & 0.43 & -2.7 & \textbf{0.007}\\
remove & -6.6 & 3.6 & -1.8 & 0.068\\
skip & -0.66 & 0.47 & -1.4 & 0.16\\
\bottomrule
\end{tabular}
\caption{Llama 70B mixed-effects model for Skip concept.}
\end{table}

\begin{table}
\centering
\begin{tabular}{lllll}
\toprule
Fixed effects & $\widehat{\beta}$ & SE & $z$ & $p$\\
\midrule
(Intercept) & -5 & 1.5 & -3.3 & \textbf{0.001}\\
cast & -0.86 & 0.39 & -2.2 & \textbf{0.028}\\
change & -1.5 & 0.85 & -1.7 & 0.087\\
convert & -0.048 & 0.27 & -0.18 & 0.86\\
type cast & -0.73 & 0.54 & -1.4 & 0.17\\
typecast & -1.4 & 0.86 & -1.6 & 0.1\\
\bottomrule
\end{tabular}
\caption{Llama 8B mixed-effects model for Typecast concept.}
\end{table}

\begin{table}
\centering
\begin{tabular}{lllll}
\toprule
Fixed effects & $\widehat{\beta}$ & SE & $z$ & $p$\\
\midrule
(Intercept) & -7.6 & 4.3 & -1.8 & 0.074\\
cast & -0.89 & 0.74 & -1.2 & 0.23\\
change & 0.27 & 0.13 & 2 & \textbf{0.045}\\
convert & 0.31 & 0.23 & 1.3 & 0.19\\
type cast & -0.61 & 0.72 & -0.85 & 0.4\\
typecast & -1 & 0.97 & -1.1 & 0.28\\
\bottomrule
\end{tabular}
\caption{Llama 70B mixed-effects model for Typecast concept.}\label{tab:mem_lt}
\end{table}

\clearpage

\begin{figure*}
\includegraphics[width=0.99\textwidth]{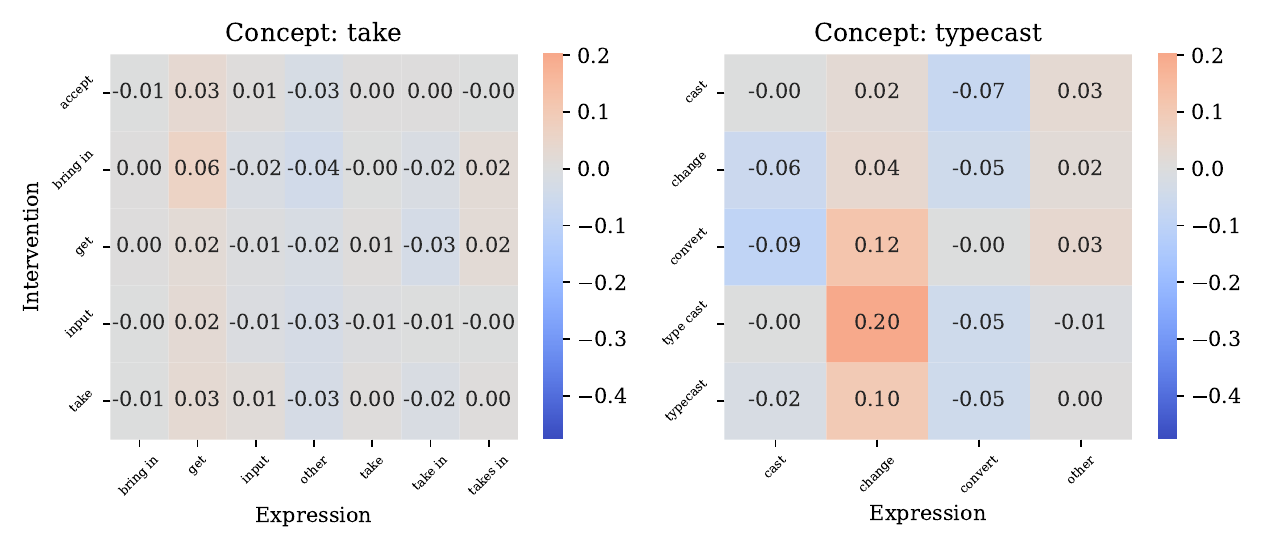}
\caption{This heatmap shows the difference in pass rates (pass@1) using Meta Llama 3.1 8B after replacing the original expression of a concept in a prompt (x-axis) with a the  expression chosen for the intervention (y-axis). We present one heatmap  per concept. We report differences on the subset of prompts that have the original expression. We group rare expressions into a single \emph{Other} class for each concept. See \cref{causal_heatmap_2,causal_heatmap_3} for more categories.}
\label{causal_heatmap_1}
\end{figure*}

\begin{figure*}
\includegraphics[width=0.99\textwidth]{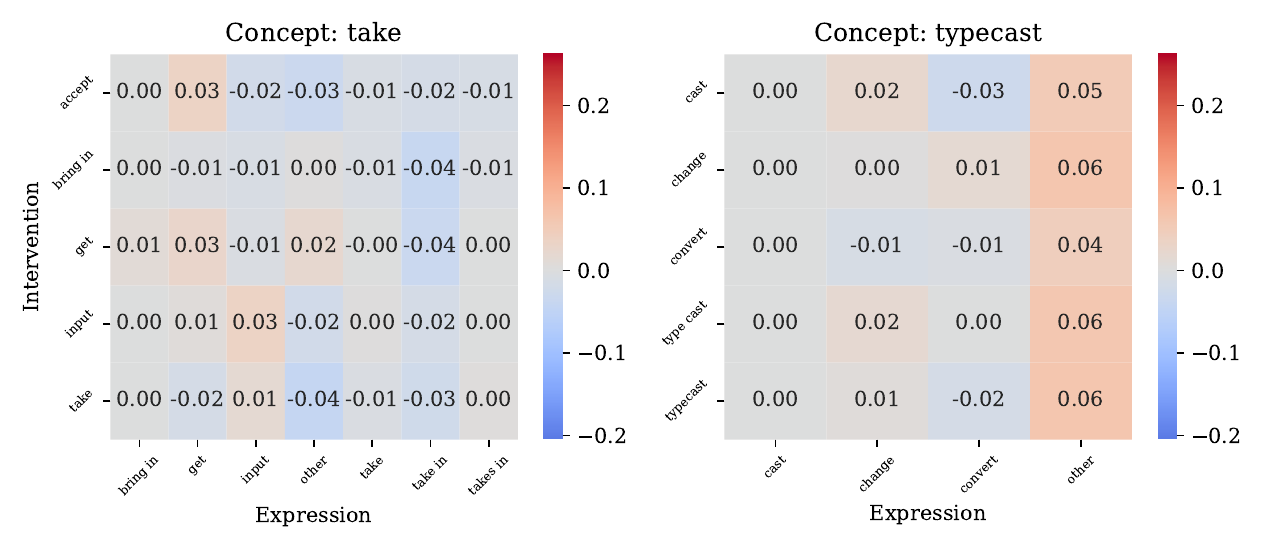}
\caption{For Meta Llama 3.1 70B. See the caption for \cref{causal_heatmap_1} for more information.}
\label{causal_heatmap_4}
\end{figure*}

\subsection{Substitution Visualizations}

\Cref{causal_heatmap_1,causal_heatmap_2,causal_heatmap_3} presents the results of causal interventions using Meta Llama 3.1 8B \cite{llama_team_llama_2024}. \Cref{causal_heatmap_4,causal_heatmap_5,causal_heatmap_6} presents the results of causal interventions using Meta Llama 3.1 70B.

\begin{figure*}
\includegraphics[width=\textwidth]{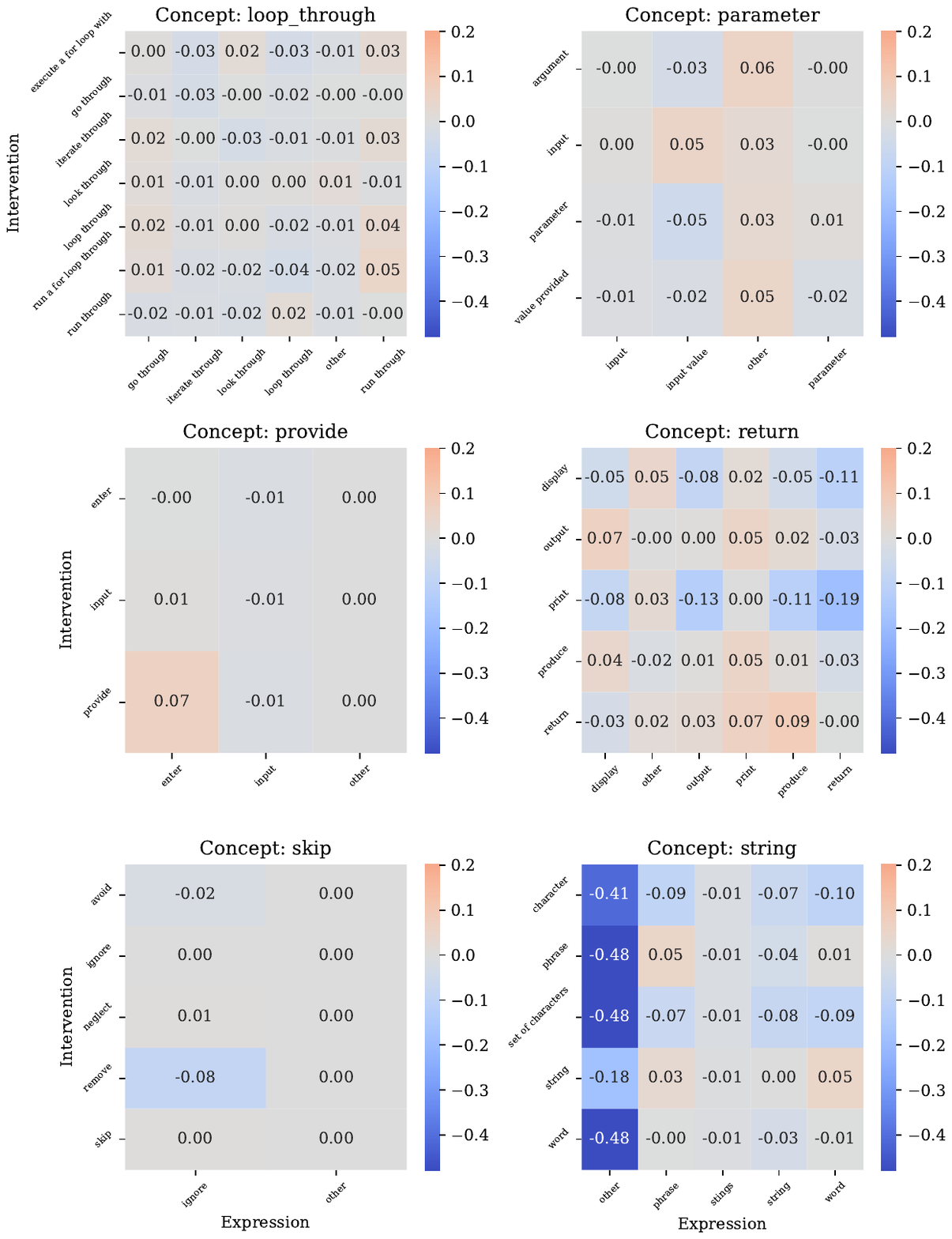}
\caption{Continuation of \cref{causal_heatmap_1}. See the caption of that figure for more information.}
\label{causal_heatmap_2}
\end{figure*}

\begin{figure*}
\includegraphics[width=\textwidth]{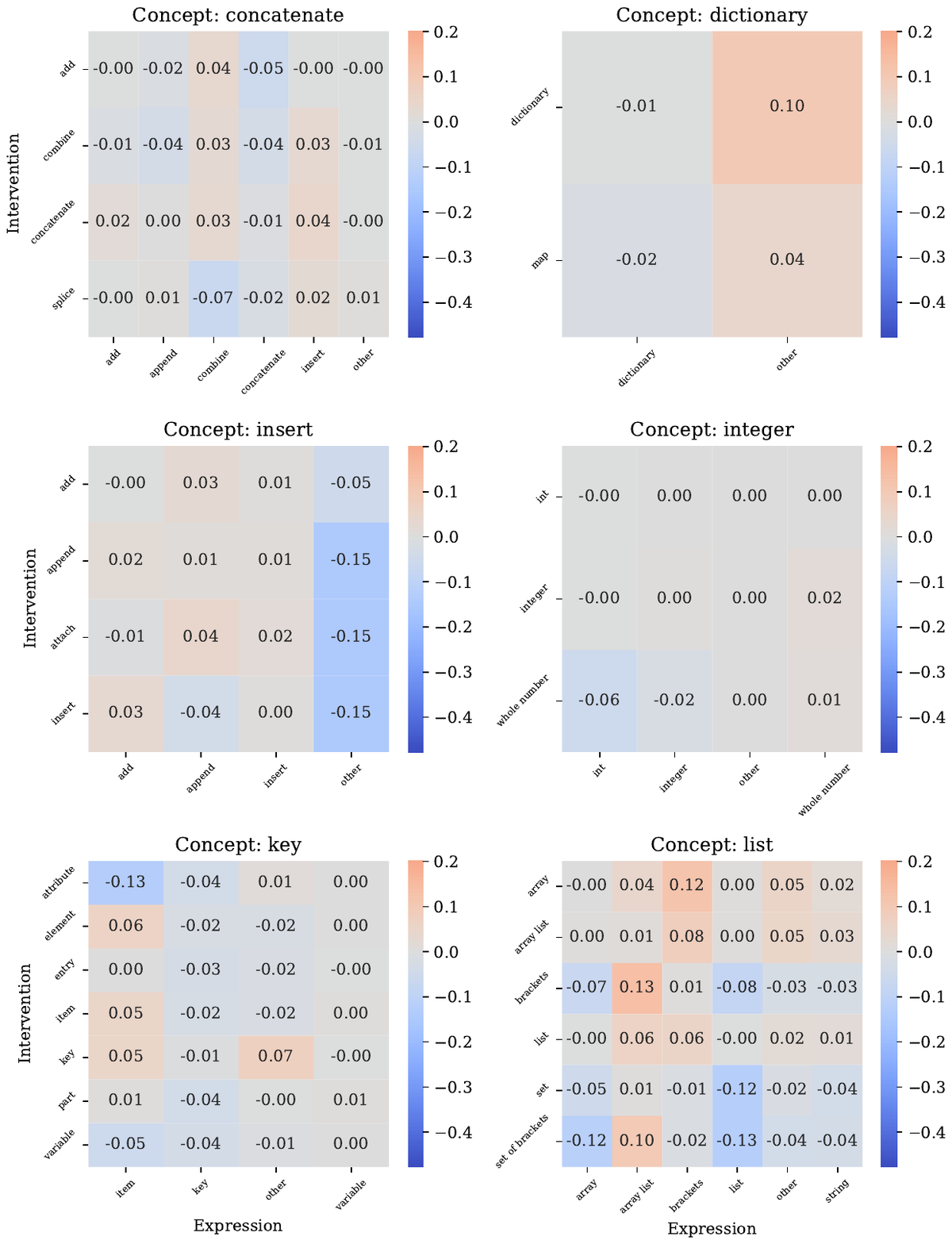}
\caption{Continuation of \cref{causal_heatmap_1}. See the caption of that figure for more information.}
\label{causal_heatmap_3}
\end{figure*}

\begin{figure*}
\includegraphics[width=\textwidth]{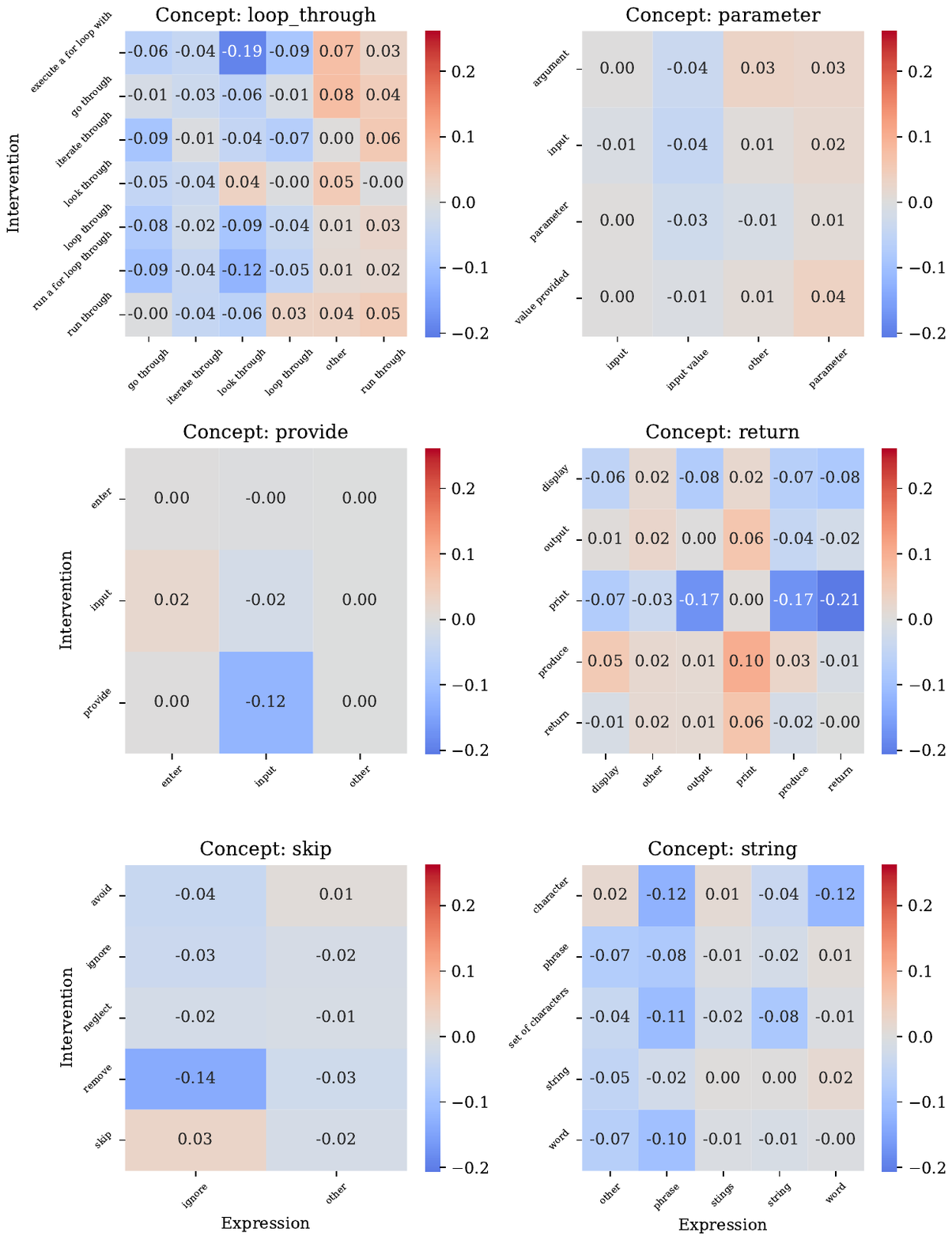}
\caption{Continuation of \cref{causal_heatmap_5}. See the caption of that figure for more information.}
\label{causal_heatmap_5}
\end{figure*}

\begin{figure*}
\includegraphics[width=\textwidth]{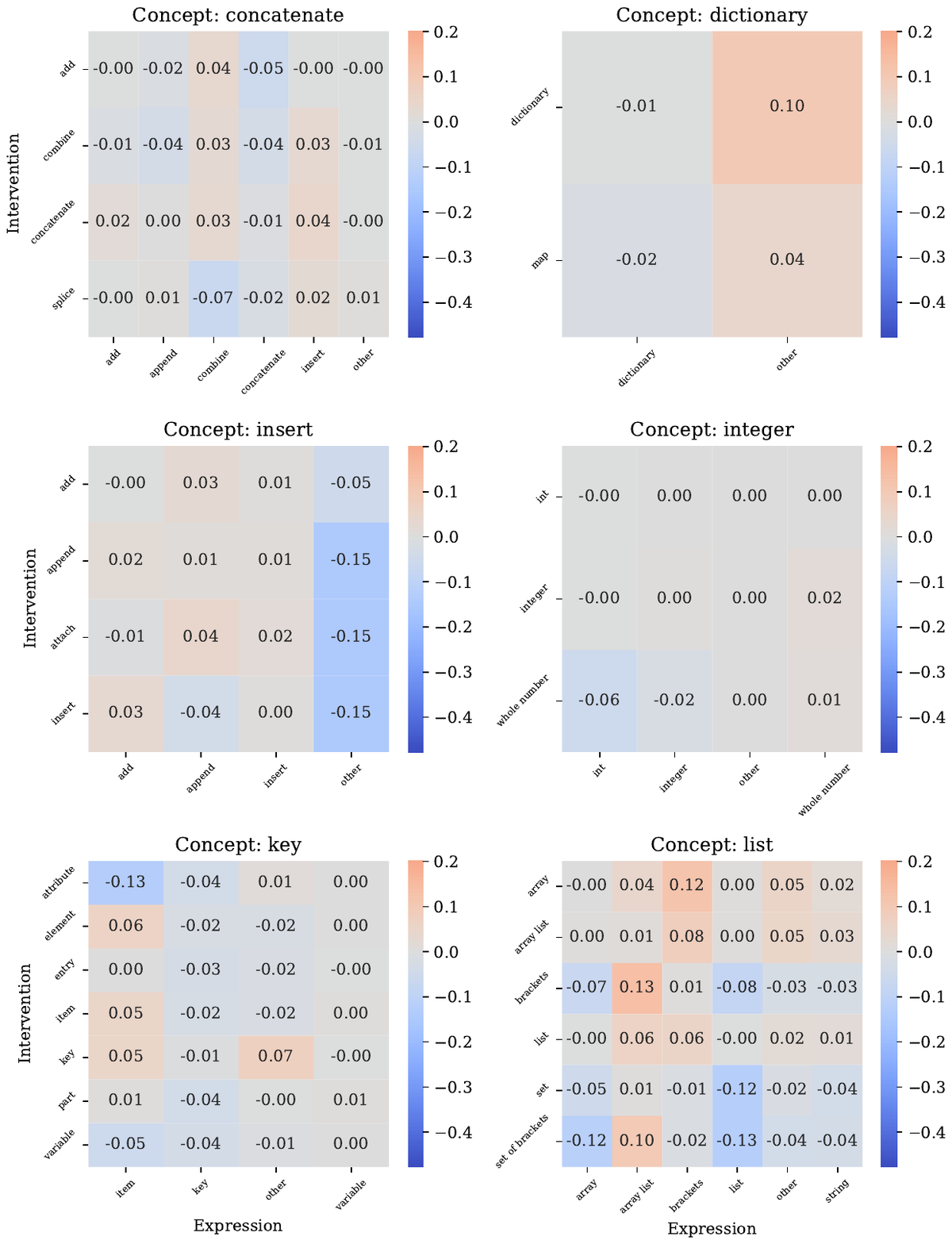}
\caption{Continuation of \cref{causal_heatmap_4}. See the caption of that figure for more information.}
\label{causal_heatmap_6}
\end{figure*}

\clearpage

\section{Analyzing Prompt Trajectories}
\label{appendix-prompting-trajectories}

\subsection{Tagging Prompt Clues}

Four expert annotators tagged the information content of the 303 prompt trajectories. All annotators have experience teaching courses in Python programming and are therefore familiar with how students talk about programming concepts. 

Annotators developed the sets of clues associated with problems by reading the successful prompts, the expert-written prompts from the original \studenteval{} dataset, and reflecting on the common information. Although information can be annotated at different levels of granularity, we strove for 3-8 clues per problem. 

After annotation, a consistency check was performed on each prompt. Any inconsistencies in tagging (e.g., tagging an ``add'' operation for an existing clue; tagging a ``modify'' operation for a clue that had not previously been tagged) were corrected at this stage.

\subsection{Additional Style Matters Examples}\label{appendix:form-beats-content}

We examine additional examples where students include all necessary clues in their prompt, but the model's generated function still fails tests.

\begin{figure}
\begin{lstlisting}[language=Python,style=codeblock]
def altText(s):
 if len(s) == 1:
     return s.upper()
 else:
     return s[::2].upper() + s[1::2]

\end{lstlisting}
\caption{Reordering letters unexpectedly.}
\label{fig:altText_23}
\end{figure}

A common model error observed across two problems (\texttt{topScores} in \Cref{fig:topScores-53} and \texttt{sort\_physicists} in \Cref{fig:sort_physicists36} and \Cref{fig:sort_physicists77}) consists of a sorting error. Both problems receive as input a nested list, with the inner lists containing fixed elements: $[[x_0, ..., x_n], ..., [x_0, ..., x_n]]$. The problems stipulate that the generated function must return one of the elements $x_i$, sorted by another elements $x_k$, where $k \neq x$. The error the model consistently makes is filtering out the key required for sorting, then subsequently attempting to sort. This however cannot be done without the sorting key. Thus, the model often simply calls \texttt{sort}, eluding the key. One plausible explanation to why this happens is that human programmers are unlikely to delete the sorting key first, then try to sort. For this reason, training data may not include many examples of how to sort in this way. Note that in all students' subsequent successful attempts, the model deletes the sorting key \emph{after} sorting.

 In a prompt from student46 for the \texttt{planets\_mass} problem (\Cref{fig:planets_mass_46}), the model conflates an extra piece of information (``first letter capitalized'') with the definition of a planet. Removing this single line leads the student to success. These examples serve to illustrate the kind of ambiguity in the wording of a prompt which can make the difference between success and fail. 

\subsection{Clue Sets}

Here we provide the clue sets for all 33 problems.

\lstset{language=Python,style=codeblock}
\noindent
\textbf{Problem:} \lstinline|add_int| \\
\textbf{Signature:} \lstinline|def add_int(lst, num):|\\
\textbf{Clues:}
\begin{enumerate}[itemsep=0mm,nosep]
\item edge case of list in list
\item concatenate num to strings
\item add num to integers
\item return list
\end{enumerate}

\noindent
\textbf{Problem:} \lstinline|add_up| \\
\textbf{Signature:} \lstinline|def add_up(arr):| \\
\textbf{Clues:}
\begin{enumerate}[itemsep=0mm,nosep]
\item 2D array
\item sum integer
\item sum float
\item return the sum of all elements
\item mention 0 base case
\item misdirection - add number within string
\end{enumerate}

\noindent
\textbf{Problem:} \lstinline|altText| \\
\textbf{Signature:} \lstinline|def altText(s):| \\
\textbf{Clues:}
\begin{enumerate}[itemsep=0mm,nosep]
\item input string
\item alternating uppercase
\item return all letters, including spaces
\item first letter upper
\end{enumerate}

\noindent
\textbf{Problem:} \lstinline|assessVowels| \\
\textbf{Signature:} \lstinline|def assessVowels(s):| \\
\textbf{Clues:}
\begin{enumerate}[itemsep=0mm,nosep]
\item argument s is a string
\item result is a list of strings
\item result is the vowels present in the argument
\item result has both upper and lower case vowels
\end{enumerate}

\noindent
\textbf{Problem:} \lstinline|changeSection| \\
\textbf{Signature:} \lstinline|def changeSection(s,i):| \\
\textbf{Clues:}
\begin{enumerate}[itemsep=0mm,nosep]
\item result is a string
\item result reverses a part of the argument 's'
\item the result reverses the first 'i' characters of the argument
\item the result also includes the remaining characters of 's', but not reversed
\end{enumerate}

\noindent
\textbf{Problem:} \lstinline|check_prime| \\
\textbf{Signature:} \lstinline|def check_prime(num):| \\
\textbf{Clues:}
\begin{enumerate}[itemsep=0mm,nosep]
\item convert input string to int
\item output bool
\item check prime
\item correct description of a procedure to check prime number
\end{enumerate}

\noindent
\textbf{Problem:} \lstinline|combine| \\
\textbf{Signature:} \lstinline|def combine(l1,l2):| \\
\textbf{Clues:}
\begin{enumerate}[itemsep=0mm,nosep]
\item input 2 lists
\item row correspondence
\item output 1 2d array
\end{enumerate}

\noindent
\textbf{Problem:} \lstinline|convert| \\
\textbf{Signature:} \lstinline|def convert(lst):| \\
\textbf{Clues:}
\begin{enumerate}[itemsep=0mm,nosep]
\item takes a list of numbers
\item maps numbers to letters
\item joins letters
\item -1 means split
\item return list of strings
\end{enumerate}

\noindent
\textbf{Problem:} \lstinline|create_list| \\
\textbf{Signature:} \lstinline|def create_list(dt, lst):| \\
\textbf{Clues:}
\begin{enumerate}[itemsep=0mm,nosep]
\item takes a dict and a list
\item looks up list items in dict
\item construct list with matching values
\item use None for items that aren't in dict
\item return list
\end{enumerate}

\noindent
\textbf{Problem:} \lstinline|fib| \\
\textbf{Signature:} \lstinline|def fib(n):| \\
\textbf{Clues:}
\begin{enumerate}[itemsep=0mm,nosep]
\item check if a Fib number
\item returns a Boolean
\item explanation of Fib
\item construct set of Fib numbers
\item hardcodes numbers
\item bound set
\end{enumerate}

\noindent
\textbf{Problem:} \lstinline|findHorizontals| \\
\textbf{Signature:} \lstinline|def findHorizontals(puzzle,wordList):| \\
\textbf{Clues:}
\begin{enumerate}[itemsep=0mm,nosep]
\item input is two lists
\item find words in second list within strings in first list
\item return dictionary
\item keys are words
\item values are indices of strings where words are found
\item words can be backwards or forwards
\end{enumerate}

\noindent
\textbf{Problem:} \lstinline|find_multiples| \\
\textbf{Signature:} \lstinline|def find_multiples(start,stop,factor):| \\
\textbf{Clues:}
\begin{enumerate}[itemsep=0mm,nosep]
\item return multiples
\item inclusive start and stop
\end{enumerate}

\noindent
\textbf{Problem:} \lstinline|generateCardDeck| \\
\textbf{Signature:} \lstinline|def generateCardDeck(suits, vals):| \\
\textbf{Clues:}
\begin{enumerate}[itemsep=0mm,nosep]
\item takes two lists
\item creates all pairs from the lists
\item sort alphabetically
\item first list item comes before second list item in pairs
\item return list
\end{enumerate}

\noindent
\textbf{Problem:} \lstinline|getSeason| \\
\textbf{Signature:} \lstinline|def getSeason(month):| \\
\textbf{Clues:}
\begin{enumerate}[itemsep=0mm,nosep]
\item input is string
\item month to season
\item return lowercase
\item explain which are which
\end{enumerate}

\noindent
\textbf{Problem:} \lstinline|increaseScore| \\
\textbf{Signature:} \lstinline|def increaseScore(score):| \\
\textbf{Clues:}
\begin{enumerate}[itemsep=0mm,nosep]
\item input integer
\item if less than 10, make 10
\item if 10 or more, add 1
\item if negative, turn positive
\item if single digit, add 0
\item return
\end{enumerate}

\noindent
\textbf{Problem:} \lstinline|laugh| \\
\textbf{Signature:} \lstinline|def laugh(size):| \\
\textbf{Clues:}
\begin{enumerate}[itemsep=0mm,nosep]
\item prefix h
\item reverse order
\item number of a's is based on size
\item space separation
\item down to 1
\item repetition
\item misdirection-print instead of return
\end{enumerate}

\noindent
\textbf{Problem:} \lstinline|pattern| \\
\textbf{Signature:} \lstinline|def pattern(value):| \\
\textbf{Clues:}
\begin{enumerate}[itemsep=0mm,nosep]
\item takes an int
\item produces a nested list
\item there are value n of inner lists
\item each inner list is from 1 to value
\item returns
\end{enumerate}

\noindent
\textbf{Problem:} \lstinline|percentWin| \\
\textbf{Signature:} \lstinline|def percentWin(guess,answers):| \\
\textbf{Clues:}
\begin{enumerate}[itemsep=0mm,nosep]
\item takes two lists
\item compares items from both lists and counts matches
\item computes percent match
\item rounds to whole percent
\item convert to string and add "
\item returns
\end{enumerate}

\noindent
\textbf{Problem:} \lstinline|planets_mass| \\
\textbf{Signature:} \lstinline|def planets_mass(planets):| \\
\textbf{Clues:}
\begin{enumerate}[itemsep=0mm,nosep]
\item takes a dictionary
\item skip Pluto
\item skip Sun
\item look up in dictionary
\item sum masses
\item return
\end{enumerate}

\noindent
\textbf{Problem:} \lstinline|print_time| \\
\textbf{Signature:} \lstinline|def print_time(day,hour):| \\
\textbf{Clues:}
\begin{enumerate}[itemsep=0mm,nosep]
\item input is a string and an int
\item how to distinguish sleeping
\item how to distinguish weekday versus weekend
\item short form of day
\item return not print
\end{enumerate}

\noindent
\textbf{Problem:} \lstinline|readingIceCream| \\
\textbf{Signature:} \lstinline|def readingIceCream(lines):| \\
\textbf{Clues:}
\begin{enumerate}[itemsep=0mm,nosep]
\item input is a list of strings
\item go through all strings
\item split on tab
\item extract last item from each string
\item convert to float
\item sum numbers
\item return total
\end{enumerate}

\noindent
\textbf{Problem:} \lstinline|remove_odd| \\
\textbf{Signature:} \lstinline|def remove_odd(lst):| \\
\textbf{Clues:}
\begin{enumerate}[itemsep=0mm,nosep]
\item takes a (potentially mixed) list of numbers
\item removes only odd numbers
\item removes only integers
\item returns list
\end{enumerate}

\noindent
\textbf{Problem:} \lstinline|reverseWords| \\
\textbf{Signature:} \lstinline|def reverseWords(words):| \\
\textbf{Clues:}
\begin{enumerate}[itemsep=0mm,nosep]
\item takes a list of strings
\item reverses each word in list
\item sorts list
\item reverse before sort
\item returns list
\end{enumerate}

\noindent
\textbf{Problem:} \lstinline|set_chars| \\
\textbf{Signature:} \lstinline|def set_chars(s,c,l):| \\
\textbf{Clues:}
\begin{enumerate}[itemsep=0mm,nosep]
\item input is described correctly
\item second argument is used to replace certain characters
\item third argument contains list of indices to replace
\item return string
\item handle indices outside string length
\end{enumerate}

\noindent
\textbf{Problem:} \lstinline|sortBySuccessRate| \\
\textbf{Signature:} \lstinline|def sortBySuccessRate(nominations):| \\
\textbf{Clues:}
\begin{enumerate}[itemsep=0mm,nosep]
\item input is list of dictionaries
\item add a key success
\item success is wins/noms
\item round success
\item sort by success
\item return
\end{enumerate}

\noindent
\textbf{Problem:} \lstinline|sort_physicists| \\
\textbf{Signature:} \lstinline|def sort_physicists(scientists):| \\
\textbf{Clues:}
\begin{enumerate}[itemsep=0mm,nosep]
\item Input is a list of lists
\item specify inner list structure
\item filter list with the right key
\item sort list with the right key
\item specify return
\item sort
\end{enumerate}

\noindent
\textbf{Problem:} \lstinline|sortedBooks| \\
\textbf{Signature:} \lstinline|def sortedBooks(books, writer):| \\
\textbf{Clues:}
\begin{enumerate}[itemsep=0mm,nosep]
\item takes a list of dictionaries
\item takes an author
\item removes books not by that author
\item sorts list
\item sorts list by year
\item returns list
\end{enumerate}

\noindent
\textbf{Problem:} \lstinline|student_grades| \\
\textbf{Signature:} \lstinline|def student_grades(students, grades):| \\
\textbf{Clues:}
\begin{enumerate}[itemsep=0mm,nosep]
\item input is two dictionaries
\item match keys to values between dictionaries
\item create a new dictionary with lists of grades
\item return
\end{enumerate}

\noindent
\textbf{Problem:} \lstinline|subtract_add| \\
\textbf{Signature:} \lstinline|def subtract_add(lst):| \\
\textbf{Clues:}
\begin{enumerate}[itemsep=0mm,nosep]
\item input is list
\item convert items to integers
\item alternate add and subtract
\item add is even indices and subtract is odd indices
\item return result
\end{enumerate}

\noindent
\textbf{Problem:} \lstinline|times_with| \\
\textbf{Signature:} \lstinline|def times_with(meetings):| \\
\textbf{Clues:}
\begin{enumerate}[itemsep=0mm,nosep]
\item input is nested list
\item describe nested list structure
\item make dictionary where keys are people and values are numbers
\item sum times
\item return dictionary
\end{enumerate}

\noindent
\textbf{Problem:} \lstinline|topScores| \\
\textbf{Signature:} \lstinline|def topScores(lst):| \\
\textbf{Clues:}
\begin{enumerate}[itemsep=0mm,nosep]
\item nested list with items in format [grade, name]
\item return only names with grade >80
\item return list of names
\item sort by grades descending
\item only top 5 scoring names
\item equal 80 edge case
\end{enumerate}

\noindent
\textbf{Problem:} \lstinline|total_bill| \\
\textbf{Signature:} \lstinline|def total_bill(grocery_list, sales_tax):| \\
\textbf{Clues:}
\begin{enumerate}[itemsep=0mm,nosep]
\item input is a list
\item list structure explained
\item input is sales tax
\item multiply item by its quantity
\item sum items
\item apply sales tax
\item round
\item return total
\end{enumerate}

\noindent
\textbf{Problem:} \lstinline|translate| \\
\textbf{Signature:} \lstinline|def translate(lexicon,sentence):| \\
\textbf{Clues:}
\begin{enumerate}[itemsep=0mm,nosep]
\item input 1 is a dictionary
\item input 2 is a string
\item extract words from string
\item look up words in dictionary
\item use UNK for words that are not in dictionary
\item function returns
\item join string
\end{enumerate}

\begin{figure}
\begin{lstlisting}[language=Python,style=codeblock]
def topScores(lst):
    """
    taken lst, loop each list and if the zeroth index is more than or equal to 80, add
    the 1st index into a list and return this new list in the order of the highest
    zeroth index. If the list is larger than 4 items, then return only 4 items with 
    the highest zeroth index. 
    """
    
    new_lst = []
    for i in lst:
        if i[0] >= 80:
            new_lst.append(i[1])
    new_lst.sort(reverse=True)
    if len(new_lst) > 4:
        return new_lst[:4]
    else:
        return new_lst

def topScores(lst):
    """
    if 0th index of the nested list of lst is greater than or equal to 80, return a 
    list of the 1st index in the order from the largest 0th index integer. Return up 
    to a list of 5 items. 
    """
    
    return [x[1] for x in sorted(lst, key=lambda x: x[0], reverse=True) if x[0] >= 80][:5]
\end{lstlisting}
\caption{Student53 example prompt for \texttt{topScores}. The second prompt is the student's final successful attempt.}
\label{fig:topScores-53}
\end{figure}

\begin{figure}
\begin{lstlisting}[language=Python,style=codeblock]
def sort_physicists(scientists):
    """
    The internal lists contain name, year graduated and subject. If the subject is 
    "Physics" return just the names in order of what year they graduated from lowest 
    to highest
    """
    
    physicists = []
    for scientist in scientists:
        if scientist[2] == "Physics":
            physicists.append(scientist[0])
    physicists.sort()
    return physicists

def sort_physicists(scientists):
    """
    The internal lists contain name, year graduated and subject. If the subject is 
    "Physics" return just the names, in order of what year they graduated from lowest 
    to highest
    """
    
    return [name for name, year, subject in sorted(scientists, key=lambda x: x[1]) if subject == "Physics"]
\end{lstlisting}
\caption{Attempt by student36 at \texttt{sort\_physicists}. The second prompt is the student's final successful attempt.}
\label{fig:sort_physicists36}
\end{figure}

\begin{figure}
\begin{lstlisting}[language=Python,style=codeblock]
def sort_physicists(scientists):
    """
    Given a set of lists with a scientist, number, and a subject, respectively, only 
    return the scientist if the subject is "Physics" in a list. Then sort the list 
    into ascending order based on the number, but don't return the number.
    """
    
    physicists = []
    for scientist in scientists:
        if scientist[2] == "Physics":
            physicists.append(scientist[0])
    physicists.sort()
    return physicists

def sort_physicists(scientists):
    """
    Given a set of lists with a scientist, number, and a subject, respectively, only 
    return the scientist if the subject is "Physics" in a list. Then sort the 
    scientists based on the numbers given in the set of lists from smallest to 
    largest. For example, if the scientist has the smallest number attached to them in 
    the set of lists, they should be placed first in the output list.
    """
    
    physicists = []
    for scientist in scientists:
        if scientist[2] == "Physics":
            physicists.append(scientist[0])
    physicists.sort(key=lambda x: x[1])
    return physicists
\end{lstlisting}
\caption{Student77's attempt at \texttt{sort\_physicists}. The second prompt is the student's final successful attempt.}
\label{fig:sort_physicists77}
\end{figure}

\begin{figure}
\begin{lstlisting}[language=Python,style=codeblock]
def planets_mass(planets):
 """
 The function will take in a dictionary 
 where the keys are strings@\textcolor{blue}{, with the first letter capitalized,}@ and integers
 are stored inside them. It will remove
 any strings that are not associated
 with planets, then it will take all 
 the integers stored inside and add 
 them together and output that.
 """
@\textbf{Completion 1}@
 for key in planets:
     if key[0].isupper() == False:
         del planets[key]
 mass = 0
 for key in planets:
     mass += planets[key]
 return mass
 
 @\textbf{Completion 2}@
 total = 0
 for planet in planets:
     if planet == "Mercury" or planet == \
     "Venus" or planet == "Earth" or \
     planet == "Mars" or planet == \ 
     "Jupiter" or planet == "Saturn" or \
     planet == "Uranus" or \ 
     planet == "Neptune":
         total += planets[planet]
 return total
\end{lstlisting}
\caption{two attempts by student46 at \texttt{planets\_mass}. Deleting the terms in \textcolor{blue}{blue} results in the second completion}
\label{fig:planets_mass_46}
\end{figure}

\subsection{All Prompt Trajectory Graphs}

\begin{figure*}[t]
\centering
\includegraphics[width=0.99\textwidth]{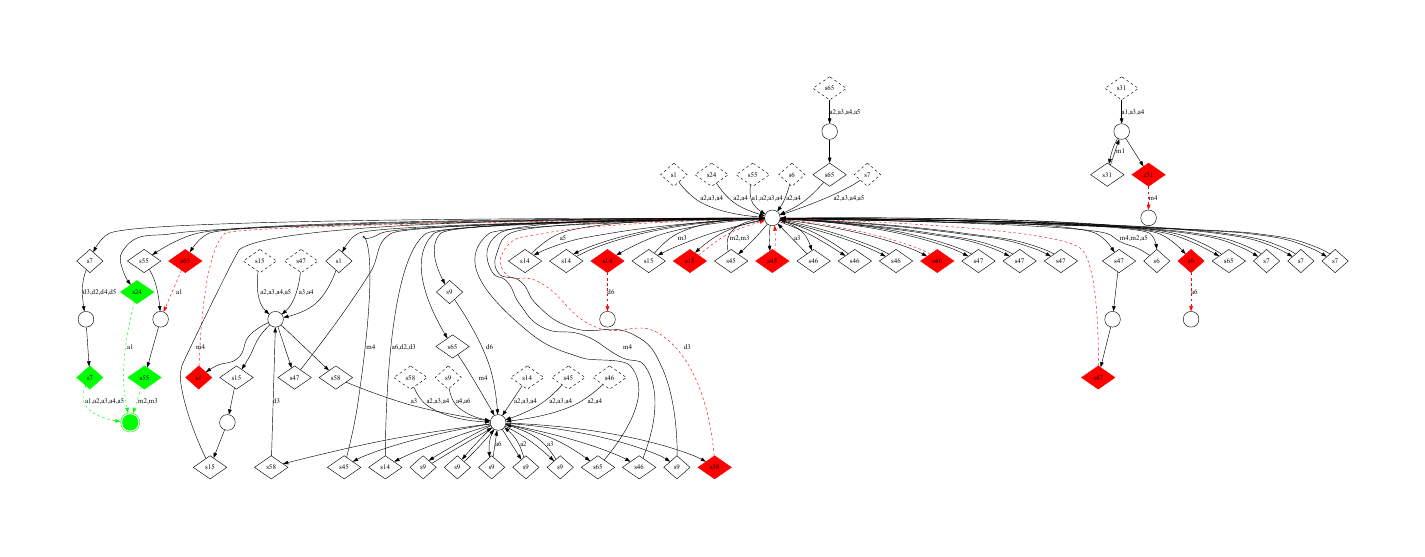}
\caption{Prompt trajectories for the ``add up'' problem.}
\label{fig:add_up}
\end{figure*}

\begin{figure*}[t!]
\centering
\includegraphics[width=0.99\textwidth]{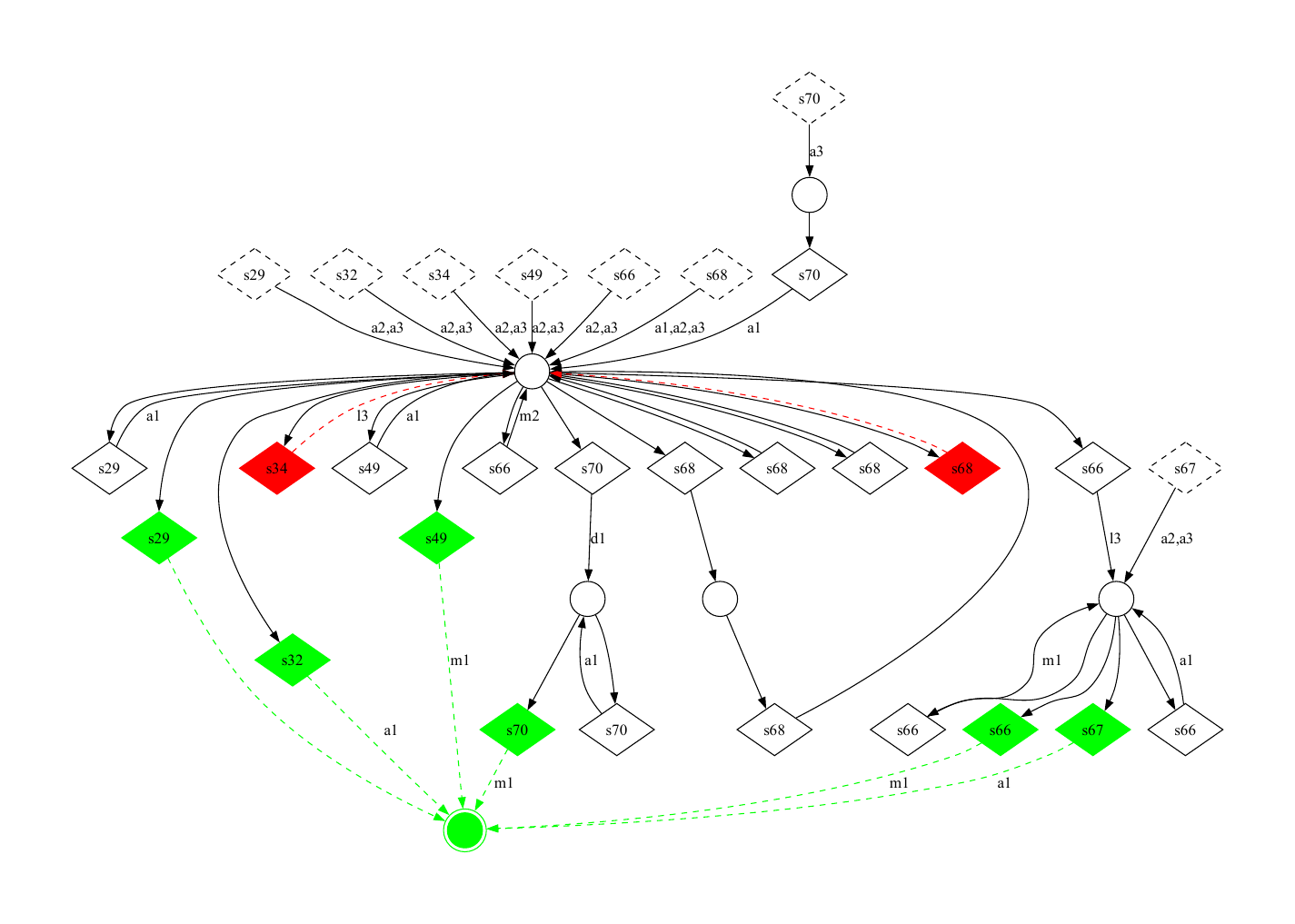}
\caption{Prompt trajectories for the ``check prime'' problem.}
\label{fig:check_prime}
\end{figure*}

\begin{figure*}[t]
\centering
\includegraphics[width=0.99\textwidth]{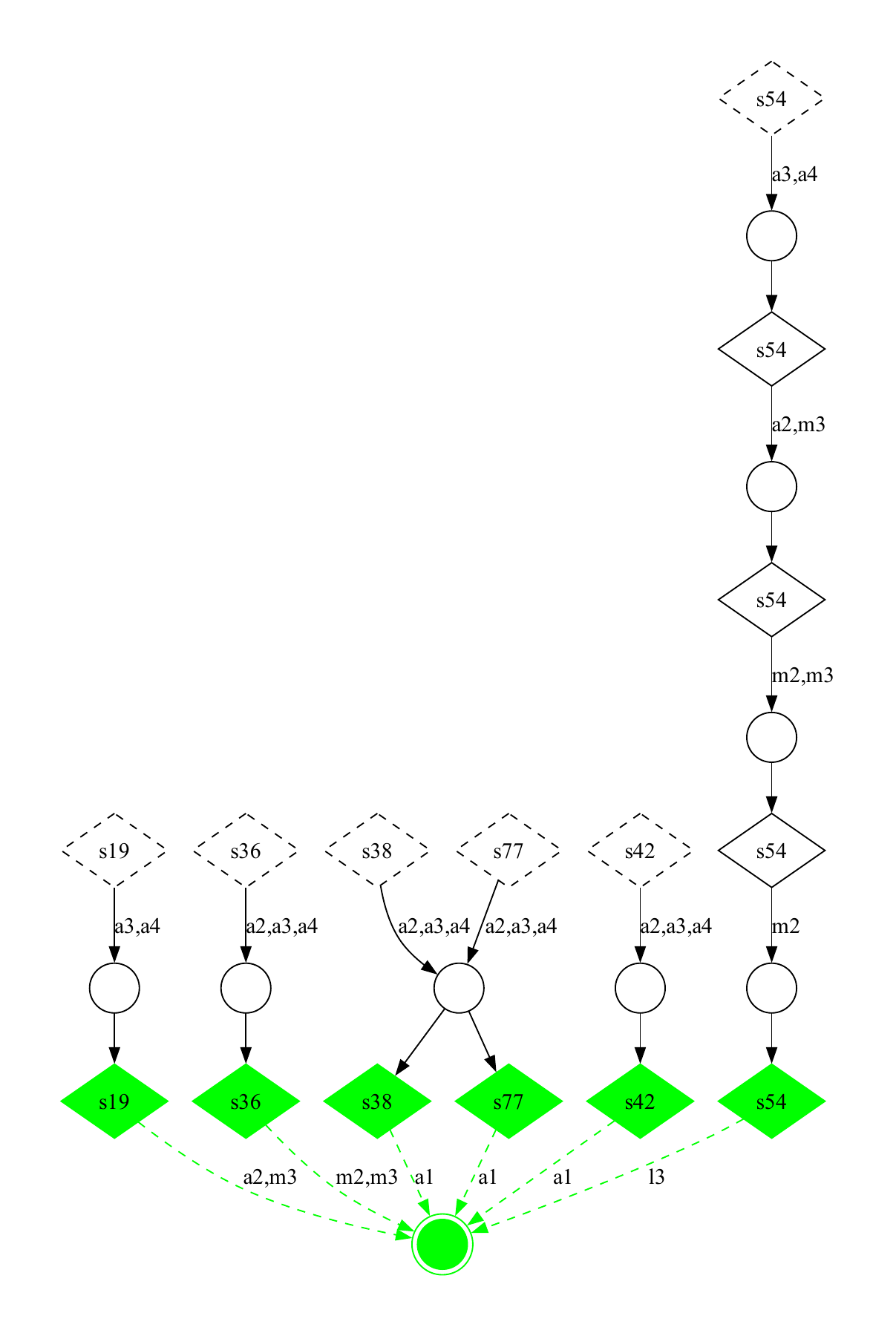}
\caption{Prompt trajectories for the ``add int'' problem.}
\label{fig:add_int}
\end{figure*}

\begin{figure*}[t]
\centering
\includegraphics[height=0.9\textheight]{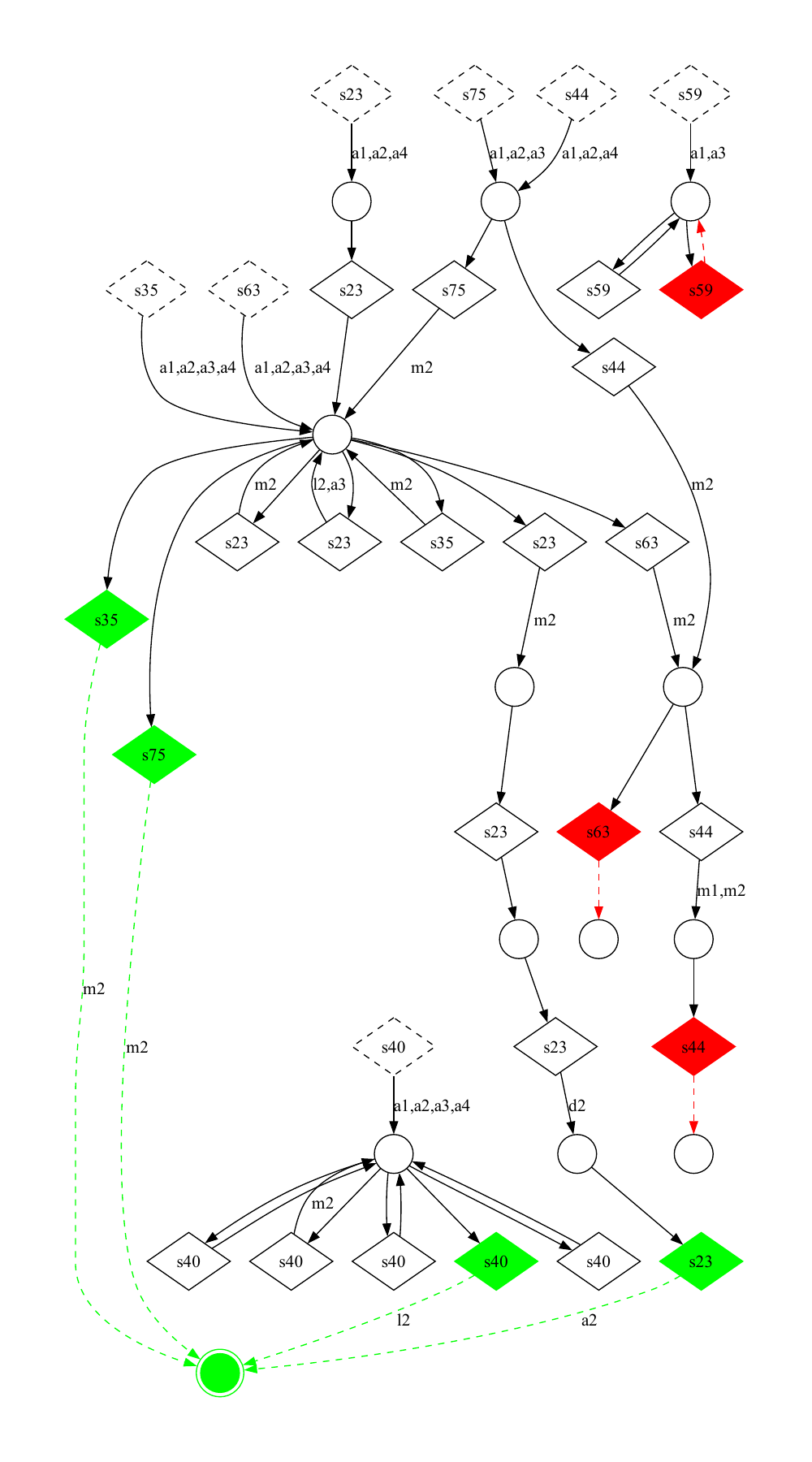}
\caption{Prompt trajectories for the ``altText'' problem.}
\label{fig:altText}
\end{figure*}

\begin{figure*}[t]
\centering
\includegraphics[height=0.7\textheight]{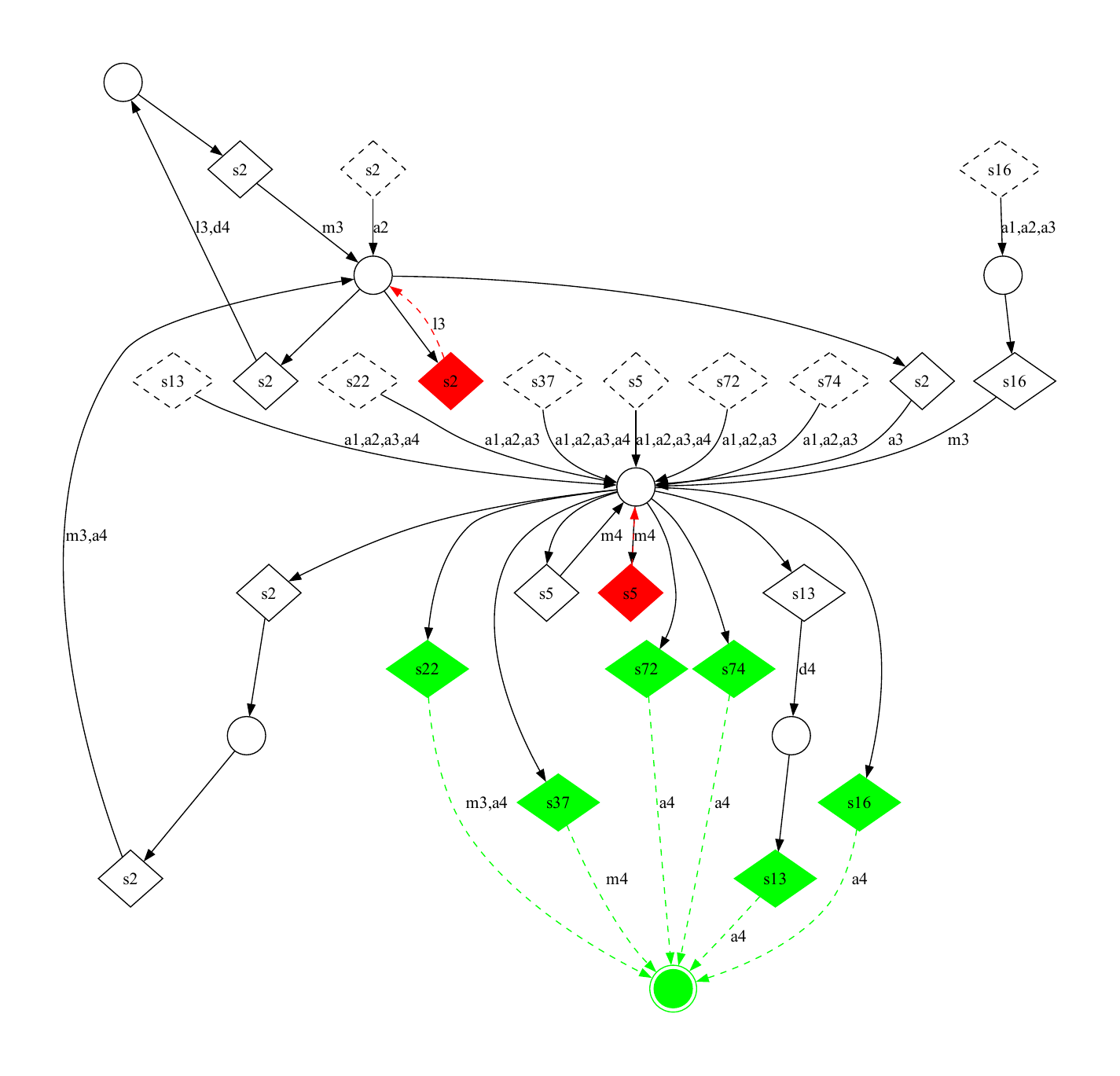}
\caption{Prompt trajectories for the ``assessVowels'' problem.}
\label{fig:assessVowels}
\end{figure*}

\begin{figure*}[t]
\centering
\includegraphics[height=0.9\textheight]{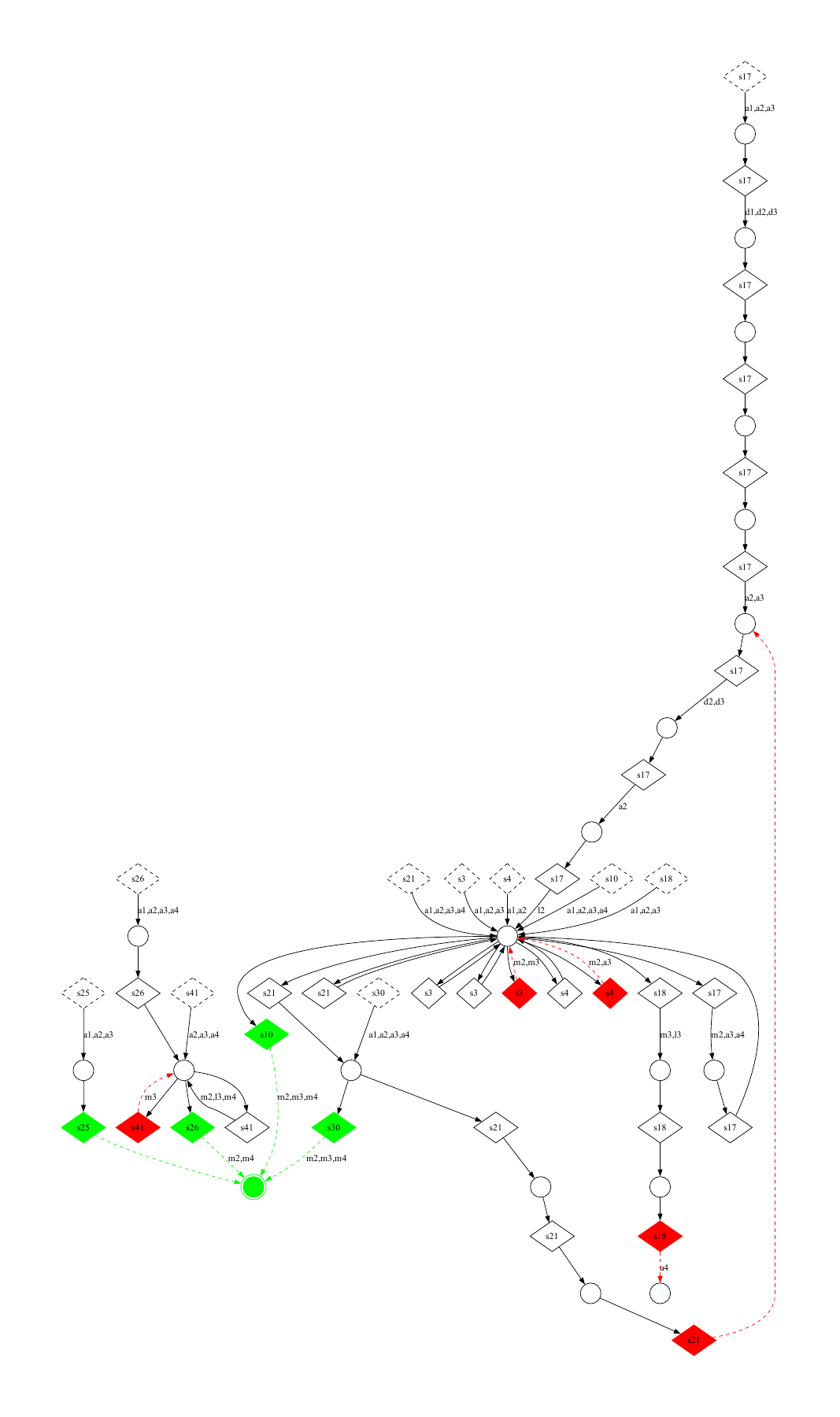}
\caption{Prompt trajectories for the ``changeSection'' problem.}
\label{fig:changeSection}
\end{figure*}

\begin{figure*}[t]
\centering
\includegraphics[width=0.99\textwidth]{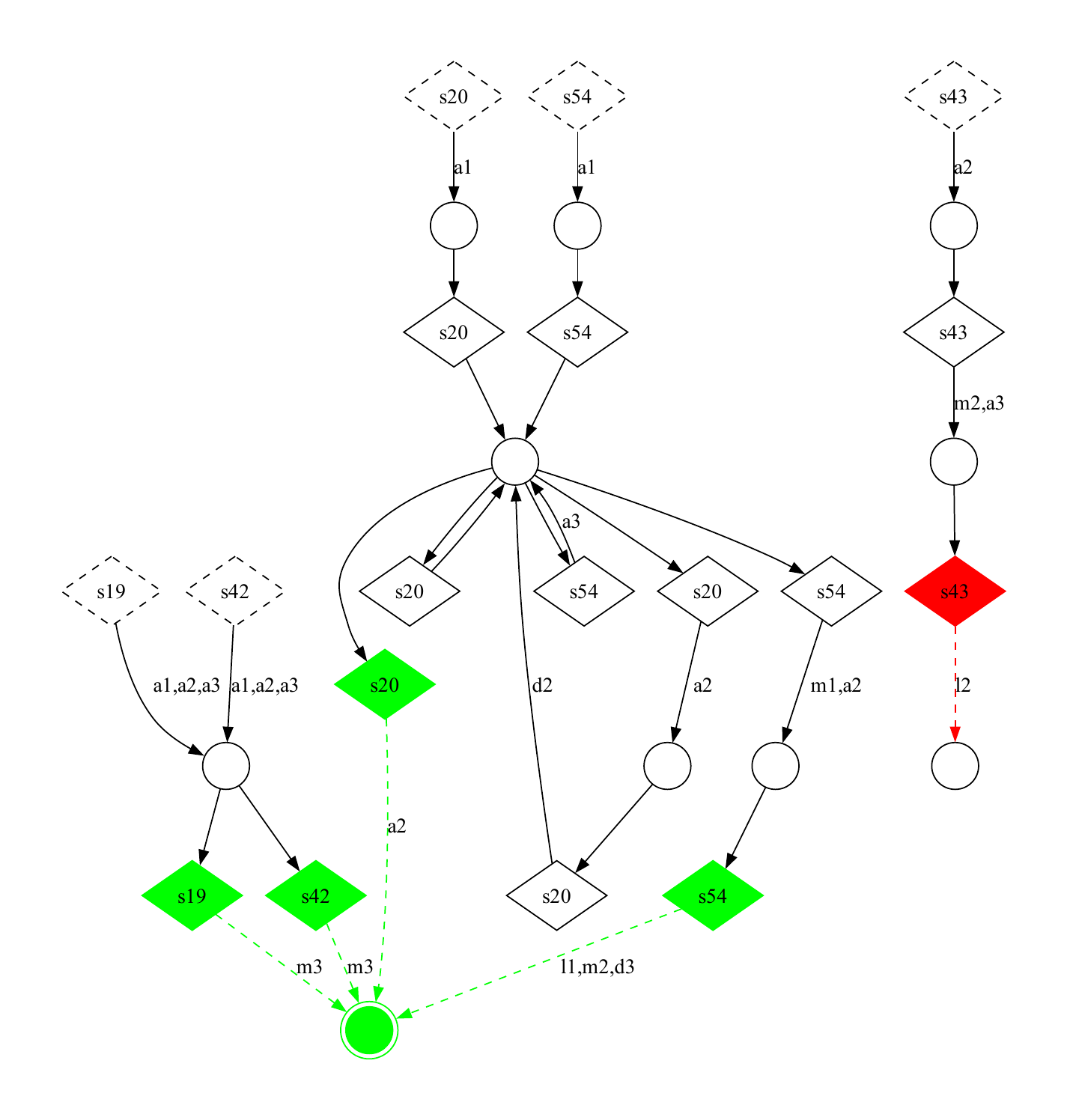}
\caption{Prompt trajectories for the ``combine'' problem.}
\label{fig:combine}
\end{figure*}

\begin{figure*}[t]
\centering
\includegraphics[height=0.9\textheight]{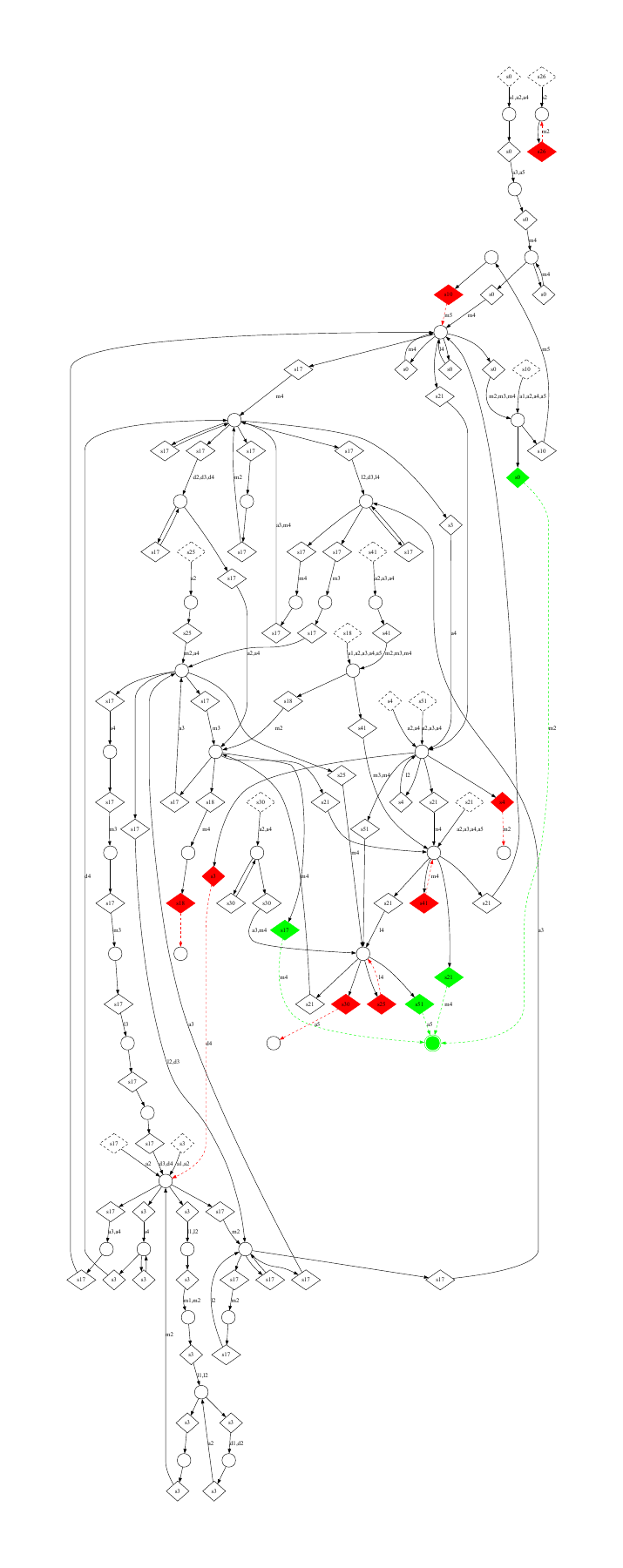}
\caption{Prompt trajectories for the ``convert'' problem.}
\label{fig:convert}
\end{figure*}

\begin{figure*}[t]
\centering
\includegraphics[width=0.99\textwidth]{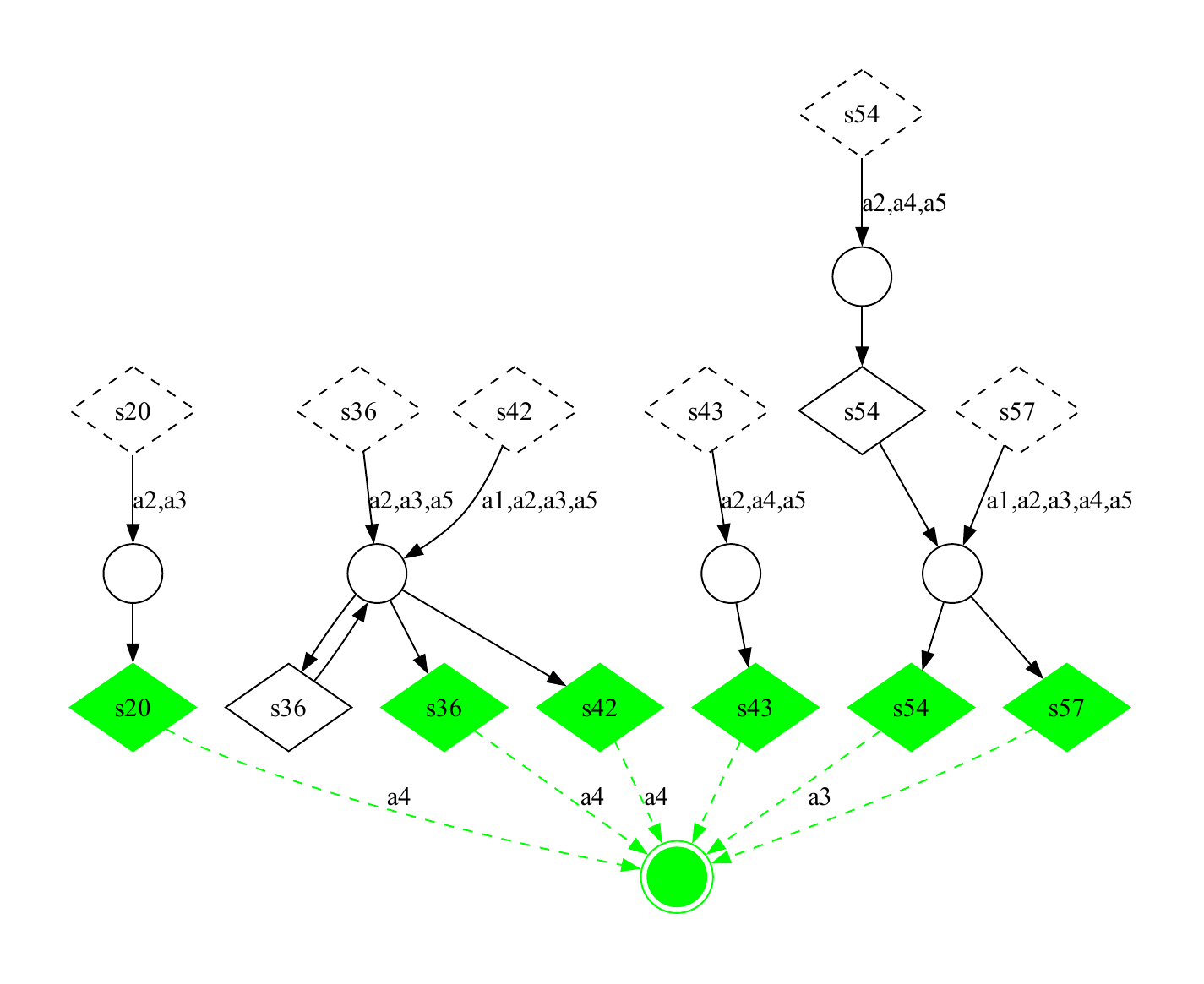}
\caption{Prompt trajectories for the ``create list'' problem.}
\label{fig:create_list}
\end{figure*}

\begin{figure*}[t]
\centering
\includegraphics[width=0.99\textwidth]{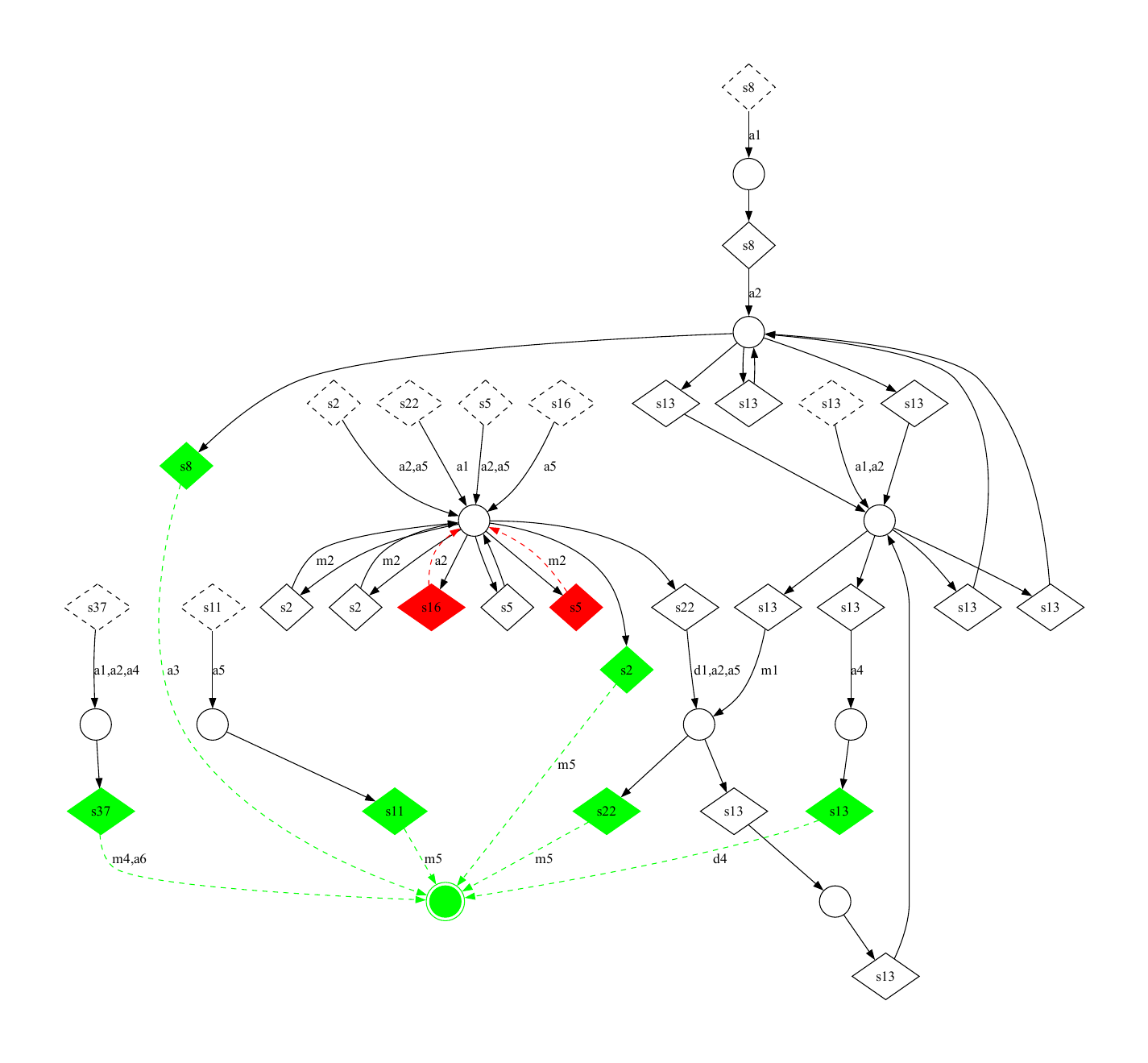}
\caption{Prompt trajectories for the ``fib'' problem.}
\label{fig:fib}
\end{figure*}

\begin{figure*}[t]
\centering
\includegraphics[width=0.99\textwidth]{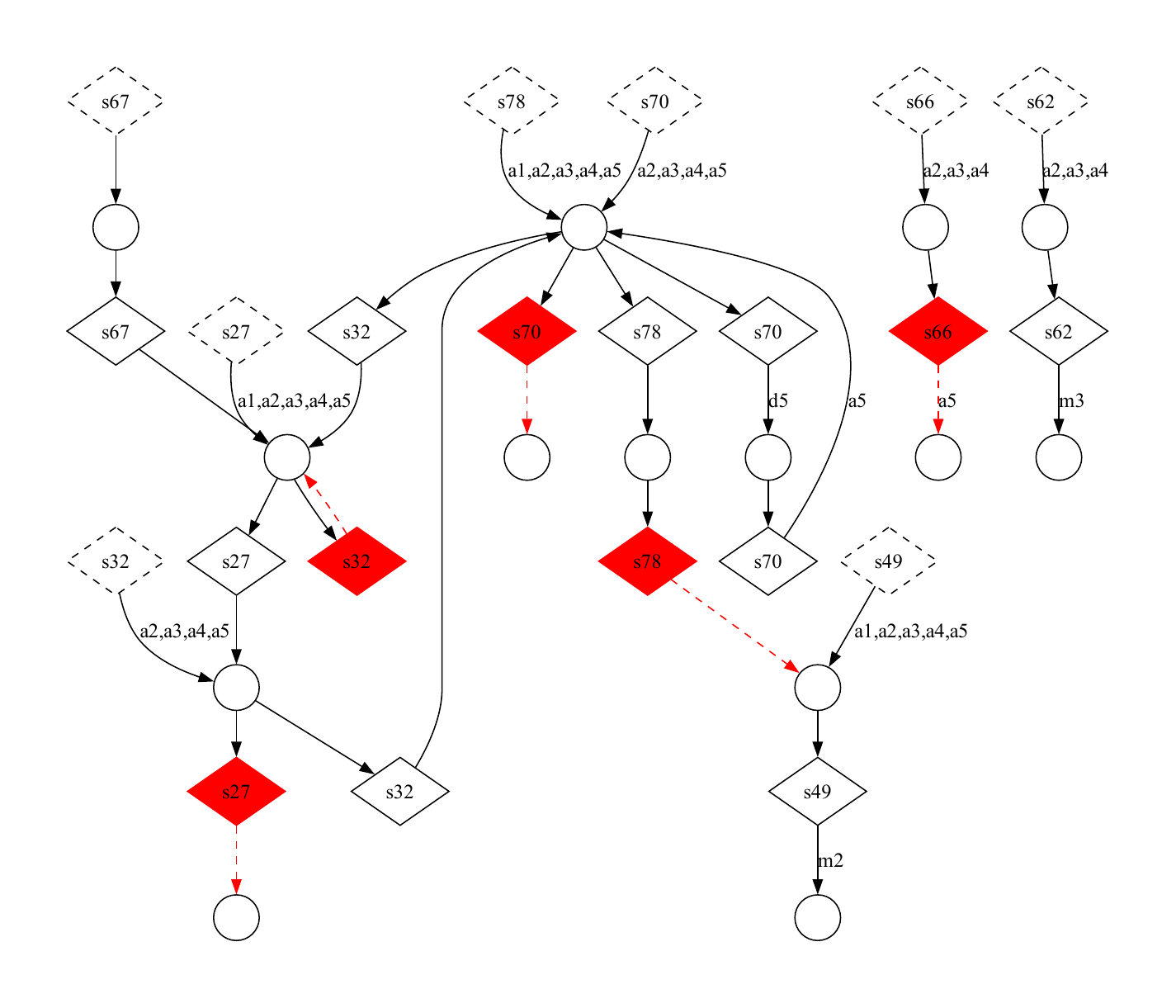}
\caption{Prompt trajectories for the ``findHorizontals'' problem.}
\label{fig:findHorizontals}
\end{figure*}

\begin{figure*}[t]
\centering
\includegraphics[width=0.99\textwidth]{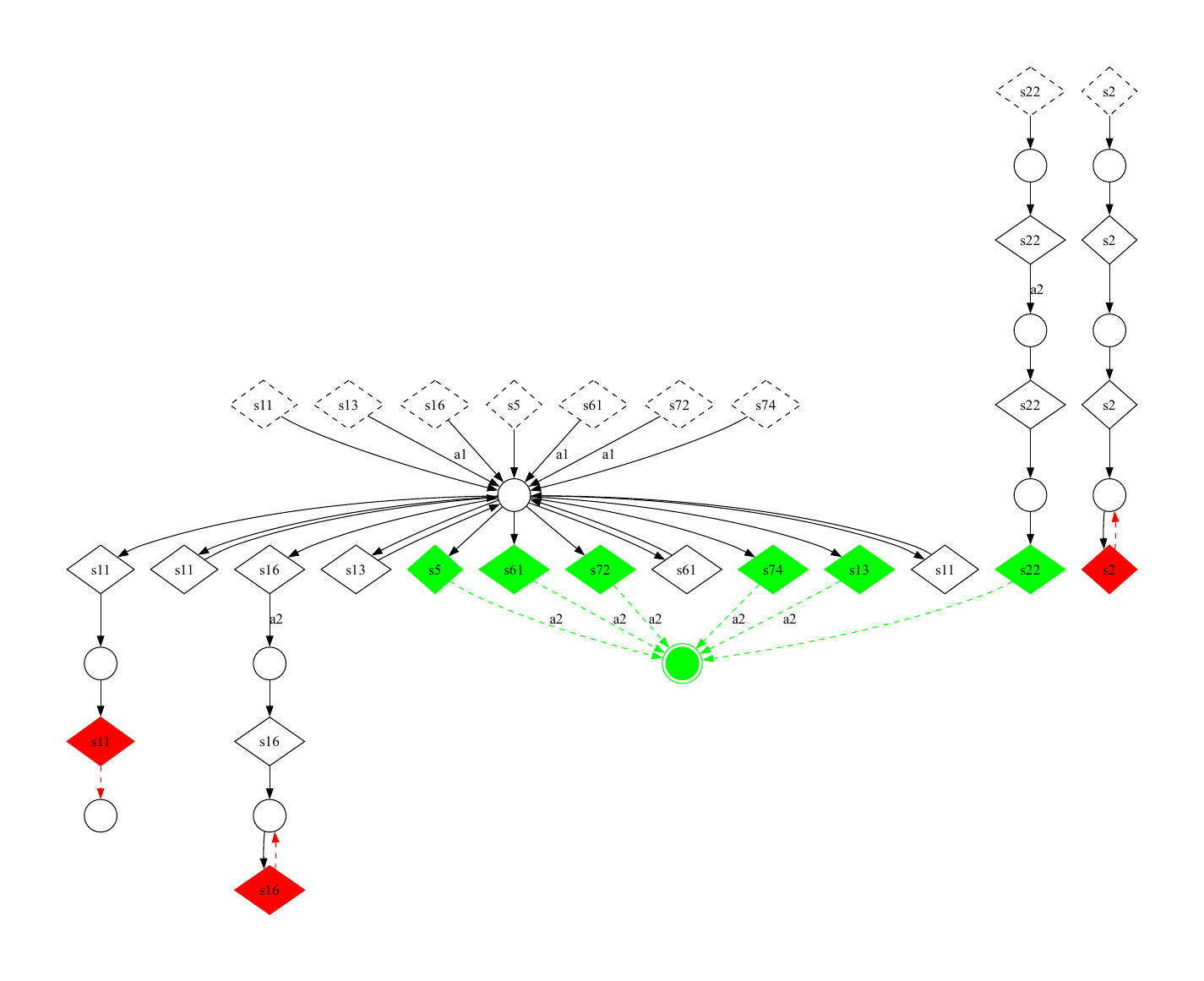}
\caption{Prompt trajectories for the ``find multiples'' problem.}
\label{fig:find_multiples}
\end{figure*}

\begin{figure*}[t]
\centering
\includegraphics[width=0.99\textwidth]{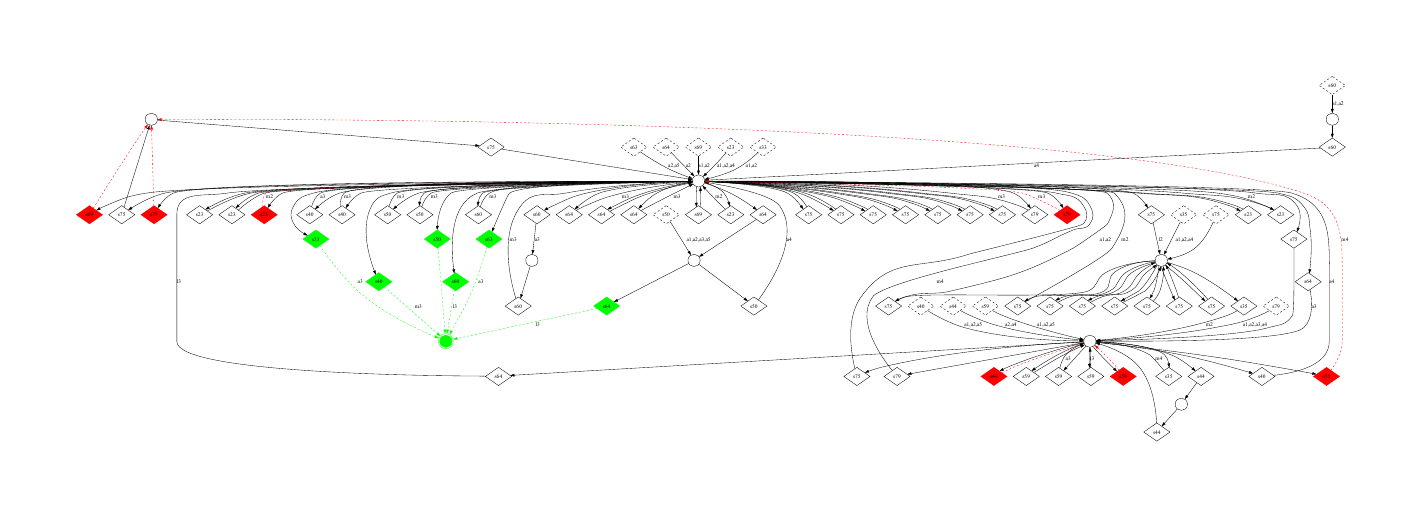}
\caption{Prompt trajectories for the ``generateCardDeck'' problem.}
\label{fig:generateCardDeck}
\end{figure*}

\begin{figure*}[t]
\centering
\includegraphics[width=0.99\textwidth]{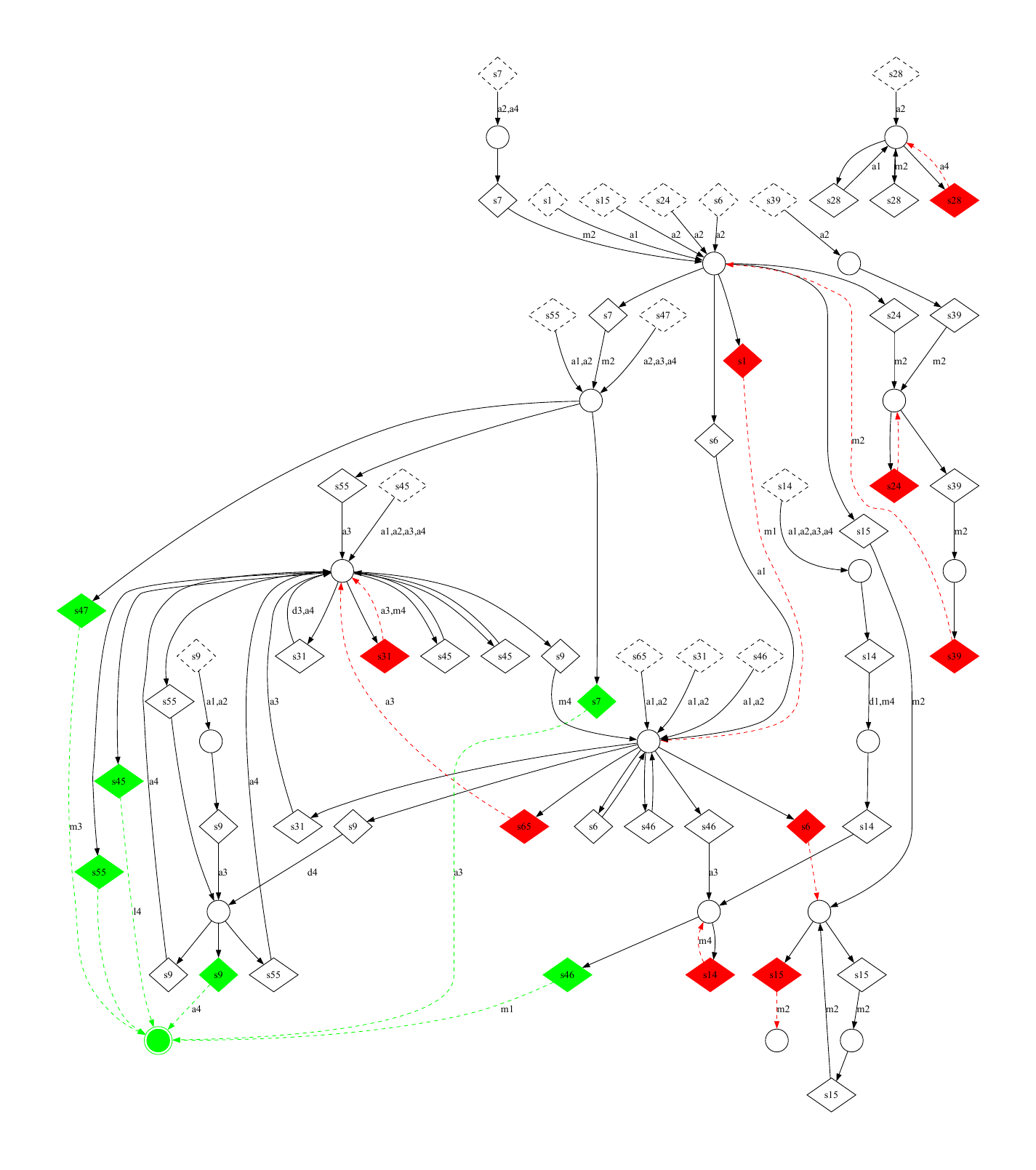}
\caption{Prompt trajectories for the ``getSeason'' problem.}
\label{fig:getSeason}
\end{figure*}

\begin{figure*}[t]
\centering
\includegraphics[width=0.99\textwidth]{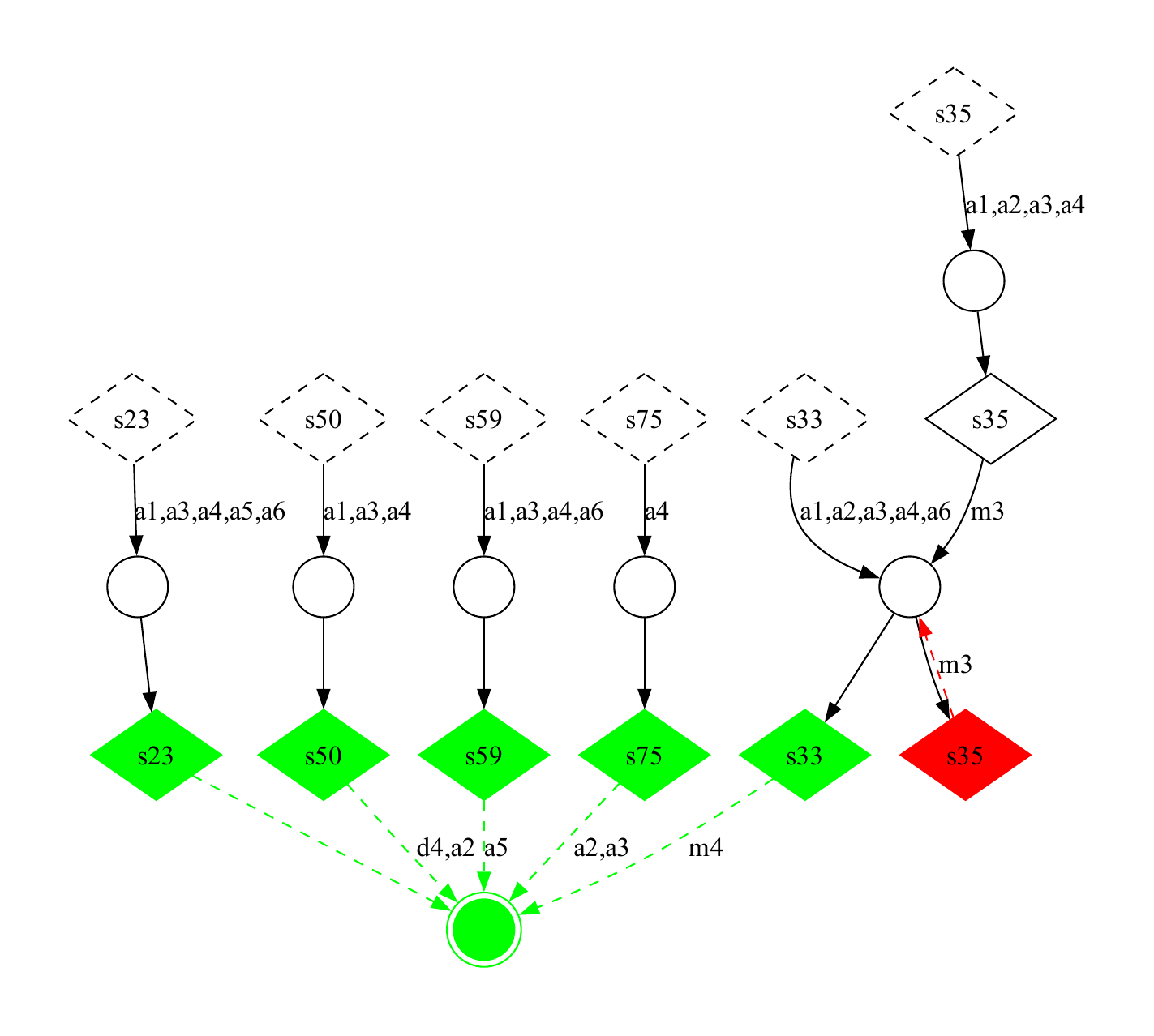}
\caption{Prompt trajectories for the ``increaseScore'' problem.}
\label{fig:increaseScore}
\end{figure*}

\begin{figure*}[t]
\centering
\includegraphics[width=0.9\textwidth]{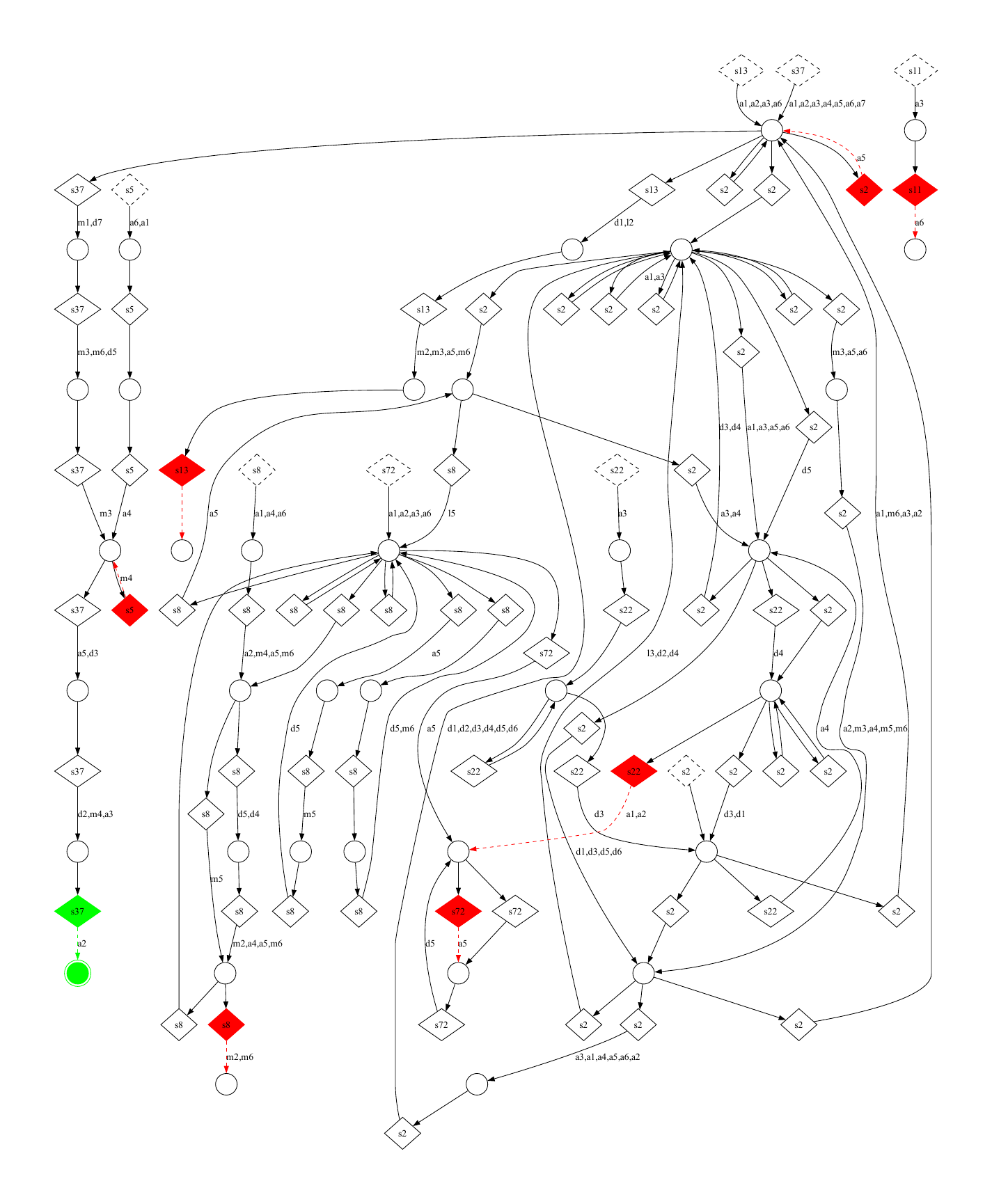}
\caption{Prompt trajectories for the ``laugh'' problem.}
\label{fig:laugh}
\end{figure*}

\begin{figure*}[t]
\centering
\includegraphics[height=0.9\textheight]{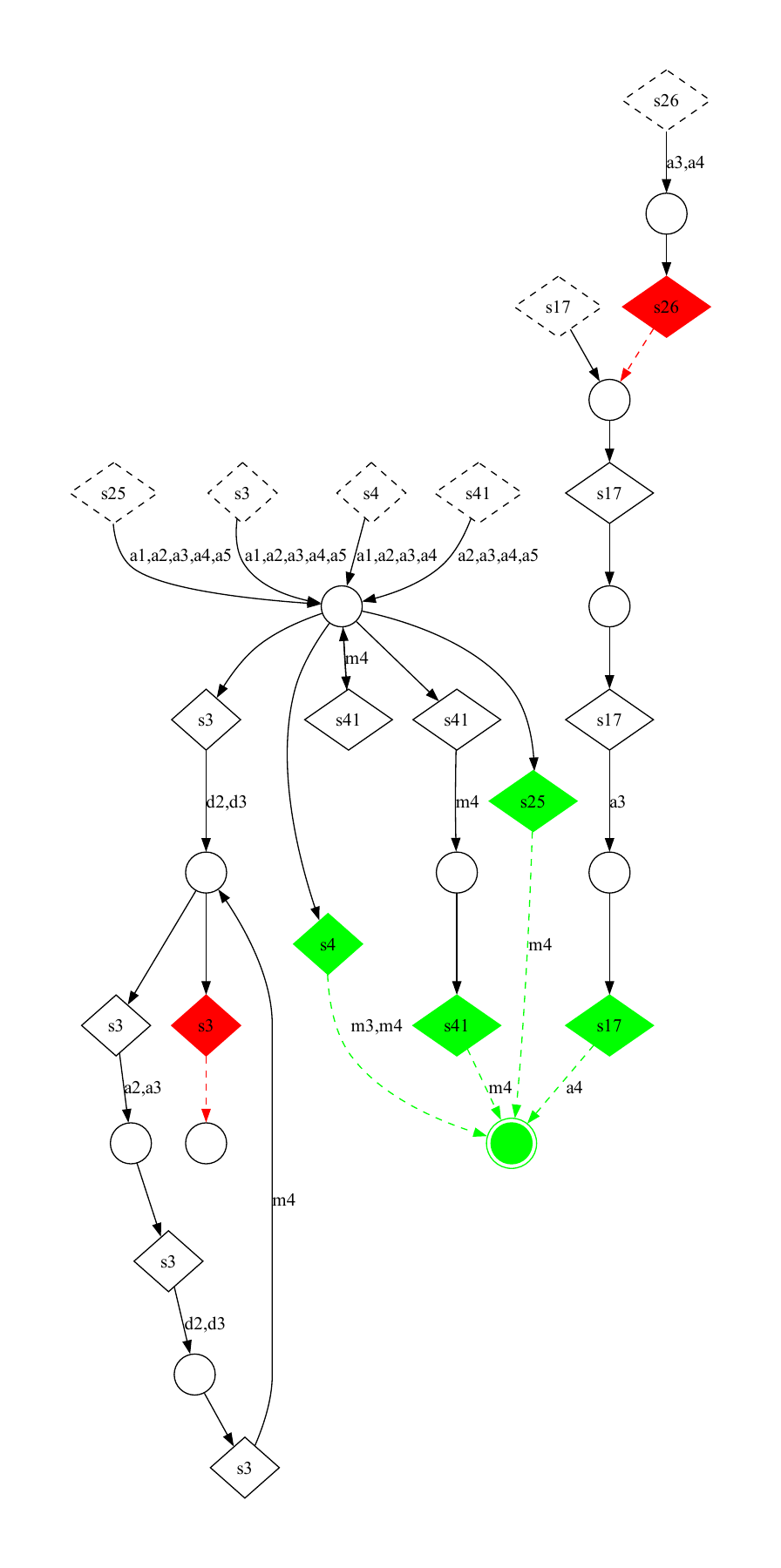}
\caption{Prompt trajectories for the ``pattern'' problem.}
\label{fig:pattern}
\end{figure*}

\begin{figure*}[t]
\centering
\includegraphics[width=0.99\textwidth]{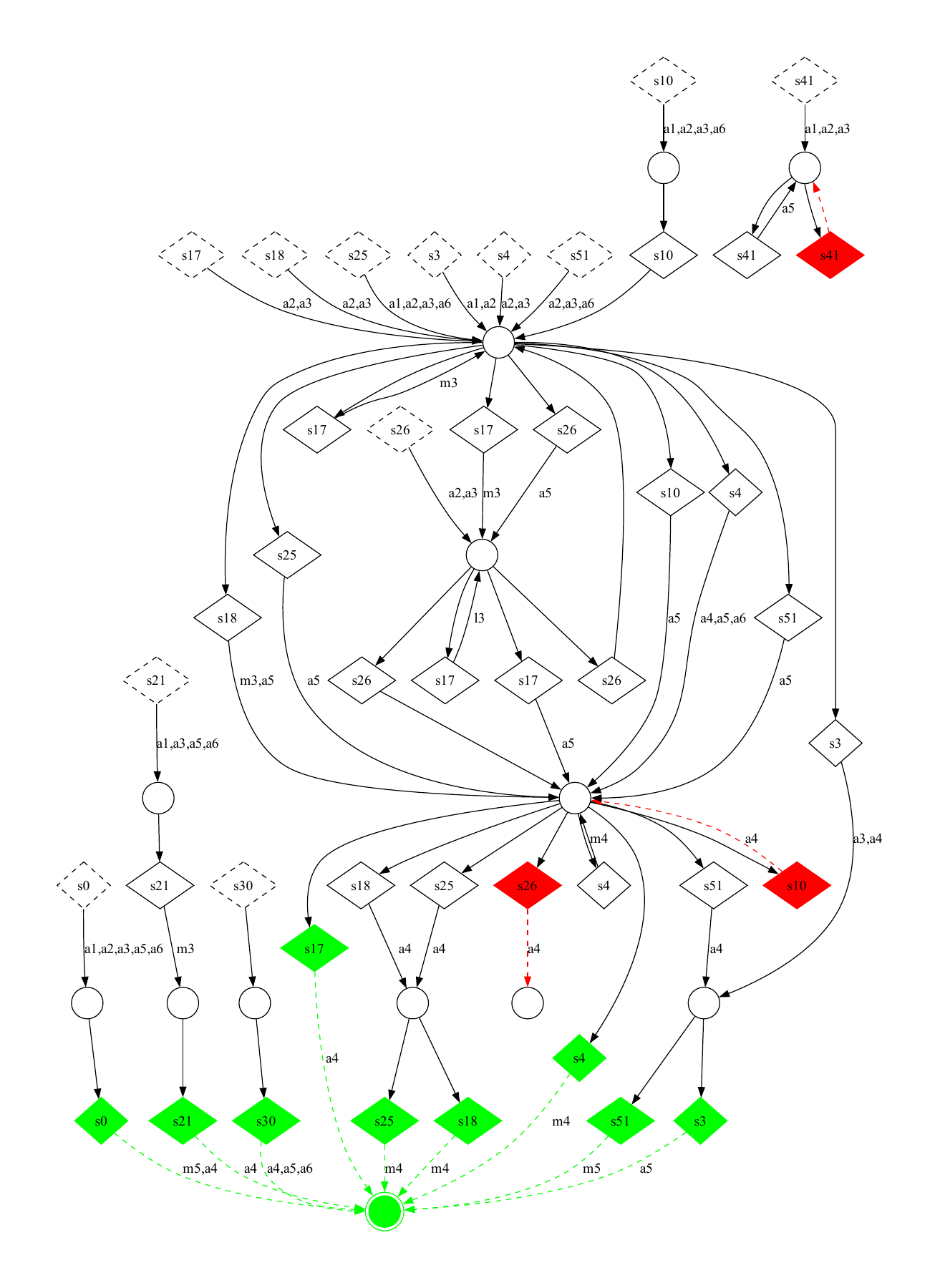}
\caption{Prompt trajectories for the ``percentWin'' problem.}
\label{fig:percentWin}
\end{figure*}

\begin{figure*}[t]
\centering
\includegraphics[width=0.99\textwidth]{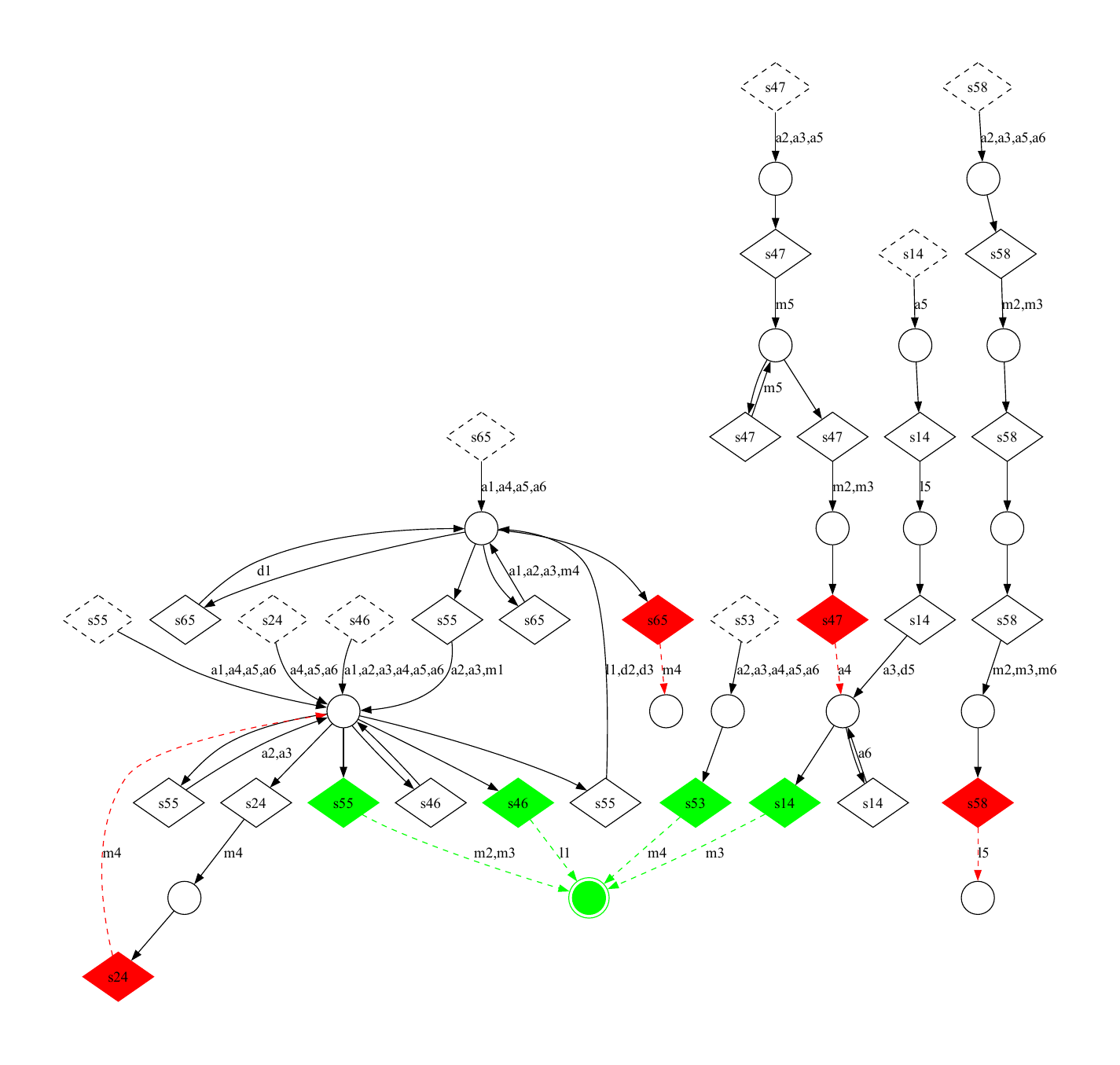}
\caption{Prompt trajectories for the ``planets mass'' problem.}
\label{fig:planets_mass}
\end{figure*}

\begin{figure*}[t]
\centering
\includegraphics[width=0.99\textwidth]{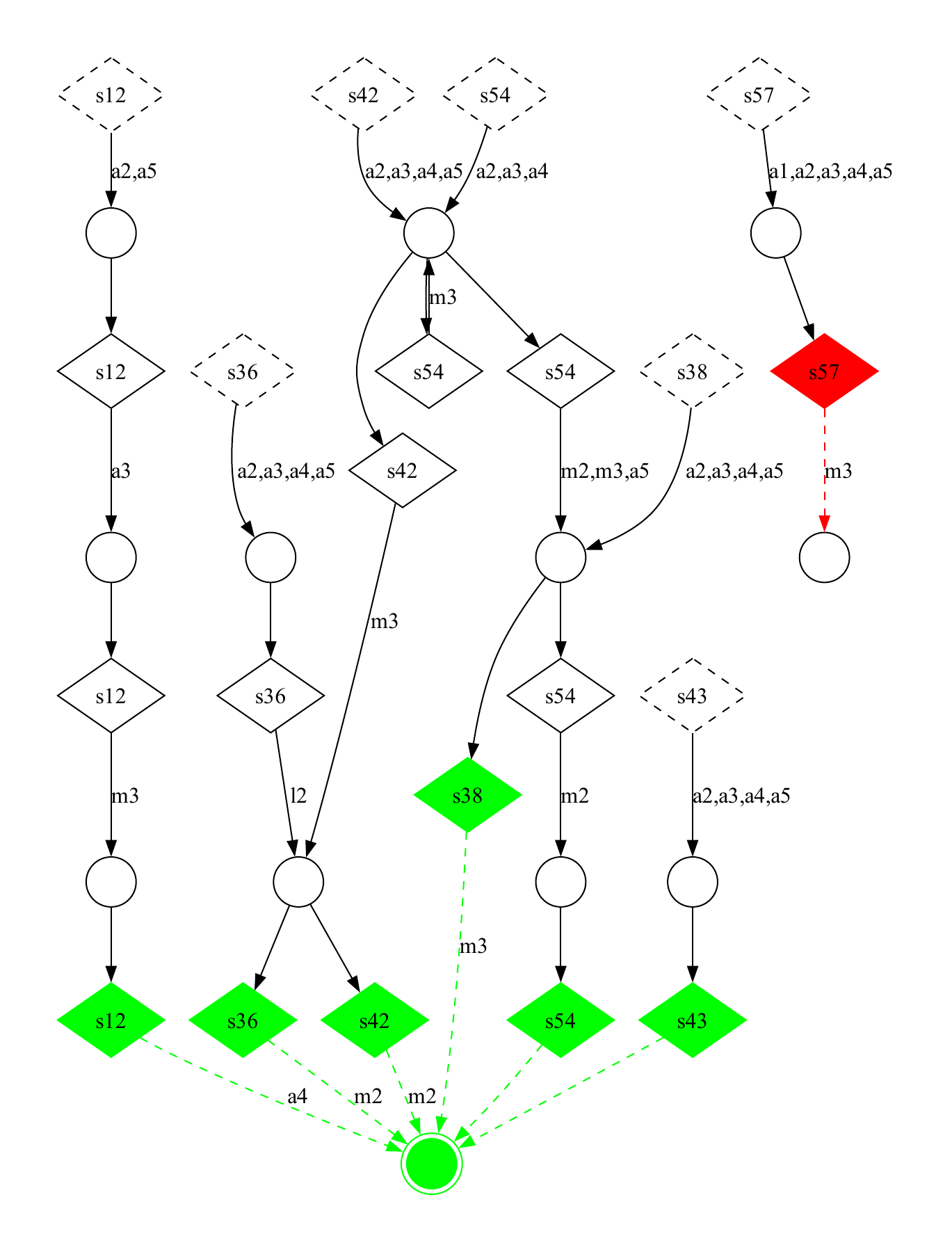}
\caption{Prompt trajectories for the ``print time'' problem.}
\label{fig:print_time}
\end{figure*}

\begin{figure*}[t]
\centering
\includegraphics[width=0.99\textwidth]{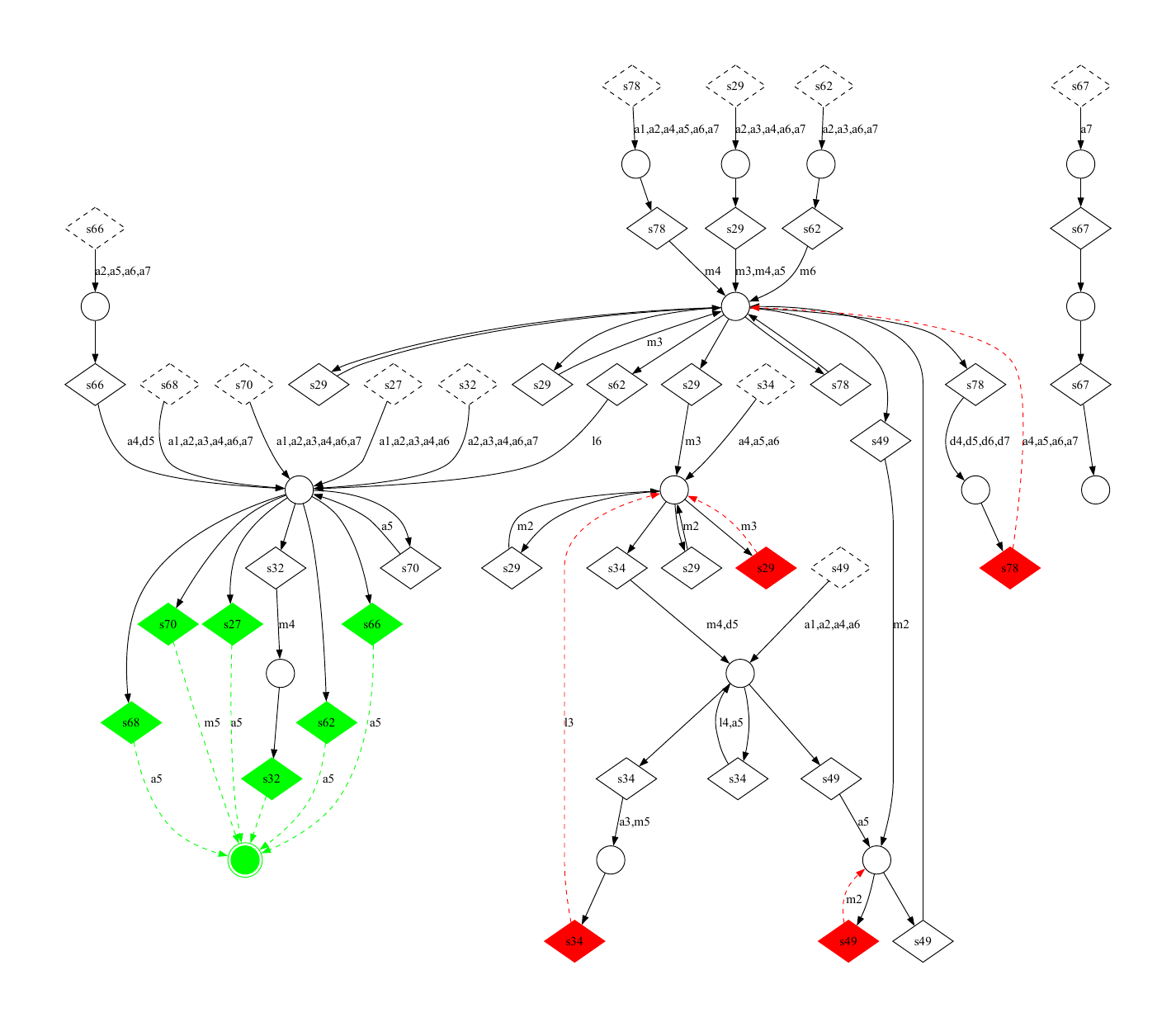}
\caption{Prompt trajectories for the ``readingIceCream'' problem.}
\label{fig:readingIceCream}
\end{figure*}

\begin{figure*}[t]
\centering
\includegraphics[width=0.99\textwidth]{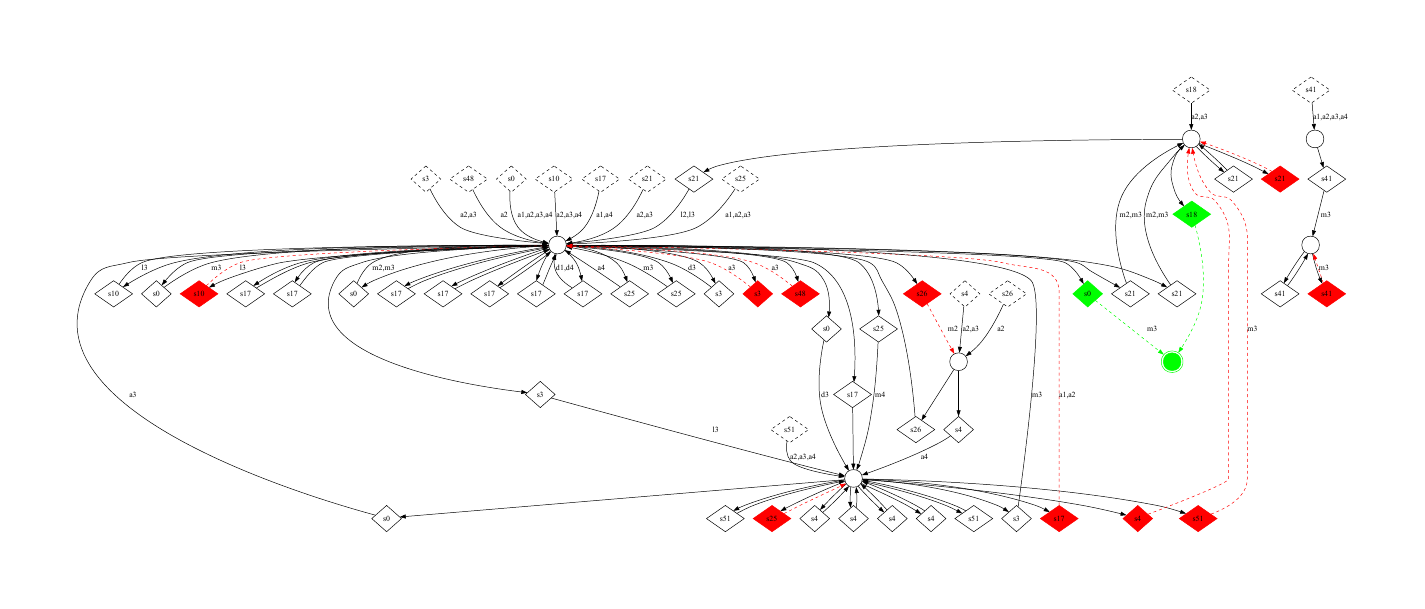}
\caption{Prompt trajectories for the ``remove odd'' problem.}
\label{fig:remove_odd}
\end{figure*}

\begin{figure*}[t]
\centering
\includegraphics[width=0.99\textwidth]{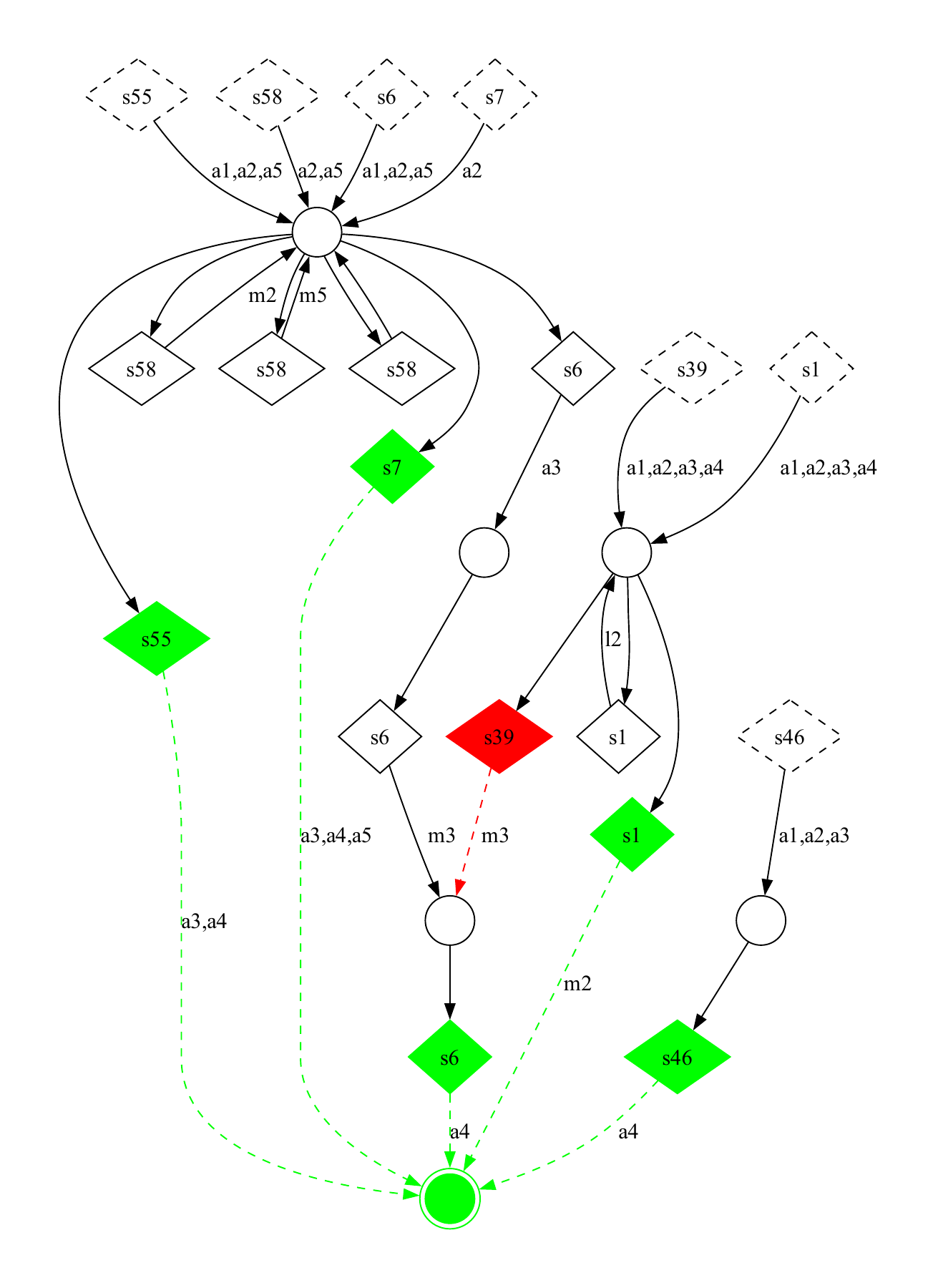}
\caption{Prompt trajectories for the ``reverseWords'' problem.}
\label{fig:reverseWords}
\end{figure*}

\begin{figure*}[t]
\centering
\includegraphics[width=0.99\textwidth]{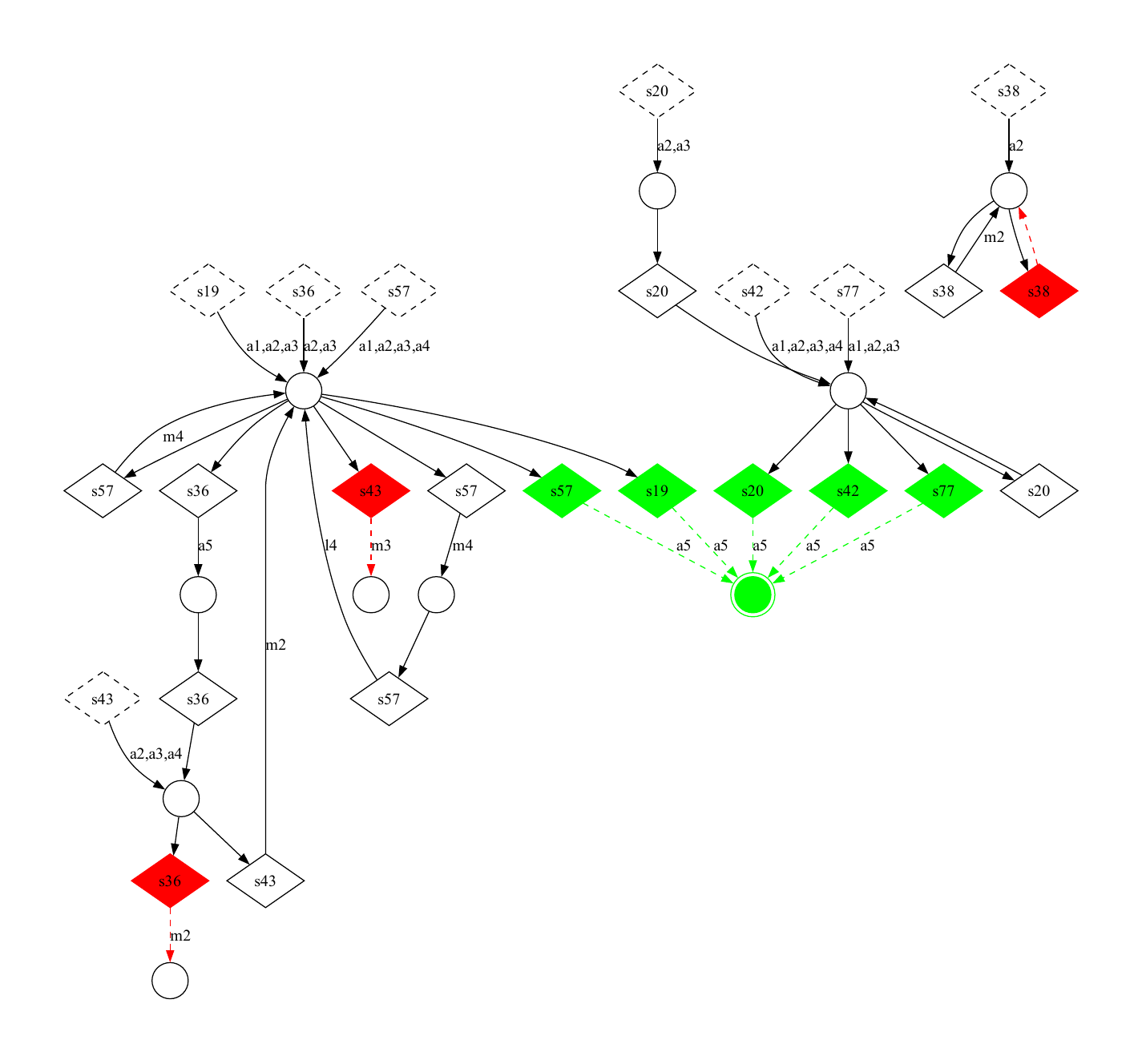}
\caption{Prompt trajectories for the ``set chars'' problem.}
\label{fig:set_chars}
\end{figure*}

\begin{figure*}[t]
\centering
\includegraphics[width=0.99\textwidth]{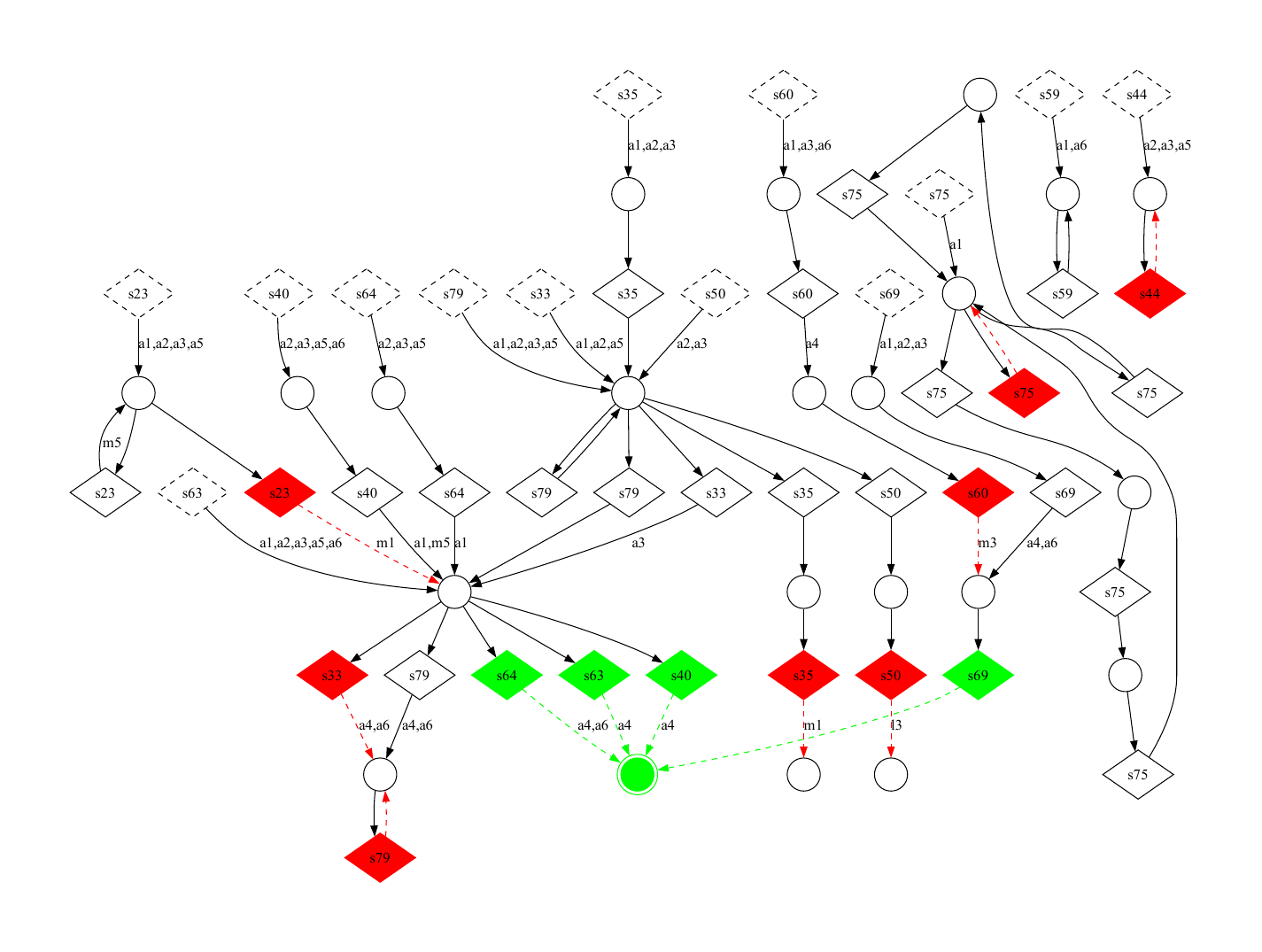}
\caption{Prompt trajectories for the ``sortBySuccessRate'' problem.}
\label{fig:sortBySuccessRate}
\end{figure*}

\begin{figure*}[t]
\centering
\includegraphics[width=0.99\textwidth]{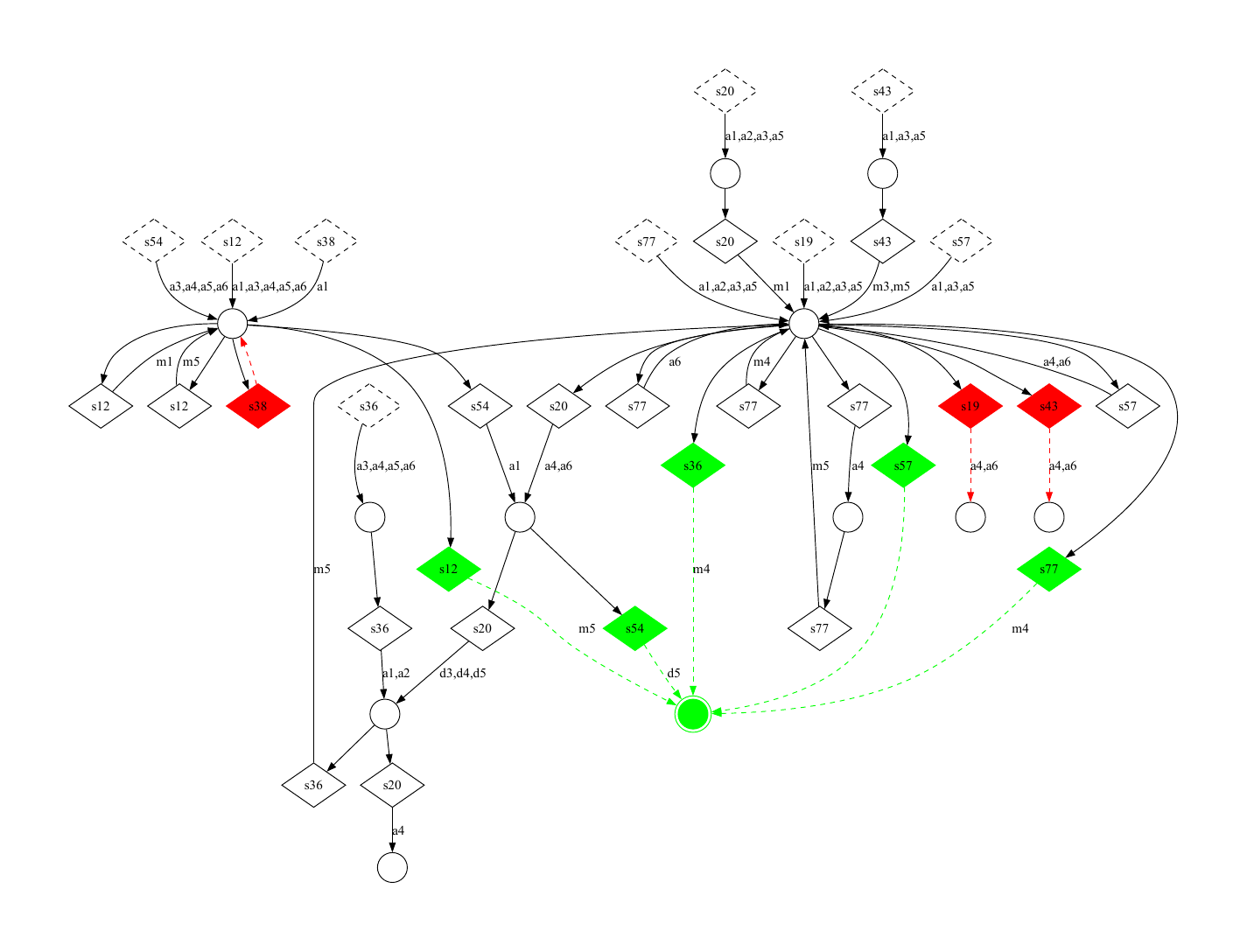}
\caption{Prompt trajectories for the ``sort physicists'' problem.}
\label{fig:sort_physicists}
\end{figure*}

\begin{figure*}[t]
\centering
\includegraphics[width=0.99\textwidth]{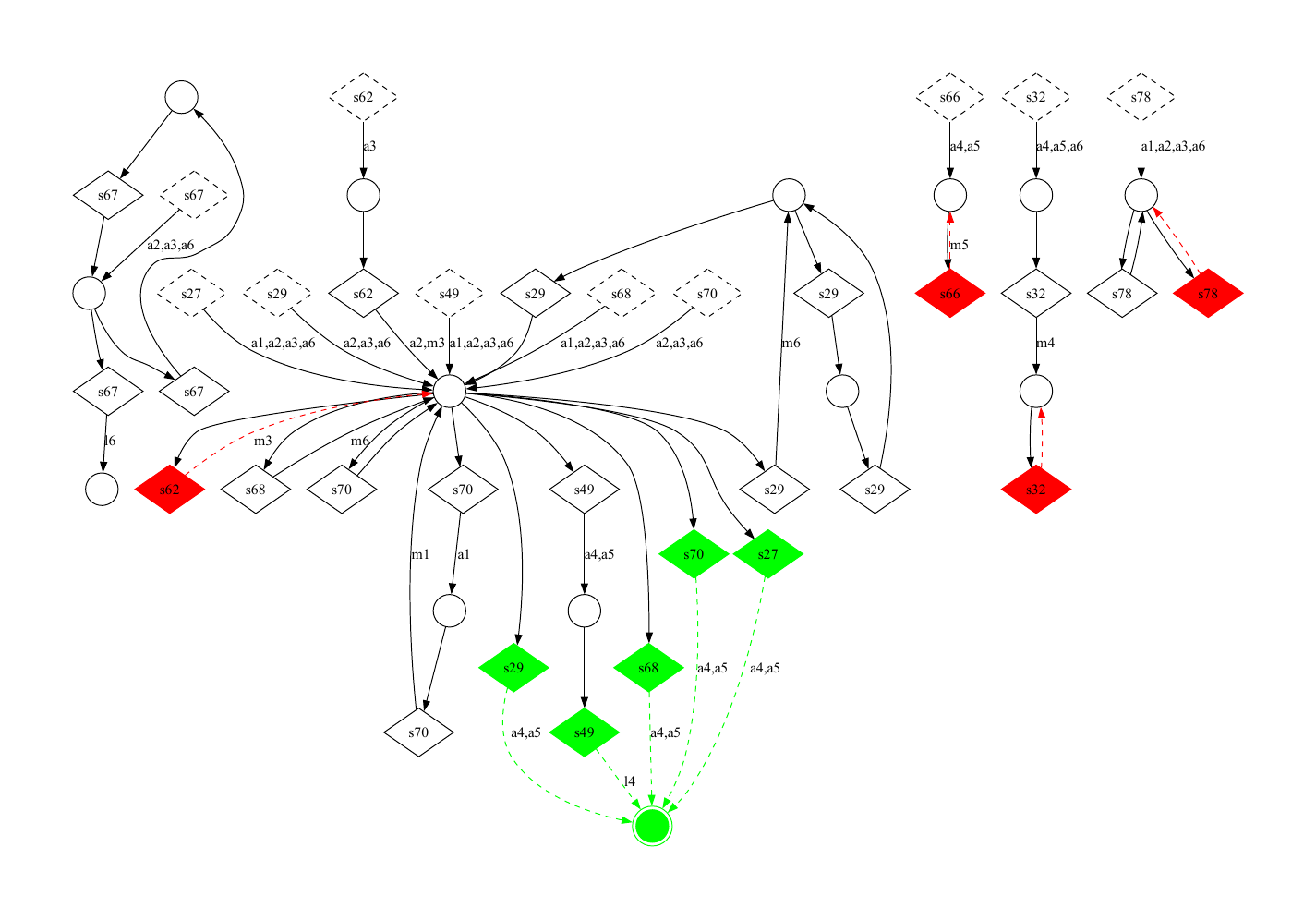}
\caption{Prompt trajectories for the ``sortedBooks'' problem.}
\label{fig:sortedBooks}
\end{figure*}

\begin{figure*}[t]
\centering
\includegraphics[width=0.99\textwidth]{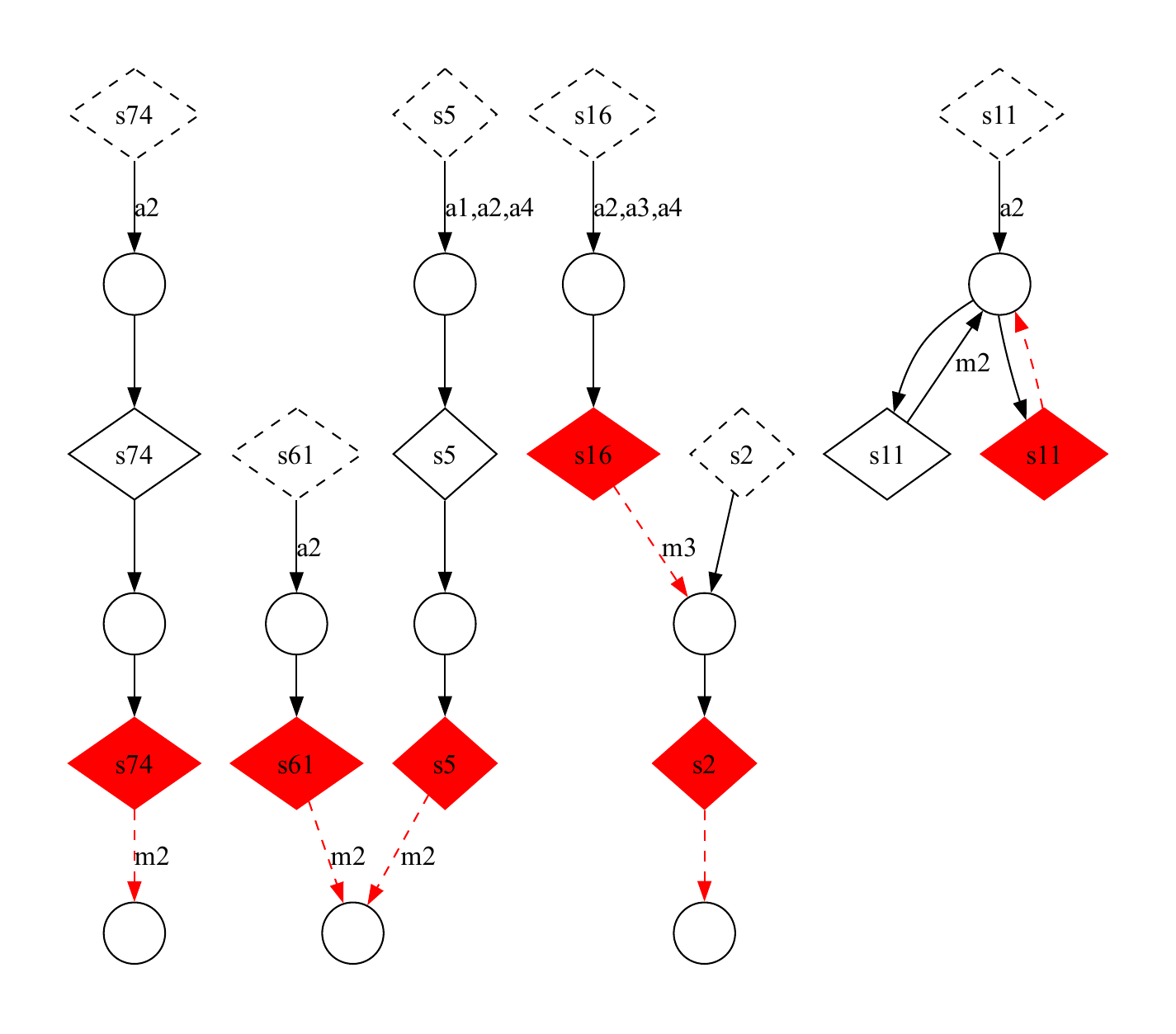}
\caption{Prompt trajectories for the ``student grades'' problem.}
\label{fig:student_grades}
\end{figure*}

\begin{figure*}[t]
\centering
\includegraphics[width=0.99\textwidth]{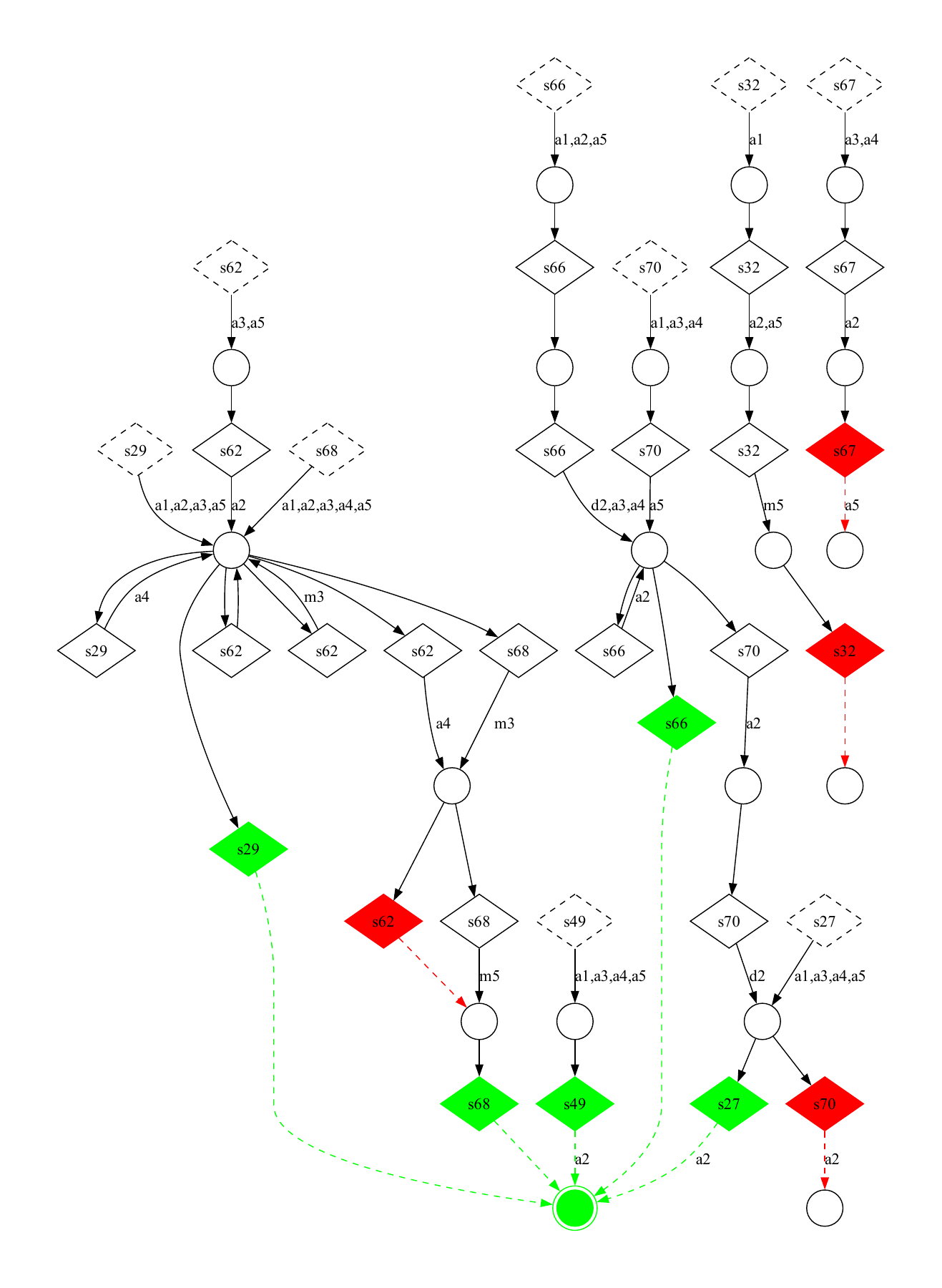}
\caption{Prompt trajectories for the ``subtract add'' problem.}
\label{fig:subtract_add}
\end{figure*}

\begin{figure*}[t]
\centering
\includegraphics[width=0.99\textwidth]{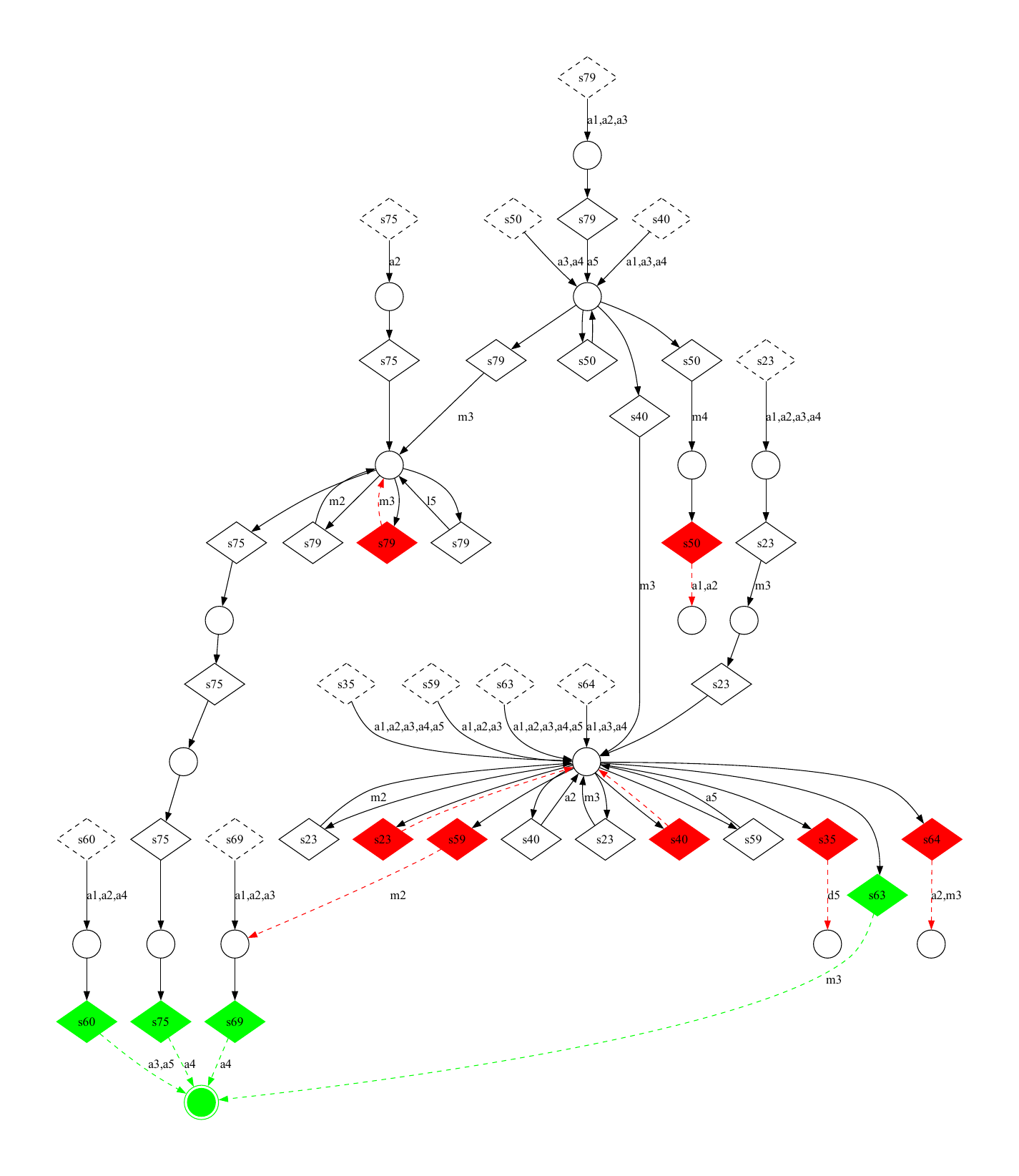}
\caption{Prompt trajectories for the ``times with'' problem.}
\label{fig:times_with}
\end{figure*}

\begin{figure*}[t]
\centering
\includegraphics[height=0.8\textheight]{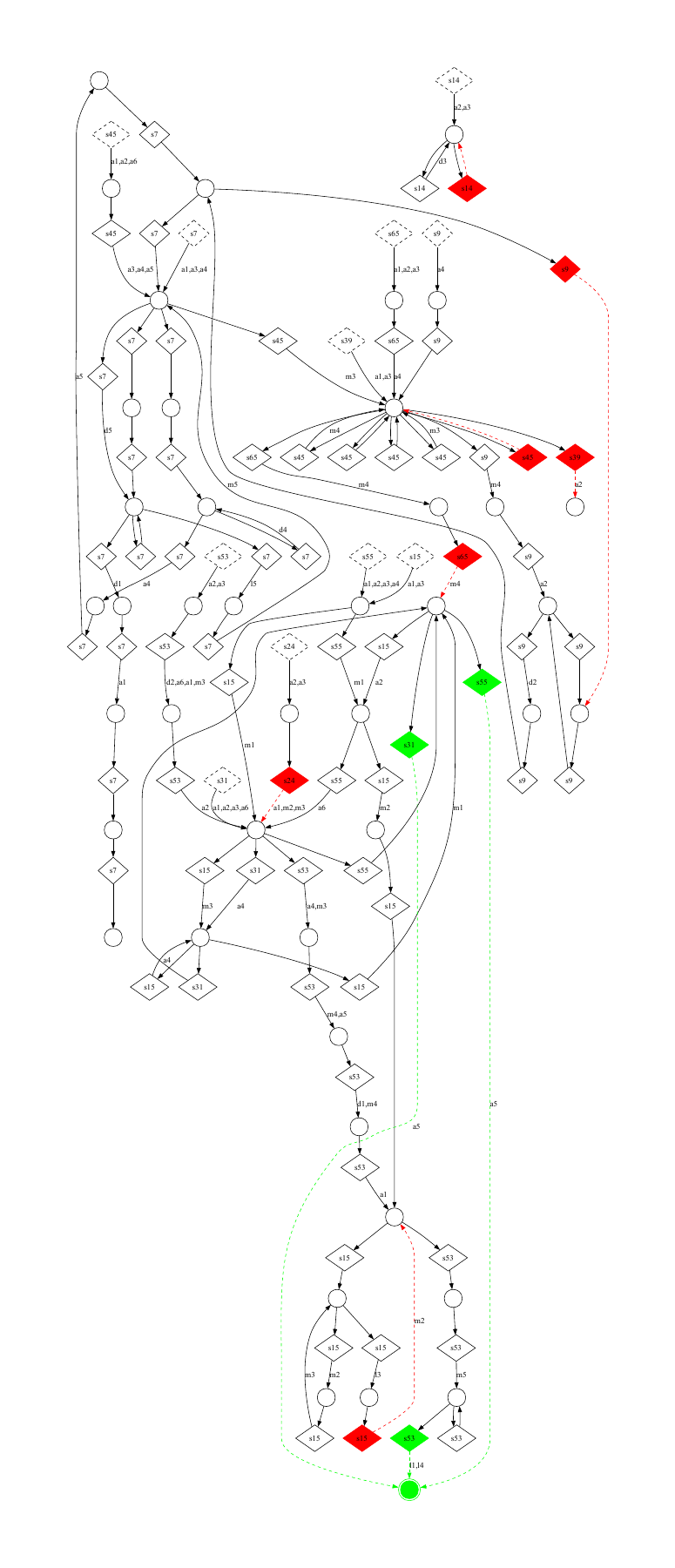}
\caption{Prompt trajectories for the ``topScores'' problem.}
\label{fig:topScores}
\end{figure*}

\begin{figure*}[t]
\centering
\includegraphics[width=0.99\textwidth]{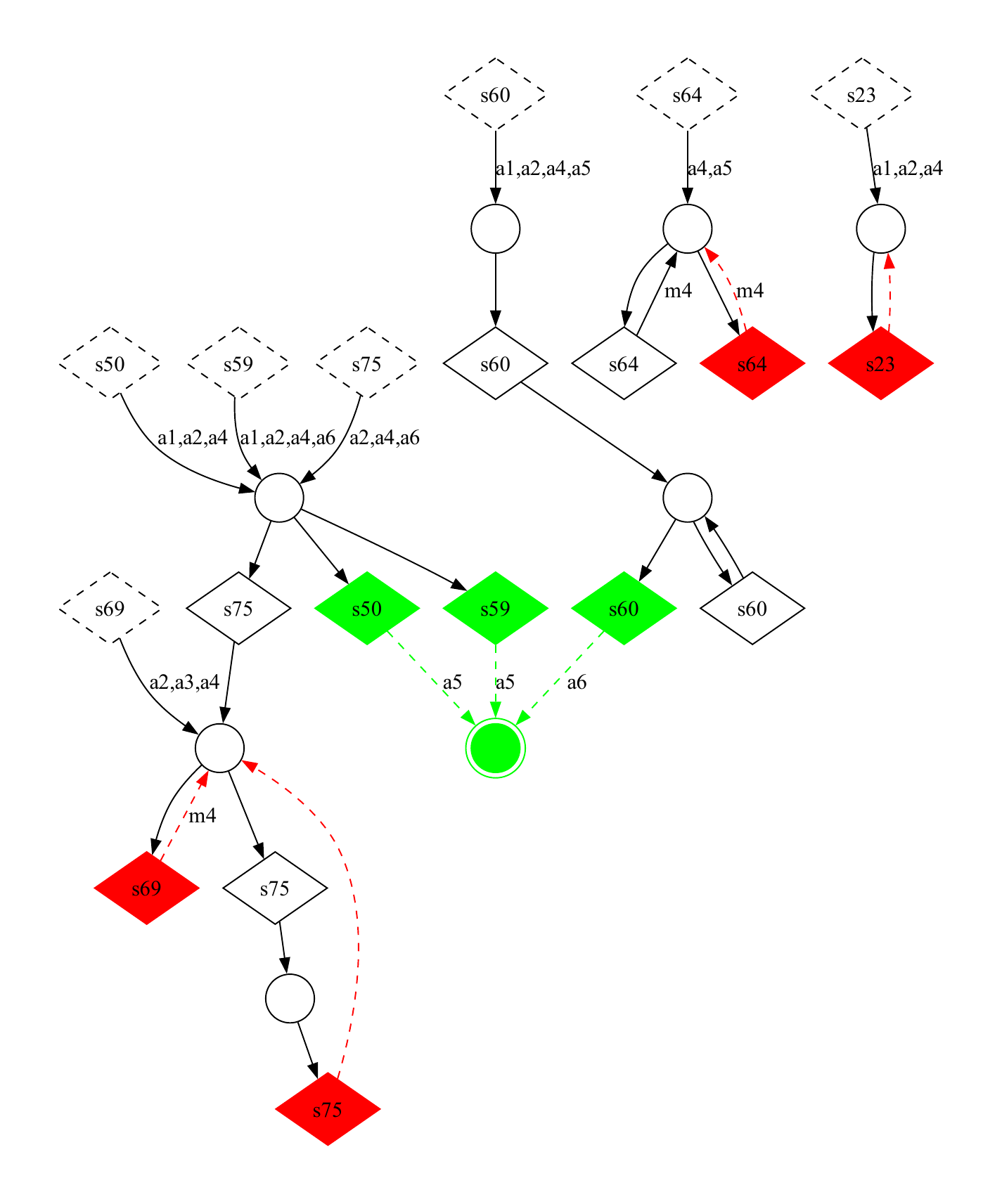}
\caption{Prompt trajectories for the ``translate'' problem.}
\label{fig:translate}
\end{figure*}

The prompt trajectories for all remaining problems are in \cref{fig:add_up}---\cref{fig:translate}.

\ifaclanonsubmission
\clearpage
\section{Checklist}

A2 Potential Risks: 
Did you discuss any potential risks of your work? [Yes/No/NA]
\textbf{No}

A2 Elaboration:
For yes, provide a section number. For no, justify why not.
\textbf{The risks of this work lie in the original dataset creation and collection, as described by \cite{nguyen_how_2024} and \cite{babe_studenteval_2024}.}

B Use Or Create Scientific Artifacts:
Did you use or create scientific artifacts? [Yes/No]
\textbf{Yes}

B1 Cite Creators Of Artifacts:
Did you cite the creators of artifacts you used? [Yes/No/NA]
\textbf{Yes}

B1 Elaboration:
For yes, provide a section number. For no, justify why not.
\textbf{\Cref{dataset}}

B2 Discuss The License For Artifacts:
Did you discuss the license or terms for use and/or distribution of any artifacts? [Yes/No/NA]
\textbf{Yes}

B2 Elaboration:
For yes, provide a section number. For no, justify why not. \textbf{\Cref{artifact-url-and-license}}

B3 Artifact Use Consistent With Intended Use:
Did you discuss if your use of existing artifact(s) was consistent with their intended use, provided that it was specified? For the artifacts you create, do you specify intended use and whether that is compatible with the original access conditions? [Yes/No/NA]
\textbf{Yes}

B3 Elaboration:
For yes, provide a section number. For no, justify why not.
\textbf{Ethics Statement}

B4 Data Contains Personally Identifying Info Or Offensive Content:
Did you discuss the steps taken to check whether the data that was collected/used contains any information that names or uniquely identifies individual people or offensive content, and the steps taken to protect/anonymize it? [Yes/No/NA]
\textbf{NA}

B4 Elaboration:
For yes, provide a section number. For no, justify why not.

B5 Documentation Of Artifacts:
Did you provide documentation of the artifacts, e.g., coverage of domains, languages, and linguistic phenomena, demographic groups represented, etc.? [Yes/No/NA]
\textbf{No}

B5 Elaboration:
For yes, provide a section number. For no, justify why not.
\textbf{This paper analyzes an existing dataset, which documents the artifact per ACL guidelines.}

B6 Statistics For Data:
Did you report relevant statistics like the number of examples, details of train/test/dev splits, etc. for the data that you used/created? [Yes/No/NA]
\textbf{Yes}

B6 Elaboration:
For yes, provide a section number. For no, justify why not.
\textbf{\cref{dataset}}

C Computational Experiments:
Did you run computational experiments? [Yes/No/NA]
\textbf{Yes}

C1 Model Size And Budget:
Did you report the number of parameters in the models used, the total computational budget (e.g., GPU hours), and computing infrastructure used? [Yes/No/NA]
\textbf{Yes}

C1 Elaboration:
For yes, provide a section number. For no, justify why not.
\textbf{\Cref{computing-resources}}

C2 Experimental Setup And Hyperparameters:
Did you discuss the experimental setup, including hyperparameter search and best-found hyperparameter values?
[Yes/No/NA]
\textbf{Yes}

C2 Elaboration:
For yes, provide a section number. For no, justify why not.
\textbf{\Cref{hyperparameters}}

C3 Descriptive Statistics:
Did you report descriptive statistics about your results (e.g., error bars around results, summary statistics from sets of experiments), and is it transparent whether you are reporting the max, mean, etc. or just a single run? [Yes/No/NA]
\textbf{Yes}

C3 Elaboration:
For yes, provide a section number. For no, justify why not.
\textbf{\Cref{sec:stats} and \Cref{app:stats}}

C4 Parameters For Packages:
If you used existing packages (e.g., for preprocessing, for normalization, or for evaluation, such as NLTK, SpaCy, ROUGE, etc.), did you report the implementation, model, and parameter settings used? [Yes/No/NA]
\textbf{Yes}

C4 Elaboration:
For yes, provide a section number. For no, justify why not.
\textbf{\Cref{packages}}

D Human Subjects Including Annotators:
Did you use human annotators (e.g., crowdworkers) or research with human subjects? [Yes/No/NA]
\textbf{No}

D1 Instructions Given To Participants:
Did you report the full text of instructions given to participants, including e.g., screenshots, disclaimers of any risks to participants or annotators, etc.? [Yes/No/NA]
\textbf{NA}

D1 Elaboration:
For yes, provide a section number. For no, justify why not.

D2 Recruitment And Payment:
Did you report information about how you recruited (e.g., crowdsourcing platform, students) and paid participants, and discuss if such payment is adequate given the participants' demographic (e.g., country of residence)? [Yes/No/NA]
\textbf{NA}

D2 Elaboration:
For yes, provide a section number. For no, justify why not.

D3 Data Consent:
Did you discuss whether and how consent was obtained from people whose data you're using/curating (e.g., did your instructions explain how the data would be used)? [Yes/No/NA]
\textbf{NA}

D3 Elaboration:
For yes, provide a section number. For no, justify why not.

D4 Ethics Review Board Approval:
Was the data collection protocol approved (or determined exempt) by an ethics review board? [Yes/No/NA]
\textbf{NA}

D4 Elaboration:
For yes, provide a section number. For no, justify why not.

D5 Characteristics Of Annotators:
Did you report the basic demographic and geographic characteristics of the annotator population that is the source of the data? [Yes/No/NA]
\textbf{NA}

D5 Elaboration:
For yes, provide a section number. For no, justify why not.

E Ai Assistants In Research Or Writing:
Did you use AI assistants (e.g., ChatGPT, Copilot) in your research, coding, or writing? [Yes/No]
\textbf{Yes}

E1 Information About Use Of AI Assistants:
Did you include information about your use of AI assistants? [Yes/No/NA]
\textbf{Yes}

E1 Elaboration:
For yes, provide a section number. For no, justify why not.
\textbf{\Cref{use-of-ai-assistants}}

\fi

\end{document}